\documentclass[twocolumn]{aastex63}

\usepackage{epstopdf}
\usepackage{multirow}
\usepackage{float}
\usepackage{amsmath}
\usepackage{graphicx}
\usepackage[T1]{fontenc}
\usepackage{apjfonts} 
\usepackage{lineno}

\newcommand{\editone}[1]{{#1}}
\definecolor{myblue}{RGB}{0,46,122}
\definecolor{myred}{RGB}{172, 42, 88}

\newcommand{\edittwo}[1]{{#1}}

\shorttitle{Size-Mass Relation for HSC Galaxies at $z<1$}
\shortauthors{Kawinwanichakij et al.}
\begin{document}

\title{\large \textbf{Hyper Suprime-Cam Subaru Strategic Program: A Mass-Dependent Slope of the Galaxy Size-Mass Relation at $z<1$\footnote{Released on \today}}}

\correspondingauthor{Lalitwadee Kawinwanichakij}
\email{lalitwadee.kawinwanichakij@ipmu.jp}
\author[0000-0003-4032-2445]{Lalitwadee Kawinwanichakij}\altaffiliation{LSSTC Data Science Fellow}
\affiliation{Kavli Institute for the Physics and Mathematics of the Universe, The University of Tokyo, Kashiwa, Japan 277-8583 (Kavli IPMU, WPI)}
\author{John D. Silverman}
\affiliation{Kavli Institute for the Physics and Mathematics of the Universe, The University of Tokyo, Kashiwa, Japan 277-8583 (Kavli IPMU, WPI)}
\affiliation{Department of Astronomy, School of Science, The University of Tokyo, 7-3-1 Hongo, Bunkyo, Tokyo 113-0033, Japan}
\author{Xuheng Ding}
\affiliation{Kavli Institute for the Physics and Mathematics of the Universe, The University of Tokyo, Kashiwa, Japan 277-8583 (Kavli IPMU, WPI)}
\author{Angelo George}
\affiliation{Institute for Computational Astrophysics and Department of Astronomy \& Physics, Saint Mary’s University, 923 Robie Street, Halifax,
Nova Scotia, B3H 3C3, Canada}
\author{Ivana Damjanov}\altaffiliation{Canada Research Chair}
\affiliation{Institute for Computational Astrophysics and Department of Astronomy \& Physics, Saint Mary’s University, 923 Robie Street, Halifax,
Nova Scotia, B3H 3C3, Canada}

\author{Marcin Sawicki}\altaffiliation{Canada Research Chair}
\affiliation{Institute for Computational Astrophysics and Department of Astronomy \& Physics, Saint Mary’s University, 923 Robie Street, Halifax,
Nova Scotia, B3H 3C3, Canada}

\author{Masayuki Tanaka}
\affiliation{National Astronomical Observatory of Japan, 2-21-1 Osawa, Mitaka, Tokyo 181-8588, Japan}
\affiliation{Department of Astronomy, School of Science, Graduate University for Advanced Studies
(SOKENDAI), 2-21-1, Osawa, Mitaka, Tokyo 181-8588, Japan}
\author{Dan S. Taranu}
\affiliation{Department of Astrophysical Sciences, Princeton University, 4 Ivy Lane, Princeton, NJ 08544, USA}
\author{Simon Birrer}
\affiliation{Kavli Institute for Particle Astrophysics and Cosmology and Department of Physics, Stanford University, Stanford, CA 94305, USA}
\author{Song Huang}
\affiliation{Department of Astronomy and Astrophysics, University of California Santa Cruz, 1156 High St., Santa Cruz, CA 95064, USA}
\author{Junyao Li}
\affiliation{CAS Key Laboratory for Research in Galaxies and Cosmology, Department of Astronomy, University of Science and Technology of
China, Hefei 230026, China}
\affiliation{School of Astronomy and Space Science, University of Science and Technology of China, Hefei 230026, China}
\affiliation{Kavli Institute for the Physics and Mathematics of the Universe, The University of Tokyo, Kashiwa, Japan 277-8583 (Kavli IPMU, WPI)}
\author{Masato Onodera}
\affiliation{Subaru Telescope, National Astronomical Observatory of Japan, National Institutes of Natural Sciences (NINS), 650 North A'ohoku Place, Hilo, HI 96720, USA}
\affiliation{Department of Astronomical Science, The Graduate University for Advanced Studies, SOKENDAI, 2-21-1 Osawa, Mitaka, Tokyo, 181-8588, Japan}
\author{Takatoshi Shibuya}
\affiliation{Kitami Institute of Technology, 165 Koen-cho, Kitami,
Hokkaido 090-8507, Japan}
\author{Naoki Yasuda}
\affiliation{Kavli Institute for the Physics and Mathematics of the Universe, The University of Tokyo, Kashiwa, Japan 277-8583 (Kavli IPMU, WPI)}


\begin{abstract} 
We present the galaxy size-mass ($R_{e}-M_{\ast}$) distributions using a stellar-mass complete sample of $\sim1.5$ million galaxies, covering $\sim100$ deg$^2$, with $\log(M_{\ast}/M_{\odot})>10.2~(9.2)$ over the redshift range $0.2<z<1.0$~$(z<0.6)$ from the second public data release of the Hyper Suprime-Cam Subaru Strategic Program. We confirm that, at fixed redshift and stellar mass over the range of $\log(M_{\ast}/M_{\odot})<11$, star-forming galaxies are on average larger than quiescent galaxies. The large sample of galaxies with accurate size measurements, thanks to the excellent imaging quality, also enables us to demonstrate that the $R_{e}-M_{\ast}$ relations of both populations have a form of broken power-law, with a clear change of slopes at a pivot stellar mass $M_{p}$. For quiescent galaxies, below an (evolving) pivot mass of $\log(M_{p}/M_{\odot})=10.2-10.6$ the relation follows $R_{e}\propto M_{\ast}^{0.1}$; above $M_{p}$  the relation is steeper and follows $R_{e}\propto M_{\ast}^{0.6-0.7}$. For star-forming galaxies,  below $\log(M_{p}/M_{\odot})\sim10.7$ the relation follows $R_{e}\propto M_{\ast}^{0.2}$; above $M_{p}$  the relation evolves with redshift and follows $R_{e}\propto M_{\ast}^{0.3-0.6}$. The shallow power-law slope for quiescent galaxies below $M_{p}$ indicates that \emph{large} low-mass quiescent galaxies have sizes similar to those of their counterpart star-forming galaxies. We take this as evidence that large low-mass quiescent galaxies have been recently quenched (presumably through environment-specific process) without significant structural transformation. Interestingly, the pivot stellar mass of the $R_{e}-M_{\ast}$ relations for both populations also coincides with mass at which half of the galaxy population is quiescent, implied that the pivot mass represents the transition of galaxy growth from being dominated by in-situ star formation to being dominated by (dry) mergers.

\end{abstract}


\section{Introduction} \label{sec:intro}
\par The size distribution and its evolution with cosmic time provide important clues about assembly history of galaxies and the relationships with the dark matter halos in which they reside \cite[e.g.,][]{Mo1998, Kravtsov2013,Shibuya2015,Huang2017,Somerville2018}. The qualitative description of galaxy formation in $\Lambda$CDM halos is based on the idea that galaxies are formed as a result of gas cooling in a gravitational potential dominated by dark matter and form a rotating disk \cite[e.g.,][]{White1978,White1991}. Assuming that gas initially has a specific angular momentum similar to that of the dark matter and that the angular momentum is conserved, 
\cite{Fall1980} and \cite{Mo1998} presented a simple model for the formation of galactic disks within a dark matter halo. According to this model, the size of a galaxy is controlled by specific angular momentum acquired by tidal torques during cosmological collapse by both baryonic and dark matter, leading to the prediction that the size of a galaxy forming at a given redshift should be scaled with the size of its parent dark matter halo. This prediction is in remarkable agreement with \cite{Kravtsov2013}, who used the abundance matching approach to estimate the virial radius of a nearby galaxy sample from SDSS and demonstrated that half-mass radii $r_{1/2}$ of galaxies are linearly related to virial radii through $r_{1/2}\propto\lambda R_{200}$, where $\lambda$ is the halo spin parameter \citep{Bullock2001}. Similar studies for galaxies at $z<3$ \citep{Huang2017,Somerville2018} and at higher redshift \citep[$0\lesssim z\lesssim6$;][]{Shibuya2015} confirmed the results of \cite{Kravtsov2013}, strongly suggesting that at all redshifts the sizes of disk-dominated/star-forming galaxies are shaped primarily by the properties of their parent dark matter halos.
\par The picture above is further complicated by the formation of stars and black holes, and the energy which they provide back into the surrounding gas. These physical processes are imprinted on the shape (i.e., amplitude, slope, and intrinsic scatter) of the size-mass relation and its evolution with cosmic time \cite[e.g.,][]{Dutton2009,Firmani2009,Shankar2013,Noguchi2018}. \cite{Dutton2009} used a $\Lambda$CDM-based disk-galaxy evolution model to investigate the impact of feedback processes on scaling relations of disk galaxies and found that a model without feedback either from supernova (SN; which primarily effects galaxies in low-mass halos) or active galactic nuclei (AGN; which primarily effects galaxies in high-mass halos) fails to reproduce the observed size-mass relation. Including these feedback mechanisms into their models reduces the efficiency of galaxy star formation and leads to the prediction that the slope of the size-mass relation depends on stellar mass. Specifically, the shallow slope of the relation for low-mass galaxies ($M_{\ast}\lesssim10^{10.5}~M_{\odot}$) favors the model with energy-driven SN feedback, whereas the relatively steeper slope for high-mass galaxies favors the model with AGN feedback.

\par Observationally, the sizes of galaxies are known to vary with galaxy mass, color, star-formation activity, AGN activity, and cosmic time \cite[e.g.,][among many others]{Shen2003, Ferguson2004,Trujillo2006,Elmegreen2007,Williams2010,Mosleh2013,Carollo2013,Ono2013,vanderWel2014,Lange2015,Shibuya2015,Allen2017,Whitaker2017,Mowla2019b,Silverman2019,Junyao2021,Yang2021}. These studies have consistently demonstrated that: (i) the sizes of more massive galaxies tend to be larger, (ii) early-type/quiescent galaxies tend be to be smaller than late-type/star-forming ones at fixed stellar mass, and (iii) galaxies are  smaller at higher redshift than they are today. 
 
\par In the local Universe,  previous studies have consistently shown that the size of late-type/disk-dominated galaxies increases with increasing stellar mass roughly as $R_{e}\propto M_{\ast}^{0.1-0.2}$ \cite[e.g.,][]{Shen2003,Dutton2011,Mosleh2013,Lange2015,Zhang2019}, broadly in agreement with the result for  star-forming galaxies at higher redshifts \cite[e.g.,][]{vanderWel2014,Faisst2017,Mowla2019b}. Besides, those studies, which utilize a sample of nearby galaxies, consistently found evidence that the slope of the size-mass relation for late-type galaxies deviates from a single-power law relation. For example, \cite{Shen2003} showed that the size-mass relation for late-type galaxies from SDSS is shallow with $R_{e}\propto M_{\ast}^{0.15}$ below a characteristic stellar mass $M_{0}\sim10^{10.6}~M_{\odot}$ and steep above $M_{0}$, with $R_{e}\propto M_{\ast}^{0.4}$. They showed that, in order to explain the observed $R_{e}-M_{\ast}$ relation for late-type galaxies, the feedback from star formation is needed to be incorporate into their models to suppress the fraction of baryons that can form stars, consistent with the results from \cite{Dutton2009}. At higher redshifts, \cite{vanderWel2014} used the combination of the 3D-\emph{HST} survey and CANDELS data and observed evidence for a deviation from a single power-law of the relation for massive late-type galaxies at $z\lesssim1$.

\par Besides growing in mass and size, star-forming galaxies cease to form stars at a critical point in their evolution and become quiescent galaxies. In the case of quiescent population, two classes of evolutionary processes may contribute to the observed evolution in their size-mass relation.
The first process is the growth of individual galaxies that are already in the quiescent population through minor mergers \cite[e.g.,][]{vanDokkum2010,Newman2012,Greene2012,  Greene2013,McLure2013, Whitney2019,Oyarzun2019}. In this process, quiescent galaxies can grow disproportionately more in size than in stellar mass, scaling as $R_{e}\propto M_{\ast}^{2}$ \citep{Naab2009,Bezanson2009,Hopkins2010, Patel2013, vanDokkum2015}. On the other hand, the second process involves the addition of more extended,  recently quenched galaxies to the pre-existing quiescent population at later times in the sense that the apparent growth in size of quiescent galaxies is driven by change in membership of the population rather than through the size growth of individual galaxies \cite[progenitor bias; e.g.,][]{Carollo2013,Poggianti2013, Belli2015,Fagioli2016,Lilly2016,Abramson2018}.



\par Important clues regarding the evolutionary process that dominates the average size growth of quiescent galaxies are contained in the mass-dependent slope of the size-mass relation for quiescent galaxies. Low-mass quiescent galaxies exhibit a much shallower slope, $R_{e}\propto M_{\ast}^{0.1-0.2}$, than the slope for massive quiescent galaxies, $R_{e}\propto M_{\ast}^{0.5}$.
This result has been reported using samples from both the local Universe \citep{Shen2003,Bernardi2011a,Bernardi2011b,Mosleh2013,Cappellari2013,Bernardi2014,Norris2014,Lange2015,Zhang2019} and at intermediate redshift  \cite[$z\lesssim3$;][]{vanderWel2014,Whitaker2017,Huang2017,Mosleh2020,Nedkova2021}. For quiescent population at least out to $z\sim1.5$, \cite{vanderWel2014} reported evidence for deviation from a single power-law of the size-mass relation at $M_{\ast}\sim 2\times10^{10}~M_{\ast}$, where the distribution flattens below this stellar mass. They also found that the size evolution of low-mass quiescent galaxies with stellar mass below $M_{\ast}\lesssim10^{10}~M_{\ast}$ is more comparable to that of star-forming galaxies at the same stellar mass, suggesting that there is a population of low-mass quiescent galaxies that may have been formed out of star-forming galaxies without significantly transforming their morphology. 

 \par Recently, \cite{Mowla2019a} analyzed the size-mass relation of all (both quiescent and star-forming) galaxies out to $z\sim3$ using a galaxy sample from 3D-\emph{HST}, CANDELS, and COSMOS-DASH surveys and found that the relation at $z\sim0.8$ has different slopes in different mass regimes split at a pivot stellar mass of  $\log(M_{\ast}/M_{\odot})\sim10.5$, roughly corresponding to halo mass of $\log(M_{\mathrm{halo}}/M_{\odot})\sim12$. The authors found that, at fixed redshift, the pivot stellar mass of the size-mass relation coincides with that of the stellar-to-halo mass relation (SMHM; $M_{\mathrm{halo}}-M_{\ast}$), which theoretically marks the mass at which the conversion from baryons to stars become maximally efficient \cite[e.g.,][]{Behroozi2010,Leauthaud2012,Moster2013,Rodriguez-Puebla2017}. \cite{Mowla2019a} reached the conclusion that the pivot mass marks the mass above which both the stellar mass growth and the size growth of galaxies transition from being star-formation dominated to dry mergers dominated.
\par Ultimately, it could be that both the individual size growth of quiescent galaxies and progenitor bias are at work, depending on galaxy stellar mass and redshift \citep{Belli2015,Matharu2020,Damjanov2019,DiazGarcia2019}. Numerical simulations \citep[e.g.,][]{Oser2010,Wellons2015}, semi-analytic models \citep[SAM; e.g.,][]{Lee2013,Lee2017}, and subhalo abundance matching (SHAM) analyses \citep[e.g.,][]{Moster2013,Moster2018,Behroozi2013,Rodriguez2017}  generally show that the fraction of stars accreted through mergers increases with total galaxy stellar mass or halo mass \citep[e.g.,][]{Lackner2012,Cooper2013,Rodriguez2016,Qu2017,Pillepich2018,Davison2020}. This finding implies that the contribution of progenitor bias to the observed size growth must increase with decreasing stellar mass, consistent with \cite{Damjanov2019} and \cite{Belli2015}. \cite{Damjanov2019} studied the relation between average size and $D_{n}4000$ index of a spectroscopic sample at $0.1<z<0.6$ from the SHELS spectroscopic survey and found that the observed size growth of low-mass ($\log(M_{\ast}/M_{\odot})\sim10$) quiescent galaxies at this redshift is mainly a result of progenitor bias. On the other hand, for  more massive ($\log(M_{\ast}/M_{\odot})\sim10.7$) quiescent galaxies, the contributions from (predominantly minor) mergers and progenitor bias to the size growth are roughly comparable.

\par Clearly, one needs large samples of galaxies with accurate photometry and excellent imaging quality to measure accurate galaxy sizes and examine size-mass distributions over a broad range of stellar mass and redshift. The Hyper Supreme-Cam Subaru Strategic Program \citep[HSC-SSP or HSC;][]{Aihara2018,Aihara2019} with its exquisite image quality and good photometry in five bands for $\sim10^{7}$ galaxies, is providing an unprecedented database for such studies.
\par Here, we present the galaxy size-mass distribution over the redshift range  $0.2<z<1.0$ using an unprecedented sample of over $1,500,000$ galaxies with $i<24.5$ mag (corresponding to stellar mass  of $\log(M_{\ast}/M_{\odot})>10$ at $z=1$) from the second public data release (PDR2) of the HSC. The goals are to determine 1) the distribution of sizes as a function of stellar mass and 2) the redshift evolution of galaxy sizes over the redshift range $0.2<z<1.0$. These measurements provide clues to the evolutionary processes that drive the observed evolution in the size-mass relation. The plan for this paper is as follows. In Section~\ref{sec:dataset} we provide an overview of the HSC-SSP, describing the data we used for this work, the details of our sample selection, and the separation of quiescent galaxies from the star-forming galaxies. In Section~\ref{sec:lenstronomy} we present a detailed description of the size measurement of galaxies and the derivation of correction for measurement biases using simulated galaxies. Section~\ref{sec:fitting_sizemass} provides a detailed description of the analytic fits to the size-mass distribution and defines the use of Bayes factor to quantify the evidence that the size-mass relations of quiescent and star-forming galaxies have the form of broken power-law. We present our main results of the size-mass distribution and its evolution with cosmic time in Section~\ref{sec:results}. In Section~\ref{sec:discussion} we compare our results with previous studies and also discuss the implications of our findings. Finally, we provide a summary in Section~\ref{sec:conclusion}. Throughout this paper, we use the AB magnitude system \citep{Oke1983} and adopt a standard cosmology with $\Omega_{M} =0.3$, $\Omega_{\Lambda} =0.7$, and $h=0.7$ (where $H_{0}=100~h~\mathrm{km}~\mathrm{s}^{-1}~\mathrm{Mpc}^{-1}$), which is  consistent with the local distance scale of \cite{Riess2019}.
\section{Data and Sample Selection}
\label{sec:dataset}
\subsection{HSC Public Data Release 2}
In this paper we use the HSC PDR2\footnote{ https://hsc-release.mtk.nao.ac.jp/}, which is based on the prime-focus camera, Hyper Suprime-Cam \citep{Miyazaki2012,Miyazaki2018, Komiyama2018,Furusawa2018} on the 8.2-m Subaru telescope. PDR2 contains imaging data from 174 nights of observation from March 2014 through January 2018. The survey consists of three layers: Wide, Deep, and UltraDeep. The Wide layer of PDR2 covers 300 deg$^2$ in five broad-band filters  \citep[$grizy$;][]{Kawanomoto2018} down to about 26 AB mag \citep[at 5$\sigma$ for point sources;][]{Aihara2019}. The Wide layer has eight separate fields: WIDE01H, XMM-LSS, GAMA09H, WIDE12H, GAMA15H, VVDS, HECTOMAP, and AEGIS.  The Deep layer contains four separate fields: XMM-LSS, COSMOS, ELAIS-N1, and DEEP2-F3 \editone{and reaches a depth of $\sim26.5$ mag ($5\sigma$ point sources). The UltraDeep layer contains two separate fields: COSMOS and the Subaru/XMM-Newton Deep Survey (SXDS) and reaches a depth of $i\sim$ 28 AB mag}. In total, the Deep and UltraDeep layers cover $32$~deg$^2$ and contain a wealth of ancillary data (e.g., CLAUDS; \citealt{Sawicki2019}, VIDEO; \citealt{Jarvis2013}) that make them the best deep- and wide-fields for studying faint galaxies.
The HSC data are reduced with the HSC Pipeline, hscPipe \citep{Bosch2018}, which is based on the Large Synoptic Survey Telescope pipeline \citep{Ivezic2019, Axelrod2010,Juric2017}. The Pan-STARRS1 data are also used for astrometric and photometric calibrations \citep{Tonry2012, Schlafly2012,Magnier2013,Chambers2016}.


\par We use the HSC database to construct a galaxy sample for measuring galaxy structural parameters. For this paper, in addition to the Deep and UltraDeep footprints, we will focus on $\sim70$ deg$^2$ of the GAMA09h field of the Wide layer overlapping with the footprint of the eROSITA \citep{Cappelluti2011,Merloni2012,Predehl2018} Final Equatorial-Depth Survey (eFEDS). In a companion paper Kawinwanichakij et al. (in prep.), we will  combine the HSC structural parameter measurements  presented in this study with the observable properties of galaxy clusters from the eFEDS, such as X-ray luminosity and temperature, serving as mass proxy for clusters, and explore the environmental effects on galaxy structural parameters.
\par To select galaxies that were observed in all five broad-bands, we impose cuts in the number of visits for each object using the {\fontfamily{lmtt}\selectfont
countinputs} parameter, which indicates the number of images used to create a coadd image for each galaxy. Specifically, we set {\fontfamily{lmtt}\selectfont inputcount\_value} $\geq4$ for $gr-$bands and {\fontfamily{lmtt}\selectfont inputcount\_value} $\geq6$ for $izy-$bands. We only use galaxies with  $i-$band cModel magnitude brighter than $i=24.5$ AB mag with the corresponding error smaller than 0.1 AB mag, and $i-$band star-galaxy separation parameter {\fontfamily{lmtt}\selectfont extendedness\_value} $=1$. 
\edittwo{However, we found that the star-galaxy separation using the HSC extendedness flag does not work well for sources fainter than $i\sim23$ mag.} We additionally adopt the SDSS point source classification scheme\footnote{https://www.sdss.org/dr12/algorithms/magnitudes/}: an  object is classified as a  point source if the difference between its cModel and PSF magnitudes is less than 0.145 \hbox{(\textsc{psfMag}$-$\textsc{cmodelMag}$\leq0.145$)}\footnote{\textsc{psfMag} and \textsc{cmodelMag} are the magnitudes measured by fitting a PSF model and a linear combination of de Vaucouleurs and exponential models, respectively, for an object's light profile.}; otherwise, it is classified as an extended source.  \edittwo{We have visually inspected a subset of our sample and verified that the additional criteria from SDSS are effectively removing the majority of stars from our sample.} We use the $i-$band for the selection magnitude because $i-$band images are on average taken in better seeing condition with a median seeing size (FWHM) of $\sim0{\farcs}6$ compared to the other bands. This allows us to measure accurate galaxy structural parameters.
\par Additionally, we remove galaxies that can be affected by bad pixels or have poor photometric measurements, by rejecting objects with any of the following flags in any of the five broad-bands: \hbox{{\fontfamily{lmtt}\selectfont pixelflags\_edge}}, {\fontfamily{lmtt}\selectfont cmodel\_flag}, and {\fontfamily{lmtt}\selectfont sdsscentroid}{\fontfamily{lmtt}\selectfont \_flag} \citep[see][]{Aihara2018}. We then apply masks to pixels around bright stars according to the procedure of \cite{Coupon2018} by setting
{\fontfamily{lmtt}\selectfont bright\_objectcenter} to be false in any of the five broad-bands.
\subsection{Photometric redshift estimates}
\label{sec:zphot}
The photometric redshift was estimated using the HSC five-band photometry ($grizy$) with six independent codes, described in detail in \cite{Tanaka2018}. \edittwo{In this paper, we use the photometric redshifts (hereafter photo-$z$, $z_{\mathrm{phot}}$, or $z$) and stellar masses from the second data release of the HSC photo$-z$ catalog \citep{Nishizawa2020}}, which are estimated using the Bayesian template fitting-code \textsc{Mizuki} \citep{Tanaka2015}. In brief, \textsc{Mizuki} computes photo-$z$'s for objects with clean \textsc{cModel} photometry in at least 3 bands (inclusive) based on the spectral energy distribution (SED) fitting. The code uses a set of templates generated with \cite{Bruzual2003} stellar population synthesis (SPS) code assuming the \cite{Chabrier2003} initial mass function (IMF), exponentially declining star formation histories with timescale $\tau$, solar metallicity, and dust attenuation law as described in \cite{Calzetti2000}. Emission lines are added to the templates assuming solar metallicity \citep{Inoue2011}. We adopt \textsc{Mizuki} $z_{\mathrm{best}}$, defined as the redshift estimator that minimizes the loss function assuming a Lorentzian kernel in $\Delta z/(1+z)$ with a width of $\sigma=0.15$ \cite[see][for additional detail]{Tanaka2018}, as our estimated photo-$z$. To select galaxies with reliable photo-$z$ and stellar mass, we require galaxies with the reduced chi-squares of the best-fit model $\chi^{2}_{\nu}<5$. We note that the fraction of galaxies with $\chi^{2}_{\nu}>5$ is only about $3\%$. We follow \cite{Speagle2019} by imposing an additional cut of $z_{\mathrm{risk}}<0.25$ to remove sources whose photo-$z$ PDFs are overly broad or have multiple peaks with several possible redshift solutions. This cut removed $\sim20\%$ of sources from the total sample, and we have verified that this cut does not affect preferentially low-mass and/or high redshift galaxies. In section~\ref{sec:lenstronomy}, we will further exclude galaxies which have catastrophically failed surface brightness profile fits.
\par \editone{Comparing with spectroscopic redshifts (hereafter spec-$z$ or $z_{\mathrm{spec}}$), we find a photo-$z$ uncertainty of our galaxy sample at $0.2<z<1.0$ and $i<24.5$ is $\sigma_{z/(1+z)}=0.06$ with $\sim7\%$ of the sample being catastrophic redshift failures (Appendix~\ref{appendix:photoz_test}). Over the same  range of redshift and magnitude, the photometric redshifts of quiescent galaxies and star-forming galaxies are equally well constrained with the uncertainties of $\sigma_{z/(1+z)}=0.02$ and $0.04$, respectively (also see Figure~\ref{fig:zspec_zphot})}. \edittwo{These photo$-z$ uncertainties, estimated using our galaxy sample at $0.2<z<1.0$ from the HSC PDR2,  are consistent with those of \cite{Tanaka2018}, who presented the photo-$z$'s and an assessment of their uncertainty for the HSC PDR1 sample. The authors demonstrated that the HSC photo-$z$'s are most accurate at $0.2\lesssim z_{\mathrm{phot}} \lesssim1.5$, where the HSC filter bands can straddle the 4000~\AA~break. \cite{Tanaka2018} further showed that the photo-$z$ accuracy depends on survey depth: the photo-$z$ accuracy at the HSC UltraDeep depth is improved by $\sim40\%$, compared to that at the Wide depth. In this work}, we quantified the effect of catastrophic redshift failures on the size evolution and found that the scatter of intrinsically large galaxies from low redshift bins to high redshift bins due to the catastrophic redshift failures leads to $\lesssim0.05$ dex overestimation of galaxy sizes at fixed redshift. This effect is more significant toward more massive star-forming galaxies. In Appendix~\ref{appendix:photoz_test}, we derive the fraction of catastrophic redshift failures, $\eta$, as a function of galaxy magnitude. Throughout this paper, we account for the catastrophic redshift failures by incorporating the fraction of catastrophic redshift failures for quiescent galaxies ($\eta_{\mathrm{Q}}$) and for star-forming galaxies ($\eta_{\mathrm{SF}}$) into the estimation of likelihood for each population when we fit the size-mass distribution and its redshift evolution (Section~\ref{sec:fitting_sizemass}). 

\begin{figure*}
	\centering
	\includegraphics[width=1\textwidth]{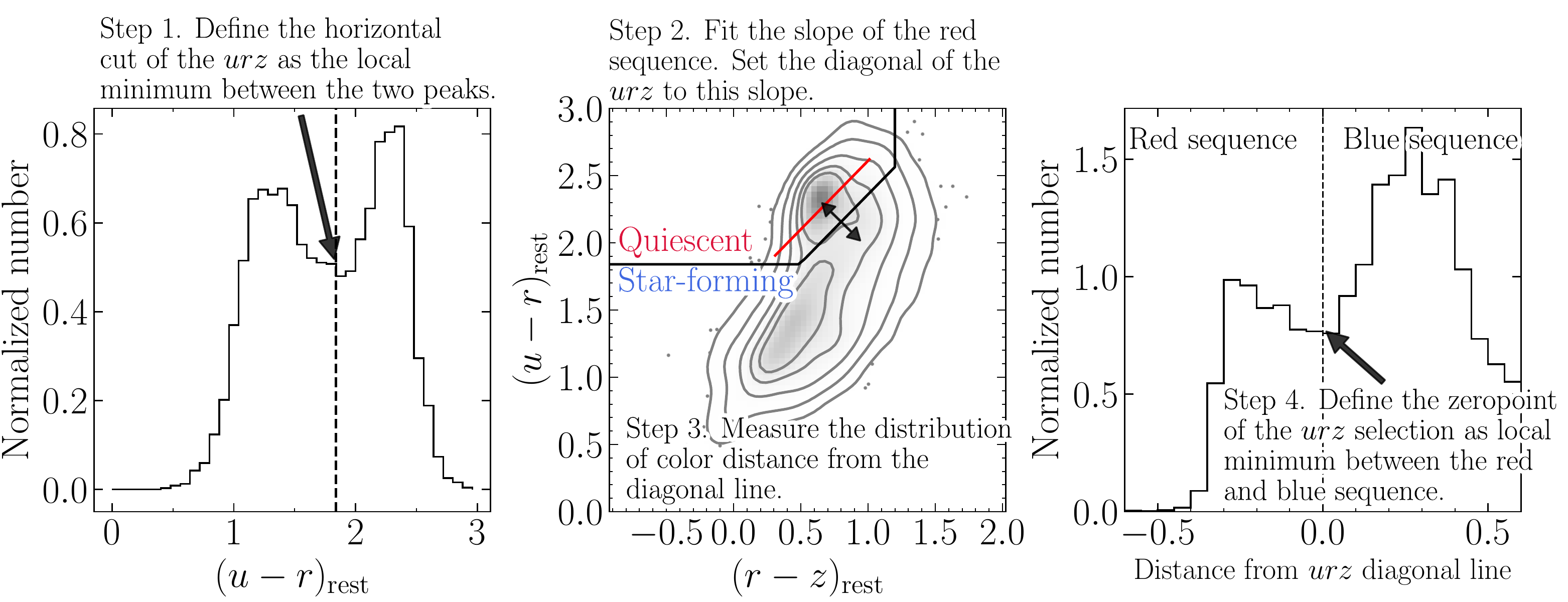}
	\caption{Demonstration of the method for self-calibrating the region delineating  the colors of HSC quiescent and star-forming galaxies. For the purpose of the calibration, we use the galaxy sample in \edittwo{COSMOS, SXDS (the Deep+UltraDeep layer) and AEGIS (the Wide layer) fields from the UltraVISTA, UKIDSS UDS, and NMBS catalogs}, which have stellar mass above the mass completeness limit at a give redshift redshift. \edittwo{In practice, we perform the calibration for galaxies in each redshift bin. In this figure we show one redshift bin ($0.6<z<0.8$) for illustration only.} Left: the distribution of rest-frame SDSS $u-r$ color. We define the horizon cut  $(u-r)^{\prime}$ of the $urz$ boundary line as the local minimum between the two peaks of the distribution, indicated by vertical dashed line. Middle: the rest-frame $u-r$ vs. $r-z$ color ($urz$ diagram). The grey contours indicate the distribution of HSC galaxies from $1\sigma$ to $3\sigma$ (with a spacing of $0.5\sigma$). The red solid line indicates the best-fitted red sequence galaxies. The galaxies in the upper left region of the plot (separated by the solid line) are quiescent. Right: the distribution of the distance (in color) from the diagonal line in $urz$ color (the slope $A$; see Equation~\ref{eq:urz_generic}) that separate the quiescent and star-forming sequence in the $urz$ color space. We define the zeropoint ($ZP$) of the $urz$ quiescent region as the local minimum in this distribution, indicated by vertical dashed line. }
	\label{fig:urz_calib}
\end{figure*}

\begin{figure}
	\centering
	\includegraphics[width=0.48\textwidth]{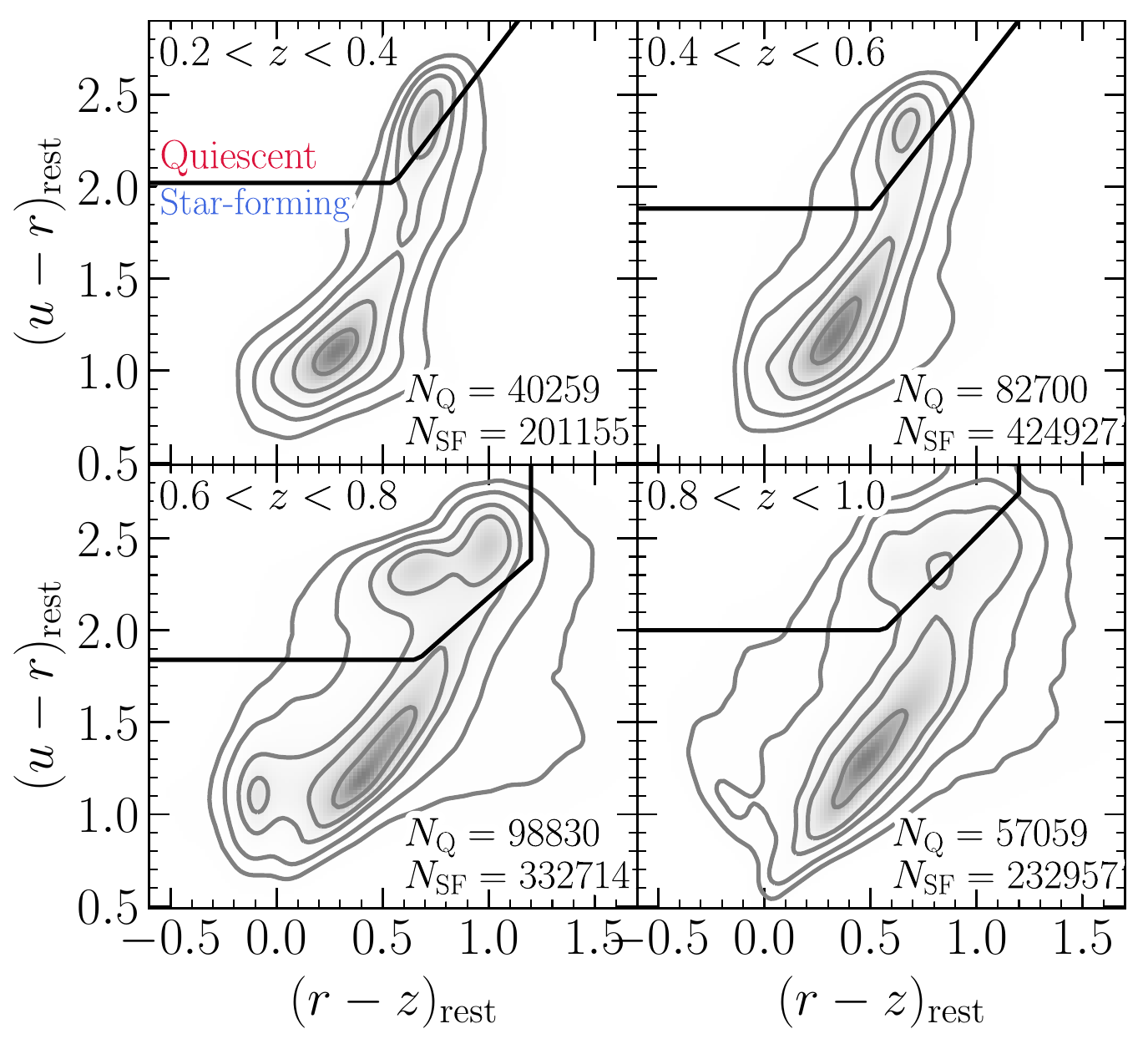}
	\caption{Rest-frame $u-r$ vs. $r-z$ relation for the HSC galaxies in four bins of redshift : $0.2<z<0.4$, $0.4<z<0.6$, $0.6<z<0.8$, and $0.8<z<1.0$.  The grey contours indicate the  distribution of all galaxies in a given redshift bin from $0.5\sigma$ to $2.5\sigma$ (with a spacing of $0.5\sigma$). The galaxies in the upper left region of the plot (separated by the thick solid line) are quiescent; galaxies outside this region are star-forming galaxies. In each redshift bin, we show the number of quiescent ($N_{\mathrm{Q}}$) and star-forming ($N_{\mathrm{SF}}$) galaxies above the stellar mass completeness limits of quiescent  ($\log(M_{\mathrm{lim,Q}}/M_{\odot})=9.0,9.2,9.8,10.2$) and star-forming galaxies ($\log(M_{\mathrm{lim,SF}}/M_{\odot})=8.5,8.9,9.3,9.7$).}
	\label{fig:urz_selection}
\end{figure}

\subsection{Rest-frame Colors and Selection of Quiescent and Star-forming Galaxies}
\label{sec:urz_selection}

\par Previous classification of quiescent and star-forming populations is commonly based on the rest-frame $U-V$ versus $V-J$ color-color diagram \citep[hereafter $UVJ$ selection; e.g.,][]{Labbe2005,Wuyts2007,Williams2009,Whitaker2011}. However, the SED fitting using our HSC five-band photometry does not allow us to obtain a robust estimate of the rest-frame $J$ magnitudes. As a result, we are unable to use $UVJ$ diagram  to distinguish between quiescent and star-forming galaxies. 
\par Instead, we classify galaxies as either quiescent or star-forming based on the rest-frame SDSS $u-r$ versus $r-z$ color-color diagram (hereafter $urz$ diagram), which is considered as an effective way for separating quiescent from star-forming galaxies similar to the $UVJ$ selection \citep[e.g.,][]{Holden2012,Robaina2012,Chang2015,Lopes2016}. \editone{We use the publicly available software package EAZY \citep{Brammer2008} to derive rest-frame colors.}  Specifically, with HSC \textsc{cModel} photometry ($grizy$), we run EAZY by fixing redshift of each galaxy to its photo-$z$ estimated from \textsc{Mizuki}. \edittwo{We then calculate the rest-frame colors ($u-r$ and $r-z$) by integrating the best-fit template SED across the redshifted rest-frame filter bandpass for each individual source \citep{Brammer2011,Whitaker2011}.}

\par To compare the $urz$ and $UVJ$ selection, we utilize the rest-frame $U-V$ and $V-J$ colors from a public K$_{s}$-selected catalog in COSMOS/UltraVISTA field \citep{Muzzin2013}, the $K$-selected catalog for the UKIRT Infrared Deep Sky Survey \citep[UKIDSS;][]{Lawrence2007}, Ultra-Deep Survey (UDS, Almaini et al., in preparation), \edittwo{and a public $K$-selected catalog for the NEWFIRM Medium-band Survey \citep[NMBS;][]{Whitaker2011}}. These datasets overlap with the HSC footprint in the COSMOS and SXDS fields (the Deep+UltraDeep layer) and \edittwo{AEGIS field (the Wide layer).} Based on this comparison, we find that applying the $urz$ selection criteria from \cite{Holden2012} directly to our HSC dataset does not optimally distinguish the quiescent from the star-forming population -- we would miss a large ($\sim75\%$) fraction of the $UVJ$ quiescent galaxies using their $urz$ selection. This could arise from systematic variations in the photometry and hence the rest-frame colors of galaxies at fixed mass and redshift in different surveys. 
\par To account for that, we follow \cite{Kawinwanichakij2016}  by implementing a method to self-calibrate the region delineating the colors of HSC quiescent and star-forming galaxies in the $urz$ color-color space. For the purpose of calibration, we again use the galaxy sample in COSMOS, SXDS, \edittwo{and AEGIS} fields with rest-frame $U-V$ and $V-J$ colors from the UltraVISTA, UKIDSS UDS, and \edittwo{NMBS} catalogs, respectively. This will allow us to test how well we can recover $UVJ$ quiescent and $UVJ$ star-forming galaxies using the $urz$ selection. We begin with defining a generic region of the $urz$ diagram for quiescent galaxies as
\begin{align}
   u-r &> A\times(r-z)+ZP \nonumber \\
    u-r &> (u-r)^{\prime} \nonumber \\
    r-z &< 1.2
    \label{eq:urz_generic}
\end{align}
\noindent where $A$, $ZP$, and $(u-r)^{\prime}$ are variables we derived as follows. We divide the galaxy sample in COSMOS, SXDS, and \edittwo{AEGIS} fields into four redshift bins we use in this study: $0.2<z<0.4$, $0.4<z<0.6$, $0.6<z<0.8$, and $0.8<z<1.0$.  \editone{For galaxies in each redshift bin with stellar masses above the completeness limit at a given redshift ($\log(M_{\ast}/M_{\odot})=8.5,8.9,9.3,9.7$; see Section~\ref{sec:derive_masscomplete}), we derive the parameters of the $urz$ selection.} We measure the horizontal cut, $(u-r)^{\prime}$, as the local minimum between the two peaks in the distribution of $u-r$ color, and we find $(u-r)^{\prime}$ of ($2.02,1.88,1.84,2.00$). Next, we fit for $A$ as the slope of the red sequence in the $urz$ plane, finding slopes of ($1.51,1.47,1.02,1.17$). Third, we measure the distribution of the distance in $urz$ color from the diagonal line defined by the slope $A$ in Equation~\ref{eq:urz_generic} (where the ``color distance'' is the distance in $urz$ color from the line). We measure the zeropoint $ZP$ as the local minimum between the two peaks in the $urz$ color distribution. We find zeropoints of ($1.18,1.14,1.34,1.16$). \edittwo{Figure~\ref{fig:urz_calib} shows a demonstration of this method for galaxy sample at $0.6<z<0.8$.} 

\par We further compare our quiescent/star-forming galaxies selection using the $urz$ diagram 
with that using the $UVJ$ diagram. \editone{We note that, even though $UVJ$ selection has been demonstrated to be a reliable method for separating quiescent galaxies from star-forming galaxies, a level of cross-contamination, particularly from dusty star-forming galaxies, is still present \citep[e.g.][]{Williams2009,Moresco2013,Fumagalli2014,Dominguez2016,DiazGarcia2019II,Steinhardt2020}. Here we focus on the differences between the $UVJ$ and the $urz$ selections to ensure consistency with previous studies utilizing the $UVJ$ selection \citep[e.g.][]{vanderWel2014,Mowla2019b}.} To do so, for each set of parameters describing the $urz$ selection boundary (Equation~\ref{eq:urz_generic}), we calculate \edittwo{the fraction of $urz$ quiescent galaxies that are classified as $UVJ$ star-forming galaxies (using the rest-frame $U-V$ and $V-J$ colors from the UltraVISTA, UKIDSS UDS, and \edittwo{NMBS} catalogs), which will be referred to as the contaminating fraction of ($urz$) quiescent galaxies ($f_{\mathrm{cont,Q}}$). We also calculate the fraction of the $UVJ$ quiescent galaxies that are classified as $urz$ quiescent galaxies, which will be referred to as the recovering fraction of quiescent galaxies ($f_{\mathrm{recov,Q}}$).} This comparison shows that, for quiescent galaxies with stellar masses above the completeness limit at a given redshift, the contaminating fraction of $urz$-selected quiescent galaxies is $f_{\mathrm{cont,Q}}\sim10\%-20\%$, while the recovering fraction of the $urz-$selected quiescent galaxies is  $f_{\mathrm{recov,Q}}\sim60\%-90\%$ over the redshift range of $0.2<z<1.0$. In Appendix~\ref{appendix:misclass}, we show the location of our quiescent and star-forming galaxies in each redshift bin (Figure~\ref{fig:urz_selection_vs_uvj} and~\ref{fig:urz_selection_vs_uvj2}) on the $UVJ$ diagram and the $urz$ diagram. 


\par Figure~\ref{fig:urz_selection} shows the $urz$ diagram for the HSC galaxy sample in each redshift bin from $z=0.2$ to $z=1.0$ and our calibrated $urz$ selection from Figure~\ref{fig:urz_calib}. Throughout this paper, we mitigate the cross-contamination between quiescent and star-forming populations by incorporating the contaminating fraction for quiescent ($f_{\mathrm{cont,Q}}$) and star-forming ($f_{\mathrm{cont,SF}}$) populations into the estimation of likelihood for each population when we model the size-mass distributions and the redshift evolution of median size (Section~\ref{sec:fitting_sizemass}). We refer the reader to Appendix~\ref{appendix:misclass} for more detail on the derivation of both $f_{\mathrm{cont,Q}}$ and $f_{\mathrm{cont,SF}}$ as a function of stellar mass and redshift. In Section~\ref{subsec:mediansize_evol}, we will demonstrate that our fitting method with cross-contamination correction allows us to effectively recover the size-mass relation and its evolution, at least for the sample with stellar mass above the completeness limit. 

\begin{figure}
	\centering
	\includegraphics[width=0.47\textwidth]{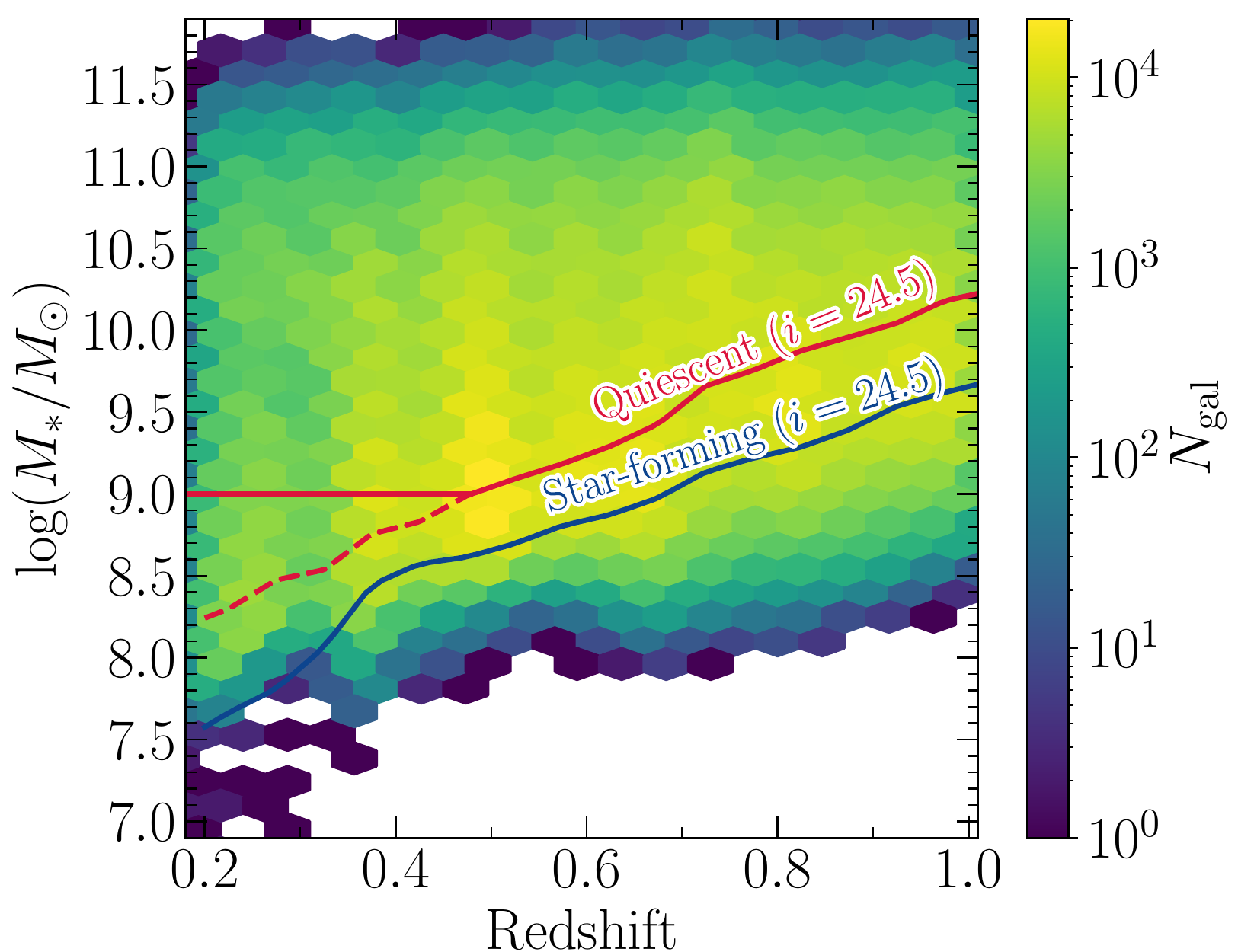}
	\caption{Distribution of stellar masses  as function of redshift for our $i$-band selected sample. The colorscale indicates the density in each bin of stellar mass and redshift. Following the approach by \cite{Pozzetti2010} and \cite{Weigel2016}, we estimate $90\%$ stellar mass completeness limit ($M_{\ast \mathrm{lim}}(z)$) for quiescent (red curves) and for star-forming (blue curves) galaxies assuming the limiting magnitude of $i_{\mathrm{lim}}=24.5$ for both HSC Wide and Deep+UltraDeep layers. For quiescent galaxies at $z<0.4$, we adopt $\log(M_{\ast \mathrm{lim}}(z)/M_{\odot})=9$ to minimize the contamination from low-mass star-forming galaxies at this redshift range (see text).}
	\label{fig:masscomp}

\end{figure}
\begin{deluxetable*}{ccccccc}
\tablecaption{Summary of the total number of galaxies sample in each survey layer and redshift bin}
\label{table:numbersample}
\tablehead{ \multirow{2}{*}{Layer} & \multirow{2}{*}{Area (deg$^2$)}  &  \multirow{2}{*}{$z_{\mathrm{med}}$} & \multicolumn{2}{c}{$\log(M_{\mathrm{lim}}/M_{\odot})$} & \multicolumn{2}{c}{$N_{\mathrm{obj}}$}   \\
& & & \colhead{Quiescent} & \colhead{Star-forming} & \colhead{Quiescent} & \colhead{Star-forming}}
\colnumbers
\startdata
  & \multirow{4}{*}{67.7} & $0.3$ & $9.0$ & $8.5$ & $25255$ & $116577$ \\
Wide & &  $0.5$ & $9.2$ & $8.9$ & $53719$& $258438$  \\
(GAMA09h)& &  $0.7$ & $9.8$ & $9.3$ & $37977$ & $224624$ \\
& &  $0.9$ & $10.2$ & $9.7$ & $31607$ & $122935$ \\
\hline
\multirow{4}{*}{Deep+UltraDeep}  & \multirow{4}{*}{32.2} & $0.3$ & $9.0$ & $8.5$ & $15004$ & $84578$  \\
&  & $0.5$ & $9.2$ & $8.9$ & $28995$ & $169948$ \\
&  & $0.7$ & $9.8$ & $9.3$ & $23509$ & $151149$ \\
&  & $0.9$  & $10.2$ & $9.7$ & $25452$ & $110022$\\
\hline
\enddata
\tablecomments{(1) the HSC survey layer, (2) the effective area, (3) the median redshift ($z_{\mathrm{med}}$) of a subsample, (4) the 90\% stellar mass completeness limit for quiescent galaxies, which is set by $i_{\mathrm{lim}}=24.5$, where we can determine galaxy sizes with high fidelity for both HSC survey layers (see Appendix~\ref{appendix:verification_sizes}),  (5) the 90\% stellar mass completeness limit for star-forming galaxies, (6) the number of quiescent galaxies ($N_{\mathrm{obj}}$) with stellar mass above the mass completeness for quiescent sample, and (7) the number of star-forming galaxies with stellar mass above the mass completeness for star-forming sample.}
\end{deluxetable*}
\subsection{Stellar masses estimates}
\label{sec:massestimate}
\subsubsection{Derivation of Stellar Mass Estimates}
\par In addition to photometric redshift, the \textsc{Mizuki} catalog provides stellar mass estimates determined from fitting the galaxy SEDs with five optical HSC bands ($grizy$). The assumptions on IMF, SPS model, attenuation curve, and star formation history are the same as those for photo-$z$ estimates. Further details on the SPS templates are provided in \cite{Tanaka2015}. In brief, SPS templates are generated for ages between 0.05 and 14 Gyr with a logarithmic grid of $\sim0.05$~dex. In addition to the single stellar population model (i.e., $\tau=0$) and constant SFR model (i.e., $\tau=\infty$), \edittwo{ \cite{Tanaka2015} assume} $\tau$ in the range of $0.1~\mathrm{Gyr} < \tau < 11~\mathrm{Gyr}$ with a logarithmic grid of 0.2 dex and optical depth in the $V$ band in the range of $0<\tau_{V}<2$ with an addition of $\tau_{V}=2.5, 3,4,5$ models to cover very dusty sources. \edittwo{In this work, we adopt the median stellar mass from the \textsc{Mizuki} catalog}\footnote{The stellar mass estimate from \textsc{Mizuki} has already included the mass returned to ISM by evolved stars via stellar winds and supernova explosions.}, which is derived by marginalizing the probability distribution of stellar mass over all the other parameters. 
\par \cite{Tanaka2018} compared the stellar masses from \textsc{Mizuki} to those from the NEWFIRM Medium Band Survey \citep[NMBS;][]{Whitaker2011} and showed that, stellar mass of galaxies at $z<1.5$ from \textsc{Mizuki} are systematically larger than those from NMBS  by 0.2 dex and have a scatter of $\sim0.25$ dex. The authors noted that the bias in stellar mass might be due to systematic differences in the data (either HSC or NMBS) and the combination of adopted template error functions and priors. Despite these differences, the stellar mass offsets of $\sim0.2-0.3$ dex among the different surveys can be expected even when both datasets have deep photometry in many filters \cite[e.g.,][]{vanDokkum2014}.
\par In Appendix~\ref{appendix:discuss_masscorr}, we further investigate the effect of ``outshining'' of the old stellar population by young stars on the SED of galaxies \citep[e.g.,][]{Sawicki1998, Papovich2001}, which could potentially lead to the underestimation of stellar masses of star-forming galaxies. In brief, we followed \cite{Sorba2018} to derive the corrections for unresolved stellar mass estimates and found the correction of $0.08-0.11$ dex for star-forming galaxies at $0.2<z<1.0$. \edittwo{After we added these corrections to the \textsc{Mizuki} stellar masses for star-forming galaxies, we re-fitted the size-mass relation. We found that the slope of size-mass relation of star-forming galaxies with $\log(M_{\ast}/M_{\odot})\gtrsim10.7$ decreases by $0.03-0.05$~dex, while there is no significant change in the slope of the relation of their lower mass counterparts.} Given that our main conclusion of a broken power-law form of size-mass relation of this population and their redshift evolution remain unchanged, throughout this analysis, we use the stellar mass estimated taken directly from the \textsc{Mizuki} catalog, and we do not apply any offset to our stellar mass estimates.
\par \edittwo{We note that the outshining mass correction of \cite{Sorba2018} only accounts for the difference between spatially resolved/unresolved SED fitting, but not galaxy sizes. We expect that outshining would also likely have an effect on the sizes as well as on masses, but it is beyond the scope of this paper to quantify these effects.}

\subsubsection{Stellar mass completeness limit}
\label{sec:derive_masscomplete}
\par \edittwo{First of all, we determine the magnitude limit for which we are able to measure accurate galaxy sizes for both Wide and Deep+UltraDeep layers.  In Appendix~\ref{appendix:verification_sizes}, we demonstrate that galaxy sizes can be determined with 5\% accuracy or better down to $i=24.5$ mag\footnote{For the same level of accuracy, we are able to recover sizes of simulated galaxies brighter than $i=23$ mag for those with intrinsic sizes smaller than $8\arcsec$.} for all HSC survey layers, at least for galaxies smaller than $\sim3\arcsec$ (corresponding to 10 kpc at $z=0.2$ and 25 kpc at $z=1.0$); at fainter magnitudes the random and systematic errors can exceed 40\% for large galaxies with high S\'{e}rsic indices. Because most galaxies in our sample have small sizes ($R_{e}=0\farcs55$ in the median) and low S\'{e}rsic indices ($n=1.5$ in the median), the magnitude limit of $i=24.5$ mag is conservative. }

\par Second, we employ the technique described by \cite{Pozzetti2010} and \cite{Weigel2016} \citep[see also][]{Quadri2012,Tomczak2014} to
estimate the 90\% mass-completeness limit as a function of redshift. We select quiescent galaxies (or star-forming galaxies) in narrow redshift bins and scale their fluxes and masses downward until they have the same magnitude as our adopted limit of $i_{\mathrm{lim}}=24.5$~mag. Then we define the mass-completeness limit as the stellar mass at which we detect 90\% of the dimmed galaxies at each redshift. Figure~\ref{fig:masscomp} shows the distribution of galaxies in our sample in the mass-redshift space and the adopted stellar mass completeness as a function of redshift ($M_{\ast\mathrm{lim}}(z)$) for the quiescent and star-forming galaxies. \editone{In the lowest redshift bin of our study ($0.2<z<0.4$), the comparison with $UVJ$ selection indicates that our $urz-$selected quiescent galaxies with $\log(M_{\ast})<9.0$ have high ($20\%-50\%$) contamination from star-forming galaxies (Figure~\ref{fig:contamfrac_vs_mass}). To minimize this effect,  we adopt  $\log(M_{\ast\mathrm{lim}}/M_{\odot})=9.0$ for quiescent galaxies, even though our sample is 90\% mass-completed down to $\log(M_{\ast}/M_{\odot})=8.8$. In Table~\ref{table:numbersample}, we provide the 90\% mass-completeness limits for quiescent and star-forming galaxies, and the number of both populations with stellar masses above the completeness limit at a given redshift.}

\section{Size Measurements}
\label{sec:lenstronomy}
\par We perform two-dimensional fits to the surface-brightness distributions of HSC $i-$band galaxy images using \textsc{Lenstronomy}\footnote{https://github.com/sibirrer/lenstronomy} \citep{Birrer2015,Birrer2018}, a multi-purpose open-source gravitational lens modeling python package. The main advantage of \textsc{Lenstronomy} is that it outputs the full posterior distribution of each parameter and the Laplace approximation of the uncertainties. \cite{Ding2020} have further developed a python front-end wrapper around \textsc{Lenstronomy} and other contemporary astronomy python packages to prepare the cutout image to be fed into \textsc{Lenstronomy}, detect extra neighboring objects in the field of view, and perform bulk structural analysis with \textsc{Lenstronomy} on an input catalog of galaxies (described in more detail below). In our study, we will utilize this python wrapper and the initial configuration of \textsc{Lenstronomy} from \cite{Ding2020}.
\par The input ingredients to \textsc{Lenstronomy} include: galaxy imaging data, noise level map, and PSF image. We provide \textsc{Lenstronomy} prior measurement of source positions of target galaxies from the HSC database, and we also use the PSF \textsc{Picker} tool\footnote{https://hsc-release.mtk.nao.ac.jp/psf/pdr2/} to generate a PSF image at a position of a target galaxy in the HSC band for which we aim to fit the surface-brightness profile. \editone{We then determine an image cutout size by estimating the root-mean-square (RMS) of the background pixels near the edge of an initial cutout size of $41 ~\mathrm{pixels}\times41~\mathrm{pixels}$ (i.e., $7 \arcsec \times 7 \arcsec$). If the fraction of outlier ($>2\sigma$) pixels exceeds $3\%$, the cutout size is increased. We iteratively perform this step until this condition is satisfied.} However, we limit the maximum allowed cutout size to $111~\mathrm{pixels}\times111~\mathrm{pixels}$ ($20\arcsec \times 20\arcsec$) to optimize the computational time. \edittwo{We find that only 4\% of our galaxy sample has a maximum allowed cutout size.} For each cutout image, we adopt the \textsc{Photutils}\footnote{https://photutils.readthedocs.io/en/stable/} \citep{Bradley2019} python package to detect neighboring sources and model the global background light in 2D based on the \textsc{SExtractor} algorithm. We remove the background light when it is not properly accounted for by the HSC pipeline\footnote{\edittwo{Typically, we find that the median local sky background of the HSC image is nearly zero ($\sim0.001$ flux/pixel), implying that the background has been properly accounted for by the HSC pipeline. However, in some cases, the local background can be as large as 0.01 flux/pixel, we account for this by subtracting it from the cutout image.}}, and we also fit a target galaxy and their neighboring sources simultaneously as described below. We use a single S\'{e}rsic profile to describe the total galaxy light distribution. We denote the ``size'' as the semimajor axis of the ellipse that contains half of the total flux of the best-fitting S\'{e}rsic model, $R_{e}$.

\par To avoid any unphysical result, \edittwo{we follow \cite{Ding2020}} to set the following upper and lower limits on the parameters: effective radius $R_{e}\in [0\farcs1,10\farcs0]$ \editone{(corresponding to $R_{e}\in [0.34~\mathrm{kpc},34~\mathrm{kpc}]$ at $z=0.2$ and  $R_{e}\in[0.8~\mathrm{kpc},82~\mathrm{kpc}]$ at $z=1$)}, S\'{e}rsic index $n\in[0.3,7]$, and projected axis ratio $q\in[0.1,1]$. \edittwo{Previous works on galaxy structural measurements commonly adopted the upper limit of S\'{e}rsic index of $n=8$ \citep[e.g.,][]{vanderWel2014,Mowla2019b}. We have tested the impact of our choice of the upper limit of $n$ on our results. To do so, we re-measure structural parameters for a subset of our sample and extend the upper limit of $n$ to 8. For galaxies which have new $n$ larger than 7, we find that their sizes increase but no more than 0.05 dex and 0.1 dex for those with sizes smaller than 1\arcsec and 2\farcs5, respectively. Nevertheless, most of our galaxies have small sizes ($R_{e}=0\farcs55$ in the median) and low S\'{e}rsic indices ($n=1.5$ in the median). Also, the fraction of sources with the best-fit S\'{e}rsic index larger than $n=7$ is only $\sim0.02$\% of the entire sample. Therefore, our choice of the upper limit should have no impact on our conclusions.}

\par With the input ingredients, \textsc{Lenstronomy} convolves the theoretical models (i.e., a single S\'{e}rsic profile) with the PSF before fitting them to the galaxy images. Finally, \textsc{Lenstronomy} finds the maximum likelihood of the parameter space by adopting the Particle Swarm Optimizer \citep[PSO;][]{Kennedy1995}.



\par \editone{In Appendix~\ref{appendix:size_uncertainties} we demonstrate using a set of simulated galaxies and a comparison with the $HST-$based size measurements that we typically obtain accurate size measurement with $5\%-10\%$ level accuracy or better for HSC galaxies in both Wide and Deep+UltraDeep layers, at least for galaxies smaller than $\sim3\arcsec$ (down to $i=24.5$) to $\sim8\arcsec$ (down to $i=23$).} We therefore adopt the limiting magnitude of $i=24.5$ throughout our analysis of the size-mass relation. \edittwo{In the following analysis, we only use galaxies which have reasonable fitting result}, defined as those fits with median and scatter of the one-dimensional residual (the difference between model and observed light profile) less than 0.05 mag pixel$^{-2}$ and 0.03 mag pixel$^{-2}$, respectively. The galaxies with reasonable fits accounts for $\sim70\%$ of the original galaxy samples in both Wide and Deep+UltraDeep layers. \edittwo{Finally, a total of 1,513,932 galaxies passed all of the quality requirements thus form the sample for our analysis.}

\par Additionally, we note that  our structural parameters measured using \textsc{Lenstronomy} are consistent with those using \textsc{GALFIT} (George et al. in prep.). \editone{In Appendix~\ref{appendix:size_uncertainties}, we investigate possible observational biases of our structural parameter measurements due to PSF bluring and surface brightness dimming in the outskirts of galaxies by using a set of simulated galaxies and then derive a correction to account for these biases, following the same technique as in \cite{Carollo2013} and \cite{Tarsitano2018}.  For our galaxy sample with $i<24.5$ mag, the correction applied to the effective radius is on average less than 5\%.}

\edittwo{In addition, we estimate random uncertainties on size measurements using a MCMC technique for a subset of our galaxy sample and then assign those to all galaxies in our sample (Appendix~\ref{appendix:mcmc}), while we estimate the systematic uncertainties using the set of simulated galaxies (Appendix~\ref{appendix:sizecorr}). The total uncertainty on the size of each galaxy is the quadrature sum of these uncertainties (Equation~\ref{eq:total_re_err})}. In the following section, we further investigate the wavelength dependence on the effective radius and correct our size measurements in HSC$-i$ band images to the rest-frame $5000$~\AA.  

	\label{fig:sizedist1D}
\begin{figure*}
\centering
\includegraphics[width=1\textwidth]{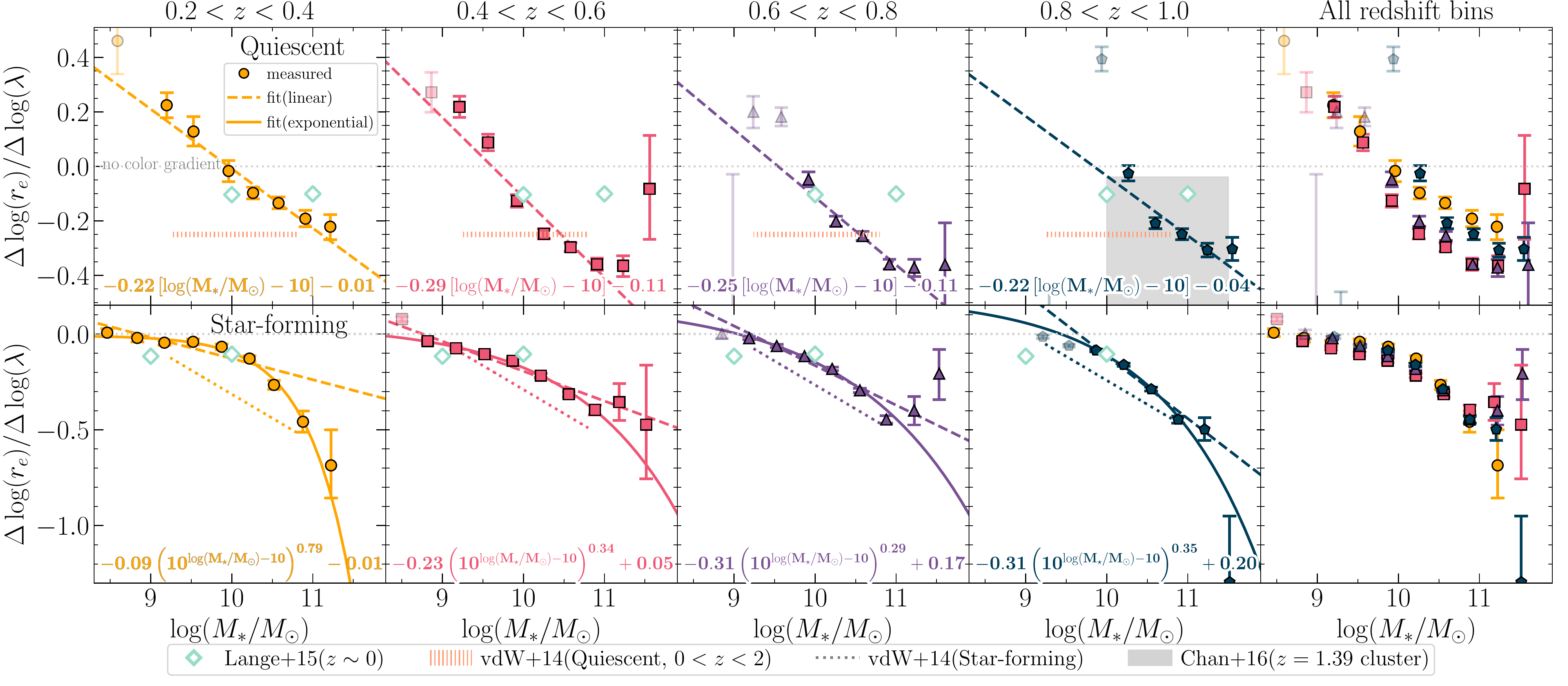}
\caption{Wavelength dependence of effective radius (color gradient, $\Delta \log( r_{e})/\Delta \log (\lambda)$) as a function of stellar mass for HSC quiescent (top) and star-forming (bottom) galaxies in four redshift bins (from left to right).  The best-fit linear model (Equation~\ref{eq:linear_sizecorr}) describing $\Delta \log( r_{e})/\Delta \log (\lambda)$ as a function of $M_{\ast}$ is indicated and plotted as dashed line (for both galaxy populations). The best-fit of exponential model (Equation~\ref{eq:expo_sizecorr}) is indicated and plotted as solid curve (for star-forming galaxies only). \edittwo{Symbols with lighter colors indicate the results in mass bins below $M_{\ast\mathrm{lim}}$ thus are excluded from the fitting.} The color gradients from previous works for the redshift range similar to our study and at $z\sim0$ are shown. The last columns show the HSC color gradients from all redshift bins.  The horizontal grey dotted lines in all panels indicates $\Delta \log (r_{e})/\Delta \log(\lambda)=0$ (no color gradient). The negative values imply that the galaxy sizes are smaller at longer wavelengths (i.e., galaxies are bluer in the outer parts). Overall, the color gradient is stronger for more massive galaxies.}
\label{fig:colorgradients_mass}
\end{figure*}
\subsection{Wavelength Dependence of Galaxy Sizes}
\label{sec:colorgrad}
\par \editone{Previous studies of galaxy half-light radii in multiple imaging bands  at low-redshift \citep[e.g.,][]{LaBarbera2010,Kelvin2012,Vulcani2014,Lange2015} and at higher redshifts \citep[e.g.,][]{vanderWel2014,Chan2016} generally showed that the observed galaxy half-light radii are smaller if they are measured in longer wavelength bands, which implies negative color gradients in galaxies \citep[e.g., outskirts of galaxies are bluer than the central regions;][]{Tortora2010, Wuyts2010, Guo2011,Szomoru2013,Mosleh2017,Suess2019ApJ}. As a consequence, the color gradients and their evolution affect the galaxy size measurements. Here we follow \cite{vanderWel2014} to probe the effect of color gradients by quantifying the relation between galaxy sizes and wavelength at which they are measured. Later, we will use this relation to correct the HSC size measurement to a common rest-frame wavelength of 5000~\AA. }
\par We begin by measuring the effective radii of a subset of HSC galaxies in the five-HSC band ($grizy$) images. In each of the four redshift bins ($\Delta z=0.2$) from $z=0.2$ to $z=1.0$, we divide our subsample of quiescent (star-forming) galaxies into $0.4$ dex bins in $\log(M_{\ast}/M_{\odot})$. We then fit the relation between effective radii and rest-frame wavelength of the form: 
\begin{equation}
    \log (r_{e}/\mathrm{kpc}) = A_{\lambda}\log \lambda_{\mathrm{rest}}+ B_{\lambda},
    \label{eq:size_lambda}
\end{equation}
\noindent \editone{where $A_{\lambda}$ and $B_{\lambda}$ are the slope and intercept. $\lambda_{\mathrm{rest}}$ is the rest-frame wavelength of a galaxy observed at a redshift $z$ in band $b$ and corresponds to $\lambda_{\mathrm{rest}}=\lambda_{\mathrm{obs},b}/(1+z)$, where $b$ is for each HSC bandpass we measure a galaxy size: 4798.2~\AA~($g$), 6218.4~\AA~($r$), 7727.0~\AA~($i$), 8908.2~\AA~($z$), and 9775.1~\AA~($y$). We present the best-fit slopes ($A_{\lambda}$) and intercepts ($B_{\lambda}$) for these fits, as well as their error bars in Table~\ref{table:size_lambda_parameters}. At a given stellar mass and redshift bin, we take the slope $A_{\lambda}$ as a color gradient, $\Delta \log( r_{e}) / \Delta \log (\lambda)$. }

\startlongtable
\begin{deluxetable*}{cccccc}
\tablecaption{Best-fit parameters of Linear Relation Describing the Effective Radii ($r_{e}$) as a Function of Rest-Frame Wavelength of the Form $\log (r_{e}/\mathrm{kpc}) = A_{\lambda}\log \lambda_{\mathrm{rest}}+ B_{\lambda}$}
\label{table:size_lambda_parameters}
\tablehead{\colhead{\multirow{3}{*}{$z$}} & \colhead{\multirow{3}{*}{ $\log(M_{\ast,\mathrm{med}}/M_{\odot}$)}} & \multicolumn{2}{c}{Quiescent}  &\multicolumn{2}{c}{Star-forming} \\
\cline{3-6}
  & & \colhead{$A_{\lambda}$} & \colhead{$B_{\lambda}$} & \colhead{$A_{\lambda}$} & \colhead{$B_{\lambda}$} } 
\startdata
\multirow{9}{*}{$0.2<z<0.4$} & $8.5$ & $-$ & $-$ & $0.01\pm0.02$ & $0.28\pm0.08$  \\
 & $8.8$ & $-$ & $-$ & $-0.02\pm0.02$ & $0.51\pm0.08$  \\
 & $9.2$ & $0.22\pm0.05$ & $-0.69\pm0.17$ & $-0.05\pm0.01$ & $0.68\pm0.05$  \\
 & $9.5$ & $0.13\pm0.05$ & $-0.25\pm0.20$ & $-0.04\pm0.02$ & $0.72\pm0.06$  \\
 & $9.9$ & $-0.02\pm0.04$ & $0.31\pm0.14$ & $-0.07\pm0.02$ & $0.86\pm0.06$  \\
 & $10.2$ & $-0.10\pm0.02$ & $0.68\pm0.08$ & $-0.13\pm0.02$ & $1.12\pm0.06$  \\
 & $10.6$ & $-0.13\pm0.02$ & $0.98\pm0.08$ & $-0.27\pm0.02$ & $1.70\pm0.09$  \\
 & $10.9$ & $-0.19\pm0.03$ & $1.43\pm0.11$ & $-0.46\pm0.05$ & $2.55\pm0.21$  \\
 & $11.3$ & $-0.22\pm0.05$ & $1.73\pm0.17$ & $-0.69\pm0.18$ & $3.60\pm0.67$  \\
\hline
\multirow{10}{*}{$0.4<z<0.6$} & $8.5$ & $-$ & $-$ & $-$ & $-$  \\
 & $8.8$ & $-$ & $-$ & $-0.04\pm0.01$ & $0.48\pm0.05$  \\
 & $9.2$ & $0.22\pm0.04$ & $-0.66\pm0.14$ & $-0.07\pm0.01$ & $0.71\pm0.03$  \\
 & $9.5$ & $0.09\pm0.03$ & $-0.15\pm0.11$ & $-0.11\pm0.01$ & $0.91\pm0.03$  \\
 & $9.9$ & $-0.13\pm0.02$ & $0.69\pm0.09$ & $-0.14\pm0.01$ & $1.11\pm0.04$  \\
 & $10.2$ & $-0.25\pm0.02$ & $1.20\pm0.07$ & $-0.22\pm0.01$ & $1.44\pm0.04$  \\
 & $10.6$ & $-0.30\pm0.02$ & $1.54\pm0.06$ & $-0.31\pm0.01$ & $1.89\pm0.05$  \\
 & $10.9$ & $-0.36\pm0.02$ & $1.99\pm0.08$ & $-0.40\pm0.03$ & $2.29\pm0.10$  \\
 & $11.3$ & $-0.36\pm0.04$ & $2.20\pm0.14$ & $-0.36\pm0.10$ & $2.25\pm0.36$  \\
 & $11.7$ & $-0.08\pm0.19$ & $1.14\pm0.71$ & $-0.47\pm0.30$ & $2.83\pm1.08$  \\
\hline
\multirow{9}{*}{$0.6<z<0.8$} & $8.8$ & $-$ & $-$ & $-$ & $-$  \\
 & $9.2$ & $-$ & $-$ & $-0.02\pm0.01$ & $0.48\pm0.03$  \\
 & $9.5$ & $0.18\pm0.03$ & $-0.47\pm0.12$ & $-0.06\pm0.01$ & $0.71\pm0.03$  \\
 & $9.9$ & $-0.05\pm0.03$ & $0.40\pm0.10$ & $-0.12\pm0.01$ & $0.98\pm0.03$  \\
 & $10.2$ & $-0.20\pm0.02$ & $1.00\pm0.07$ & $-0.18\pm0.01$ & $1.30\pm0.03$  \\
 & $10.6$ & $-0.26\pm0.02$ & $1.30\pm0.06$ & $-0.29\pm0.01$ & $1.79\pm0.04$  \\
 & $10.9$ & $-0.36\pm0.02$ & $1.90\pm0.07$ & $-0.45\pm0.02$ & $2.42\pm0.07$  \\
 & $11.3$ & $-0.37\pm0.03$ & $2.16\pm0.11$ & $-0.40\pm0.07$ & $2.36\pm0.27$  \\
 & $11.7$ & $-0.36\pm0.16$ & $2.00\pm0.57$ & $-0.21\pm0.13$ & $1.77\pm0.47$  \\
\hline
\multirow{8}{*}{$0.8<z<1.0$} & $9.2$ & $-$ & $-$ & $-$ & $-$  \\
 & $9.5$ & $-$ & $-$ & $-0.06\pm0.01$ & $0.65\pm0.03$  \\
 & $9.9$ & $0.39\pm0.04$ & $-1.13\pm0.16$ & $-0.09\pm0.01$ & $0.84\pm0.03$  \\
 & $10.2$ & $-0.03\pm0.03$ & $0.38\pm0.10$ & $-0.16\pm0.01$ & $1.20\pm0.03$  \\
 & $10.6$ & $-0.21\pm0.02$ & $1.11\pm0.07$ & $-0.29\pm0.01$ & $1.74\pm0.04$  \\
 & $10.9$ & $-0.25\pm0.02$ & $1.40\pm0.07$ & $-0.45\pm0.02$ & $2.40\pm0.06$  \\
 & $11.3$ & $-0.31\pm0.03$ & $1.84\pm0.09$ & $-0.50\pm0.06$ & $2.71\pm0.21$  \\
 & $11.7$ & $-0.30\pm0.04$ & $2.01\pm0.15$ & $-1.30\pm0.33$ & $5.76\pm1.18$  \\
 \hline
 \enddata
\end{deluxetable*}

\par \editone{Figure~\ref{fig:colorgradients_mass} shows $\Delta \log( r_{e}) / \Delta \log (\lambda)$ as a function of stellar mass for quiescent and star-forming galaxies in four redshift bins. First of all, $\Delta \log( r_{e}) / \Delta \log (\lambda)$ for both quiescent and star-forming galaxies in all redshift bins on average have negative values (except for quiescent galaxies with $\log(M_{\ast}/M_{\odot})<10$), which suggests that galaxies tend to have smaller sizes at longer wavelengths, consistent with negative color gradients. Second, $\Delta \log( r_{e}) / \Delta \log (\lambda)$ for both populations are correlated with stellar mass, such that more massive galaxies have more steeply negative color gradients. Also, over the redshift range of $0.2<z<1.0$, we find no significant redshift evolution in  $\Delta \log( r_{e}) / \Delta \log (\lambda)$ for both galaxy populations.} 
\par \editone{Our finding of negative color gradients is broadly consistent with \cite{vanderWel2014} and \cite{Chan2016} for the galaxy samples in the COSMOS field and in $z=1.39$ cluster, respectively. For star-forming galaxies over the same redshift range, the magnitude of the color gradients ($|\Delta \log( r_{e}) / \Delta \log (\lambda)|$) from HSC is roughly a factor $1.5-3$ lower than those of \cite{vanderWel2014}. On the other hand, due to the small sample size and mass range ($9.2<\log(M_{\ast}/M_{\odot})<10.8$) of quiescent galaxies of \cite{vanderWel2014}, they were unable to constrain the stellar mass dependence of $\Delta \log( r_{e}) / \Delta \log (\lambda)$ for this population, but instead reported the average color gradient of $\Delta \log( r_{e}) / \Delta \log (\lambda)=-0.25$ with no discernible trend with stellar mass and redshift. Also, our color gradients are broadly in agreement with the results at $z\sim0$ from \cite{Kelvin2012} and \cite{Lange2015}, particularly with \cite{Lange2015} result for galaxies with $\log(M_{\ast}/M_{\odot})=10$. The difference in the results of \cite{vanderWel2014} and \cite{Lange2015} and the results presented in this paper (Figure~\ref{fig:colorgradients_mass}) are likely due to differences in sample size, particularly for massive galaxies.}

\par \editone{Previous works have directly measured the strength of color gradients by computing the ratio of the galaxy's half-mass and half-light radius\footnote{Color gradients cause the light profile of galaxies to deviate from their underlying mass profiles. The large value of the ratio of the galaxy's half-mass, $r_{e,\mathrm{mass}}$, to the half-light radius, $r_{e,\mathrm{light}}$, ($r_{e,\mathrm{mass}}/r_{e,\mathrm{light}}$) indicates positive color gradient (center of the galaxy is bluer than the outskirt); small value of $r_{e,\mathrm{mass}}/r_{e,\mathrm{light}}$ indicates negative color gradient (the center of the galaxy is redder than the outskirts); $r_{e,\mathrm{mass}}/r_{e,\mathrm{light}}$ of unity indicates no radial color gradient.} and showed that color gradients are correlated with other galaxy properties \citep[e.g., stellar mass, rest-frame $U-V$ colors;][]{Tortora2010,Mosleh2017,Suess2019ApJ,Suess2019ApJL}. Recently, \cite{Suess2019ApJL} measured color gradients for quiescent and star-forming galaxies with $9.0<\log(M_{\ast}/M_{\odot})<11.5$ at $0<z<1.0$ and found that both populations have negative color gradients with stronger effect for more massive galaxies \citep[see also][for similar conclusion]{Mosleh2017}. The \cite{Suess2019ApJL} result is qualitatively in agreement with our result of steeper color gradients for more massive galaxies.}
 \par \editone{Interestingly, we find that color gradients for low-mass quiescent galaxies with $\log(M_{\ast}/M_{\odot})\lesssim 9.5$, particularly at $0.2<z<0.6$ become positive, which means that the observed sizes of these galaxies are on average larger at longer wavelengths. This result is similar to  \cite{Tortora2010}, who analyzed the color gradients of SDSS galaxies and found that color and metallicity gradients slopes become highly positive at the very low masses, and the transition from negative to positive occurring at $\log(M_{\ast}/M_{\odot})\sim9-9.5$. On the other hand, we find evidence that color gradients of quiescent galaxies seems to plateau at  $\log(M_{\ast}/M_{\odot})\gtrsim11$. This trend is again consistent with \cite{Tortora2010}, who reported shallow (or even flat) color gradients for massive early-type galaxies of similar stellar masses. They attributed this trend mainly to metallicity gradients, while age gradients play a smaller but still significant role. Over the redshift range overlapping with our study, \cite{Huang2018} also found evidence that color gradient of massive galaxies with $\log(M_{\ast}/M_{\odot})\ge11.5$ at $0.3<z<0.5$ does not significantly depend on total galaxy stellar mass. To summarize, the qualitative agreement between our results and those studies demonstrates that both large sample size and long lever arm on stellar mass of the HSC dataset enable us to probe the stellar mass dependence of $\Delta \log( r_{e}) / \Delta \log (\lambda)$ at $0.2<z<1.0$. In the following section, we proceed with using the derived color gradients to correct the HSC size measurements to a common rest-frame wavelength.}
\subsubsection{Analytic Fits to Color Gradients and Correction of Galaxy Size to Rest-Frame Wavelength}
\par \editone{In each redshift bin, we parameterize the $\Delta \log( r_{e}) / \Delta \log (\lambda)-\log(M_{\ast})$ relation for both quiescent and star-forming galaxies assuming a linear model \cite[e.g.,][]{vanderWel2014}:}
\begin{equation}
\begin{split}
\frac{\Delta \log( r_{e})}{\Delta \log (\lambda)} =\mathcal{A}_{\lambda} \left[\log (M_{\ast}/M_{\odot})-10\right]+\mathcal{B}_{\lambda},
    \label{eq:linear_sizecorr}
\end{split}
\end{equation}


\par \editone{For star-forming galaxies, a color gradient flattens ($\Delta \log( r_{e}) / \Delta \log (\lambda)\sim0$) as stellar mass decreases, such that the observed sizes of these low-mass star-forming galaxies are nearly independent of the wavelength they are measured at. This trend is more pronounced at $0.2<z<0.4$, and  we further parameterize the relation for star-forming galaxies using an exponential model: }
 
\begin{equation}
     \frac{\Delta \log( r_{e})}{\Delta \log (\lambda)} =\mathcal{A}_{\lambda}  \left (10^{\log (M_{\ast}/M_{\odot})-10}  \right)^{\alpha_{\lambda}}+\mathcal{B}_{\lambda},
    \label{eq:expo_sizecorr}
\end{equation}
\par \editone{We utilize leave-one-out (LOO) cross-validation to account for overfitting and evaluate how well a linear model and an exponential model fit the $\Delta \log( r_{e}) / \Delta \log (\lambda) -\log (M_{\ast}/M_{\odot})$ relation for star-forming galaxies. This approach is useful when the number of data points used to fit the relation is small ($N=6-9$). In brief, for $N$ data points of $\Delta \log( r_{e}) / \Delta \log (\lambda)$, we leave out one measurement and fit a model (linear or exponential) to the relation using the rest of the data points. For a given model, we evaluate the cross validation error by computing the mean square error (MSE) using the left-out data point. We repeat this process $N$ times and take the median value of a set of MSEs as the final cross-validation error (LOOCV). In all redshift bins, the LOOCVs for an exponential model are significantly lower than those for a linear model (by more than a factor of 2), meaning that the  $\Delta \log( r_{e}) / \Delta \log (\lambda)-\log (M_{\ast})$  relations are better described by an exponential model. As a result, we adopt an exponential model to describe $\Delta \log( r_{e}) / \Delta \log (\lambda)-\log (M_{\ast})$ relation for star-forming galaxies throughout this analysis.  In Figure~\ref{fig:colorgradients_mass}, we plot the best-fit relations for both a linear model (dashed curve) and an exponential model (solid curve) for star-forming galaxies. In the same Figure, we also indicate the best-fit parameters corresponding to the model we adopted for each galaxy population: a linear model for quiescent galaxies and an exponential model for star-forming galaxies. \edittwo{We note that we also experimented with fitting an exponential model to the $\Delta \log( r_{e}) / \Delta \log (\lambda)-\log (M_{\ast})$ of quiescent galaxies and evaluated the LOOCVs, but we do not find any significant improvement in the LOOCVs compared to those using a linear model.}}

\par Finally, we estimate the effective radius $R_{e}$ at a rest-frame of $5000$~\AA~using the form of correction given by
\begin{equation}
    R_{e}=R_{e,\mathrm{HSC-}i}\left ( \frac{1+z}{1+z_{p}} \right )^\frac{\Delta \log r_{e}}{\Delta \log \lambda},
\end{equation}
\noindent where $R_{e,\mathrm{HSC-}i}$ denotes effective radius measured in HSC $i-$band imaging, and $z_{p}$ is the ``pivot redshift'', which is equal to 0.55 for HSC $i-$band. \edittwo{This pivot redshift is the redshift at which the observed wavelength of the HSC $i$-band (7727.0~\AA) corresponds to the rest-frame wavelength of 5000~\AA.} We find that the correction ($R_{e}/R_{e,\mathrm{HSC-}i}$) applied to the observed sizes ($R_{e,\mathrm{HSC-}i}$) are mild and do not exceed a 10\% with respect to the observed ones.  We have tested this by adopting the corrections from \cite{vanderWel2014} and found that our main conclusions do not significantly change. We therefore adopt our correction derived from HSC dataset throughout the paper. Finally, we denote the observed effective radii corrected to the rest-frame $5000$~\AA~ and corrected for the observational biases as $R_{e}$.

\section{Analytic Fits to The Size-Mass Relation}
\label{sec:fitting_sizemass} 
 \par The large galaxy sample from the HSC allows us to test the shape (i.e., amplitude, slope, and the intrinsic scatter) of the observed $R_{e}-M_{\ast}$ relation. In this work, we fit both single power-law \cite[e.g.,][]{vanderWel2014, Mowla2019b} and a smoothly broken (or two-component) power-law function \cite[e.g.,][]{Shen2003,Dutton2009,Mowla2019a,Mosleh2020} to the the observed $R_{e}-M_{\ast}$ distribution.  In contrast to previous works, we are not making any a priori assumptions about which functional form provides a better fit. Importantly, we are assuming that the observed $R_{e}-M_{\ast}$ relation can be modeled by a single power-law or a smoothly broken power-law model. In Section~\ref{sec:model_selection}, we will provide the details on selecting the model that better describes the $R_{e}-M_{\ast}$ relation using Bayes-factor evidence.
 
 \par We fit the observed $R_{e}-M_{\ast}$ relation of the HSC galaxy sample by following \cite{Shen2003} and \cite{vanderWel2014}. We assume a log-normal distribution $N(\log r,\sigma_{\log r})$, where $\log r$ is the mean and $\sigma_{\log r}$ is the intrinsic scatter. Additionally, $r$ is taken to be a function of galaxy stellar mass. The single power-law function is defined as 
\begin{equation}
     r(M_{\ast}) = r_p\left ( \frac{M_{\ast}}{5\times10^{10}~M_{\odot}} \right )^{\alpha}
     \label{eq:singlepwl}
\end{equation}
where $r_{p}$ is the radius for a galaxy with $M_{\ast}=5\times10^{10}~M_{\odot}$  (or equivalently, $\log(r_p)$ is the relation's intercept), and $\alpha$ is the power-law slope.
\par The smoothly broken power-law function is defined as
\begin{equation}
    r(M_{\ast}) = r_p\left ( \frac{M_{\ast}}{M_{p}} \right )^{\alpha}\left [\frac{1}{2}\left \{ 1+\left ( \frac{M_{\ast}}{M_{p}  } \right )^{\delta}  \right \}  \right ]^{(\beta-\alpha)/\delta}
    \label{eq:doublebrokenpwl}
\end{equation}
where $M_{p}$ is the pivot stellar mass at which the slope changes, $r_p$ is the radius at the pivot stellar mass, $\alpha$ is the power-law slope for low-mass galaxies ($M_{\ast}<M_{p}$), $\beta$ is the power-law slope for high-mass galaxies ($M_{\ast}>M_{p}$), and $\delta$ is the smoothing factor that controls the sharpness of the transition from the low- to high-mass end of the relation. Following \cite{Mowla2019a}, we adopt $\delta=6$ to reduce the degeneracy between $\delta$ and the slopes and require $\alpha<\beta$. The smoothly broken power-law function can be easily reduced to a single power-law relation by setting $\beta=\alpha$.
We also assume the $\sigma_{\log r}$ is constant with $M_{\ast}$\footnote{In addition to the stellar mass-dependence of the power-law slope, one might expect that $\sigma_{\log r}$ will also depend on $M_{\ast}$\citep[see also][who fitted a broken power-law model to $R_{e}-M_{\ast}$ relation for late-type galaxies but also defined $M_0$ to be the characteristic mass  at which both $\sigma_{\log r}$ and the power-law slope significantly change]{Shen2003}. We have attempted to fit $\sigma_{\log r,1}$ and $\sigma_{\log r,2}$ separately for the relation below and above $M_{p}$. However, both of these parameters are not well constrained, which we partly attribute to our photo-$z$  uncertainties (and hence stellar mass uncertainties) and cross-contamination between quiescent and star-forming galaxies. Therefore, we will assume that $\sigma_{\log r}$ is constant with $M_{\ast}$, throughout this analysis.}


\par \editone{We follow the method described in \cite{vanderWel2014} to compute the total likelihood for a set of model parameters describing the size-mass relation. In brief, for each set of models listed in Table~\ref{table:modellist}, we first use Equation 4 of \cite{vanderWel2014} to compute the probability distribution of observed $R_{e}$ for quiescent ($P_{Q}$) and for star-forming population ($P_{SF}$), where the total uncertainty in the observed $R_{e}$ \edittwo{ is the quadrature sum of random uncertainties estimated using a MCMC technique and the systematic uncertainties estimated using the set of simulated galaxies (Equation~\ref{eq:total_re_err}).} Second, we take cross-contamination between quiescent and star-forming galaxies into account \edittwo{by incorporating the cross-contamination fraction for quiescent ($f_{\mathrm{cont,Q}}$) and star-forming ($f_{\mathrm{cont,SF}}$) populations as described in Appendix~\ref{appendix:misclass} into the likelihood of quiescent and star-forming galaxies.} Third, we assign a weight, $W$, inversely proportional to the number density, to each galaxy to avoid being dominated by the large number of low-mass galaxies and to ensure that each mass range carries equal weight in the fit. Here we take the number density from the galaxy stellar mass functions derived using data from the FourStar Galaxy Evolution Survey \cite[ZFOURGE;][]{Tomczak2014}. \edittwo{We have verified that the best-fit parameters of the $R_{e}-M_{\ast}$ relation and the results presented in this work do not significantly change if we do not assign a weight derived from the SMFs to galaxies}.  As in \cite{vanderWel2014}, we accounted for those objects with catastrophic redshift failures or misclassified stars when computing the total likelihood. In this work,  we find that a fraction of catastrophic redshift failures ($\eta$) depends on $i-$band magnitude, photo-$z$, and galaxy populations (quiescent or star-forming galaxies; Section~\ref{sec:zphot}). We therefore incorporate $\eta(i,z)$ into the likelihood estimation, instead of adopting a constant fraction (i.e., $\eta=0.01$) of those outliers as in \cite{vanderWel2014}. Finally, the likelihoods ($\mathcal{L}$) are given by}:
\begin{equation}
    \mathcal{L}_{\mathrm{Q}} = \sum\ln {\left \{ W_{\mathrm{Q}} \cdot \left [ (1-f_{\mathrm{cont,Q}} )\cdot   P_{\mathrm{Q}} + f_{\mathrm{cont,Q}} \cdot  P_{\mathrm{SF}}  + \eta_{\mathrm{Q}} \right ] \right \}}
    \label{eq:loglike_qui}
\end{equation}
\noindent for quiescent galaxies, and
\begin{equation}
    \mathcal{L}_{\mathrm{SF}} = \sum\ln {\left \{ W_{\mathrm{SF}} \cdot \left [ (1-f_{\mathrm{cont,SF}} )\cdot   P_{\mathrm{SF}} + f_{\mathrm{cont,SF}} \cdot  P_{\mathrm{Q}}  + \eta_{\mathrm{SF}}\right ] \right \}}
\label{eq:loglike_sf}
\end{equation}
\noindent for star-forming galaxies. $W_{\mathrm{Q}}$ ($W_{\mathrm{SF}}$) and $f_{\mathrm{cont,Q}}$ ($f_{\mathrm{cont,SF}}$) are the weight and contaminating fraction for quiescent (star-forming) galaxies. Each of these parameters is a function of redshift and stellar mass. The $\eta_{\mathrm{Q}}$ and $\eta_{\mathrm{SF}}$ are the fractions of catastrophic redshift failures for quiescent and star-forming galaxies, respectively. The summation is for all galaxies in a given subsample of stellar mass and/or redshift. We obtain the best-fitting parameters by finding the model with the maximum total likelihood, $\mathcal{L}=\mathcal{L}_{\mathrm{Q}}+\mathcal{L}_{\mathrm{SF}}$.


\begin{deluxetable}{cccc}
\tablecaption{Power-Law Models (Equation~\ref{eq:singlepwl} and~\ref{eq:doublebrokenpwl}) and the Corresponding Parameters for Fitting the $R_{e}-M_{\ast}$ Distributions of Quiescent and Star-Forming Galaxies}
\label{table:modellist}
\tablehead{ \multicolumn{2}{c}{Power-law model} & \multicolumn{2}{c}{Parameter}\\
\colhead{Quiescent} & \colhead{Star-forming} & \colhead{Quiescent} & \colhead{Star-forming} }  
\startdata
Single & Single & $\alpha,r_{p},\sigma_{\log r}$ & $\alpha,r_{p},\sigma_{\log r}$\\ Single & Broken & $\alpha,r_{p},\sigma_{\log r}$ &  $\alpha$,
$\beta$, $M_{p}$,  $r_{p}$,   $\sigma_{\log r}$\\
Broken & Single & $\alpha,
\beta, M_{p},  r_{p}, \sigma_{\log r}$ & $\alpha,r_{p},\sigma_{\log r}$\\
Broken & Broken & $\alpha,
\beta, M_{p},  r_{p}, \sigma_{\log r}$ & $\alpha$,
$\beta$, $M_{p}$,  $r_{p}$,   $\sigma_{\log r}$\\
 \hline
 \enddata
\tablecomments{ $M_{p}$ is the pivot stellar mass of the $R_{e}-M_{\ast}$ relation where the power-slope changes, $\alpha$ is the power-law slope of the relation for galaxies with stellar mass below the pivot stellar mass ($M_{\ast}<M_{p}$), $\beta$ is the power-law slope for galaxies with $M_{\ast}>M_{p}$, and $r_{p}$ is the radius at the pivot stellar mass $M_{p}$, and $\sigma_{\log r}$ is the intrinsic scatter of the relation. We obtain the best-fitting parameters by finding the power-law model with the maximum total likelihood of quiescent and star-forming galaxies, $\mathcal{L}=\mathcal{L}_{\mathrm{Q}}+\mathcal{L}_{\mathrm{SF}}$, given by Equation~\ref{eq:loglike_qui} and ~\ref{eq:loglike_sf}.} 
\end{deluxetable}
    
    

\par In this study, we use  \textsc{dynesty}\footnote{https://dynesty.readthedocs.io/en/latest/} nested sampling \citep{Skilling2004,Skilling2006,Speagle2020} to fit the observed $R_{e}-M_{\ast}$ relation and estimate posterior distribution functions for parameters of the relation such as power-law slopes ($\alpha,\beta$), pivot stellar mass ($M_{\ast}$), and the intrinsic scatter ($\sigma_{\log r}$). The  advantage of using \textsc{Dynesty} is that it efficiently samples multi-modal distributions and has well-defined stopping criteria based on evaluations of Bayesian evidence ensuring model convergence. Specifically, we use \textsc{Dynesty} to fit the $R_{e}-M_{\ast}$ distributions of quiescent and star-forming galaxies down to the stellar mass completeness limit at a given redshift for each population. To obtain sufficient coverage of posterior samples, we choose 1024 live points from the prior distribution. \editone{We assumed flat priors on all parameters, within the ranges specified in Table~\ref{table:sampling_parameters}}. In the following section, we will compare the Bayes-factor evidence output from \textsc{Dynesty} for each set of model parameters listed in Table~\ref{table:modellist} and select the power-law model which best characterizes the $R_{e}-M_{\ast}$ relation of galaxies.

\begin{deluxetable}{ccc}
\tablecaption{Parameter Boundaries for Fitting $R_{e}-M_{\ast}$ Distribution}
\label{table:sampling_parameters}
\tablehead{\colhead{Power-law model} & \colhead{Parameter} & \colhead{Prior}  }
\colnumbers
\startdata
\multirow{3}{*}{Single (Equation~\ref{eq:singlepwl})}  & $\alpha$ & $[0-1]$ \\
& $r_{p}$ & [0.5-10] \\ 
 & $\sigma_{\log r}$ & [0.05,0.5]\\
 \hline
 \multirow{5}{*}{Broken (Equation~\ref{eq:doublebrokenpwl})} & $\alpha$ & $[0-1]$ \\
  & $\beta$ & $[0-1]$ \\
  & $\log M_{p}$ & $[8.6-11.5]$\\
  & $r_{p}$ & [0.5-10] \\
   & $\sigma_{\log r}$ & [0.05,0.5]\\
  \hline
 \enddata
\tablecomments{(1) Name of the power-law models for fitting the median $R_{e}-M_{\ast}$ relation, (2) parameters of each power-law functions, (3) the prior on a given parameter. Priors of the form [A,B] are flat with minimum and maximum given by A and B.  For a smoothly broken power-law model, we are constraining $\alpha<\beta$. } 
\end{deluxetable}

\begin{deluxetable*}{ccccccccc}
\tablecaption{Best-fit Parameters of Power-law Fit to the  Size-Mass Relation for Quiescent and Star-forming Galaxies of the Form Given in Equation~\ref{eq:singlepwl} and \ref{eq:doublebrokenpwl}.}
\label{table:resultfitparams}
\tablehead{ \colhead{ Power-law model} & \colhead{ Subsample} & \colhead{$z_{\mathrm{med}}$} &   \colhead{$\alpha$} &  \colhead{$\beta$} &  \colhead{$r_{p}$(kpc)} & \colhead{ $\log(M_{p}/ M_{\odot})$} &   \colhead{$\sigma_{\log r}$ } & \colhead{Bayes-factor Evidence ($\zeta$)}}
\colnumbers
\startdata
\multirow{4}{*}{Double} & \multirow{4}{*}{Quiescent} & $0.3$ & $0.14\pm0.01$ &  $0.63\pm0.01$ & $2.1\pm0.0$ & $10.2\pm0.0$ & $0.15\pm0.01$ & $8448.5$  \\
  &  & $0.5$ & $0.10\pm0.01$ &  $0.63\pm0.01$ & $1.9\pm0.0$ & $10.3\pm0.0$ & $0.17\pm0.01$ & $45109.2$ \\
  &  & $0.7$ & $0.10\pm0.01$ &  $0.67\pm0.01$ & $2.0\pm0.0$ & $10.5\pm0.0$ & $0.19\pm0.01$ & $9037.7$ \\
  &  & $0.9$ & $0.13\pm0.01$ &  $0.68\pm0.01$ & $2.1\pm0.0$ & $10.6\pm0.0$ & $0.19\pm0.01$ & $3889.0$ \\
\hline
\multirow{4}{*}{Double} & \multirow{4}{*}{Star-forming} & $0.3$ & $0.17\pm0.01$ &  $0.62\pm0.02$ & $5.4\pm0.1$ & $10.6\pm0.0$ & $0.17\pm0.01$ & $8448.5$  \\
  &  & $0.5$ & $0.18\pm0.01$ &  $0.39\pm0.01$ & $5.3\pm0.1$ & $10.6\pm0.0$ & $0.17\pm0.01$ & $45109.2$ \\
  &  & $0.7$ & $0.16\pm0.01$ &  $0.27\pm0.01$ & $4.9\pm0.1$ & $10.7\pm0.0$ & $0.17\pm0.01$ & $9037.7$ \\
  &  & $0.9$ & $0.19\pm0.01$ &  $0.29\pm0.02$ & $5.0\pm0.1$ & $10.8\pm0.0$ & $0.17\pm0.01$ & $3889.0$ \\
\hline
\multirow{4}{*}{Single} & \multirow{4}{*}{Quiescent} & $0.3$ & $0.34\pm0.01$ &  $-$ & $3.9\pm0.0$ & $-$ & $0.18\pm0.01$ \\
  &  & $0.5$ & $0.36\pm0.01$ &  $-$ & $3.6\pm0.0$ & $-$ & $0.19\pm0.01$ &  $-$ \\
  &  & $0.7$ & $0.41\pm0.01$ &  $-$ & $3.1\pm0.0$ & $-$ & $0.21\pm0.01$ &  $-$\\
  &  & $0.9$ & $0.44\pm0.01$ &  $-$ & $2.7\pm0.0$ & $-$ & $0.21\pm0.01$ &  $-$ \\
\hline
\multirow{4}{*}{Single} & \multirow{4}{*}{Star-forming} & $0.3$ & $0.17\pm0.01$ &  $-$ & $5.4\pm0.0$ & $-$ & $0.17\pm0.01$ &  $-$\\
  &  & $0.5$ & $0.20\pm0.01$ &  $-$ & $5.5\pm0.0$ & $-$ & $0.17\pm0.01$ &  $-$\\
  &  & $0.7$ & $0.17\pm0.01$ &  $-$ & $5.0\pm0.0$ & $-$ & $0.16\pm0.01$ &  $-$\\
  &  & $0.9$ & $0.20\pm0.01$ &  $-$ & $4.8\pm0.0$ & $-$ & $0.16\pm0.01$ &  $-$ \\
\hline
\enddata
\tablecomments{(1) Power-law model  used to fit the $R_{e}-M_{\ast}$ relation (2) subsample of HSC galaxies with $i<24.5 $, (3) $z_{\mathrm{med}}$ is the median redshift, (4) $\alpha$ is the power-law slope of the $R_{e}-M_{\ast}$ relation for galaxies with mass below the pivot stellar mass ($M_{\ast}<M_{p}$), (5) $\beta$ is the power-law slope for high-mass galaxies with $M_{\ast}>M_{p}$, (6) $r_{p}$ is the radius at the pivot stellar mass $M_{p}$, (7) $M_{p}$ is the pivot stellar mass, and (8) $\sigma_{\log r}$ is the intrinsic scatter of the $R_{e}-M_{\ast}$ relation, \edittwo{(9) Bayes-factor evidence as described in Section~\ref{sec:sizemass_quisf}.} We fit the $R_{e}-M_{\ast}$ relation of each  population down to its corresponding  $M_{\ast \mathrm{\lim}}$ at a given redshift bin. We obtain the best-fitting parameters by finding the model with the maximum total likelihood, $\mathcal{L}=\mathcal{L}_{\mathrm{Q}}+\mathcal{L}_{\mathrm{SF}}$ (see Equation~\ref{eq:loglike_qui} and~\ref{eq:loglike_sf}) and with strong Bayes-factor evidence ($\zeta>6$; see Section~\ref{sec:model_selection}) in preference of that model. When estimating the total likelihood, we also take the cross-contamination between quiescent and star-forming galaxies into account (see Section~\ref{sec:fitting_sizemass}). At all redshift bins, the $R_{e}-M_{\ast}$ distributions for both quiescent and star-forming populations show strong Bayesian evidence in preference of smoothly broken power-law fits. The uncertainties on parameters $R_{p}$ and $\log(M_{p}/M_{\odot})$ which are less than 0.1 are indicated as 0.}
\end{deluxetable*}
\subsection{Bayesian Model Selection}
\label{sec:model_selection}
 \par For each of our subsamples, we want to know if the shape of the observed $R_{e}-M_{\ast}$ relation is better fitted by a single or smoothly broken power-law function. A smoothly broken power-law function is however non-linear, implying that their number of degrees of freedom cannot be estimated. Therefore, we cannot simply compare their reduced $\chi^{2}$ values \citep{Andrae2010}. \editone{Instead, we use \textsc{Dynesty} to estimate the evidence $\mathcal{Z}_{M}$ (i.e., marginal likelihood) for the data given each pair of power-law models listed in Table~\ref{table:modellist}. We then use this evidence to compute the Bayes-factor evidence $\zeta$, which is defined as twice the natural log ratio of evidence of one model versus another model (i.e., $\zeta\equiv 2\times\ln(\mathcal{Z}_{1}/\mathcal{Z}_{2})$).} We adopt the significance criteria of \cite{Kass1995} who define the evidence to be ``very strong'' ($\zeta > 10$), ``strong'' ($6< \zeta<10$), ``positive'' ($2<\zeta<6$), or ``weak'' ($0<\zeta<2$) toward a power-law model $M_{1}$ (and equivalent negative values for evidence toward a power-law model $M_{2}$).  \edittwo{In this paper}, we refer to the best-fit $R_{e}-M_{\ast}$ relation for a galaxy sample with high $|\zeta|$ as having strong Bayes-factor evidence ($|\zeta|>6$) toward a given power-law model. However, we caution that Bayes-factor evidences do not necessarily mandate which of two models is correct but instead describe the evidence against the opposing models. For instance, the  $R_{e}-M_{\ast}$ relation of a galaxy sample with very strong evidence toward model 1 (e.g., $\zeta=8448$  promotes the rejection of model 2. Formally, it does not say that model 1 is the correct model (and vice versa).

\begin{figure*}
	\centering
	\includegraphics[width=1\textwidth]{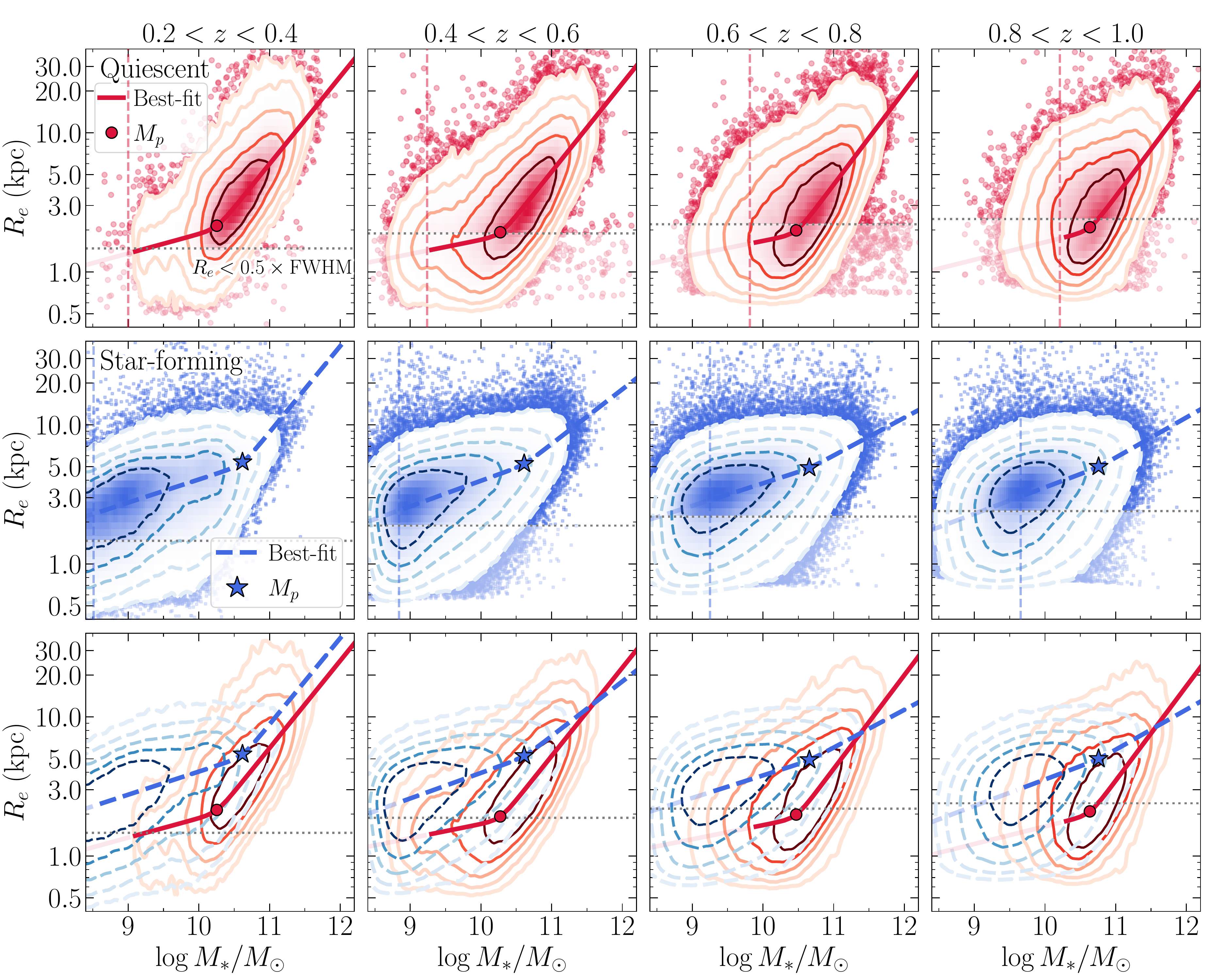}
	\caption{The observed size-mass ($R_{e}-M_{\ast}$) distribution of quiescent (red; top row) and star-forming galaxies (blue; middle row) in four redshift bins (from left to right). The contours indicate the $R_{e}-M_{\ast}$ distribution of galaxies from $1\sigma-3\sigma$ with the spacing of $0.5\sigma$. In all redshift bins, the $R_{e}-M_{\ast}$ distributions of both populations show strong Bayes-factor evidence ($|\zeta|>6$) in preference of smoothly broken power-law fits given by Equation~\ref{eq:doublebrokenpwl}, and are indicated as thick solid and thick dashed curves for quiescent and star-forming populations, respectively. We fit the relations for both populations down to the  $M_{\ast \mathrm{lim}}$ of each subsample in a given redshift bin (as indicated by vertical red and blue dashed line in the top and middle panels). We also take the cross-contamination between quiescent and star-forming populations and the catastrophic redshift failures into account (see Section~\ref{sec:fitting_sizemass} and Appendix~\ref{appendix:misclass}). The pivot stellar masses $M_{p}$  at which the slopes of the $R_{e}-M_{\ast}$ relations change for quiescent and star-forming galaxies are indicated as red circles and blue stars, respectively.  The bottom row shows the $R_{e}-M_{\ast}$ distributions of both populations together for comparison. The horizontal grey dotted line in each panel indicates where HSC size measurements are below the resolution limit of the HSC $i-$band image ($R_{e}<0.5\times$FWHM). At fixed stellar mass and redshift, star-forming galaxies are on average larger than quiescent galaxies, consistent with previous studies. }
	\label{fig:smfitbkpwlquisf}
\end{figure*}

\section{RESULTS}
\label{sec:results}
\subsection{The Size-Mass Relation for Quiescent and Star-Forming Galaxies at $0.2<z<1.0$}
\label{sec:sizemass_quisf}
\par
\begin{figure}
	\centering
	\includegraphics[width=0.47\textwidth]{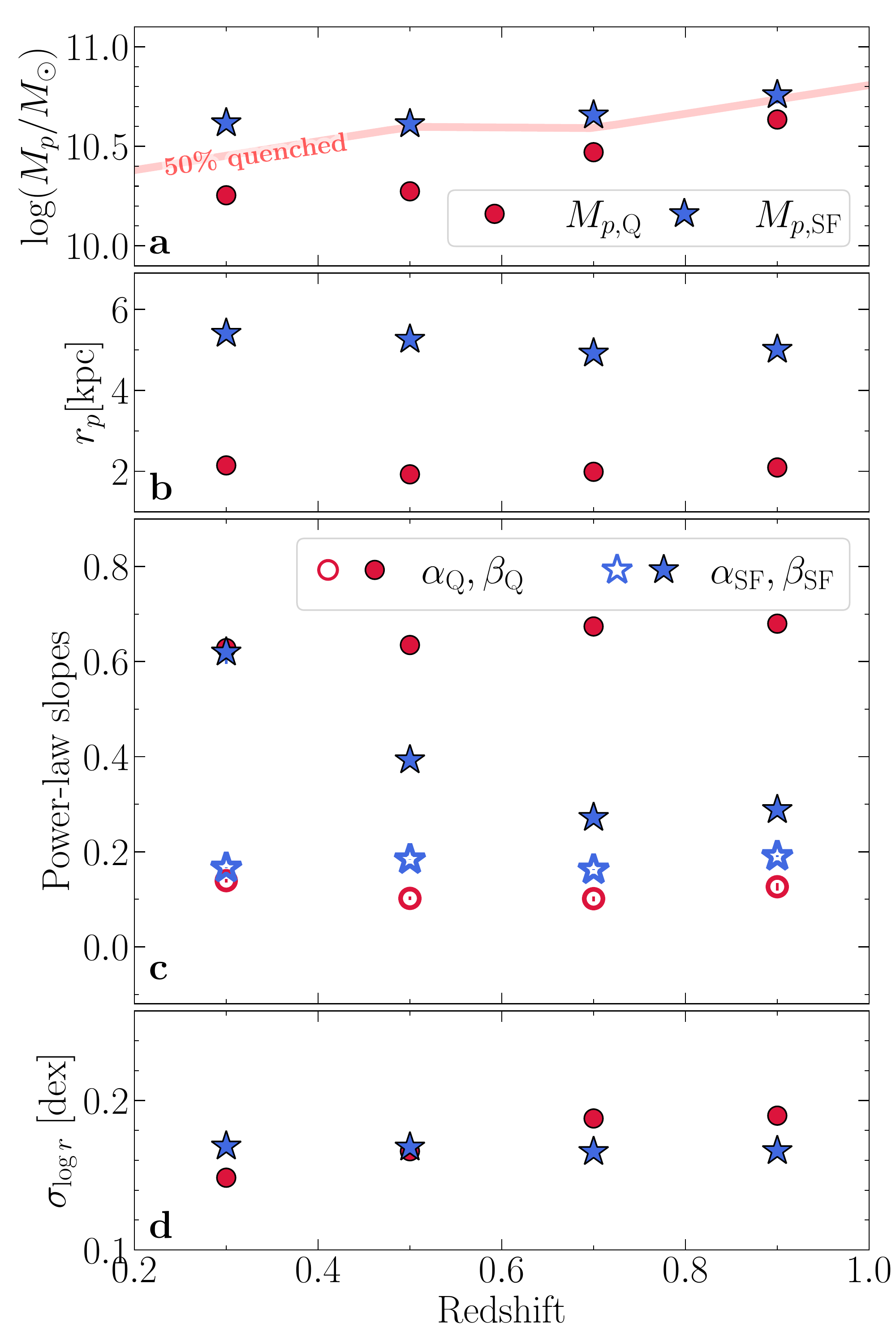}
	\caption{Redshift evolution of the best-fit parameters of the $R_{e}-M_{\ast}$ relations  for quiescent (red) and star-forming (blue) galaxies in four redshift bins from $z=1$ to $z=0.2$. As in Figure~\ref{fig:smfitbkpwlquisf}, the $R_{e}-M_{\ast}$ distributions for both quiescent and star-forming populations at all redshift bins are described by a smoothly broken power-law model. (a) shows the redshift evolution of the pivot stellar mass $M_{p}$ of the broken power-law fits for quiescent (circle) and star-forming (star) galaxies. The red line marks the stellar mass at which 50\% of the HSC galaxies are quenched, which coincides with the pivot stellar mass at a given redshift. The $M_{p}$ for quiescent galaxies decreases by 0.4 dex from $z=1.0$ to $z=0.2$, consistent with downsizing trend. (b) to (d) show the redshift evolution of pivot radius $r_{p}$, the power-law slope for low-mass galaxies with ($M_{\ast}<M_{p}$; $\alpha$; open symbols), the slope for high-mass galaxies ($M_{\ast}>M_{p}$; $\beta$; filled symbols), and the intrinsic scatter of the relation ($\sigma_{\log r}$). The error bars of the parameters are smaller than the symbols.}
	\label{fig:brokenpwl_evol}
\end{figure}

 In Figure~\ref{fig:smfitbkpwlquisf} we show the distributions of  $R_{e}-M_{\ast}$ for quiescent and star-forming galaxies, classified by using the $urz$ selection (see Section~\ref{sec:urz_selection}). The contours indicate the distribution of galaxies from $1\sigma$\footnote{In two dimensions, a Gaussian probability density function in polar coordinates is given by $f(r) =\frac{1}{2\pi \sigma^{2}} \exp\left ( -\frac{r^2}{2\sigma^2} \right) $, and the integral under this density is given by $\Phi(x)=\int_{0}^{x} f(r)dr= 1-\exp\left ( -\frac{x^2}{2\sigma^2} \right )$. Therefore, the $1\sigma$ level in the two-dimensional (2D) histogram corresponds to the Gaussian contains $1-\exp(-0.5)\sim39.3$ or 39.3\% of the volume of the 2D histogram.} to $3\sigma$ with the spacing of $0.5\sigma$.  First of all, over the stellar mass range of $\log(M_{\ast}/M_{\odot})\lesssim11$, the sizes of star-forming galaxies are on average larger than those of quiescent galaxies of similar stellar mass and redshift, confirming the results from previous works \citep[e.g.,][]{Trujillo2006,Williams2010,vanderWel2014,Faisst2017,Mowla2019b}. Furthermore, we find that the low-mass ($
\log(M_{\ast}/M_{\odot})\lesssim 10.3$) tail of the distribution of quiescent galaxies becomes shallow with a mild dependence of stellar mass on $R_{e}$, whereas the relation is relatively steeper for high-mass systems. This trend persists at all redshifts over the range $0.2<z<1.0$. However, the stellar mass-dependence of the slope of the $R_{e}-M_{\ast}$ relation is less clear for  star-forming galaxies, particularly at higher redshifts. We therefore fit both single and broken power-law models to the $R_{e}-M_{\ast}$ distribution for quiescent and star-forming galaxies and select the model with strong Bayesian evidence.
\edittwo{In Table~\ref{table:resultfitparams}, we provide parameters of the best-fitting relations for each subsample and power-law models. The best-fit parameter is obtained by taking the median of the parameter's marginalized posterior distribution, with the uncertainties quoted as the $16^{\mathrm{th}}$ through $84^{\mathrm{th}}$ percentiles. In the table, we also present
Bayes-factor evidence, estimated as twice the natural log ratio of evidences of a smoothly broken power-law model fitted to the $R_{e}-M_{\ast}$ relations of both quiescent and star-forming galaxies and compared to a single power-law model fitted to the relations of both populations.} In all redshift bins, we find that the $R_{e}-M_{\ast}$ relations for both quiescent and star-forming galaxies show very strong Bayes-factor evidences ($\zeta\gg100$) promoting a smoothly broken power-law model over a single power-law model. \editone{To better compare the best-fit parameters of the $R_{e}-M_{\ast}$ (power-law slopes $\alpha$, $\beta$, pivot stellar mass $M_{p}$, pivot radius $r_{p}$, and intrinsic scatter $\sigma_{\log r}$) for both populations and illustrate their redshift evolution, we plot the best-fit parameters for quiescent and star-forming galaxies as a function of redshift in Figure~\ref{fig:brokenpwl_evol}. }



\par Focusing on the quiescent population, at all redshifts, the $R_{e}-M_{\ast}$ relation  exhibits a change in slope at the pivot stellar mass of $\log(M_{p}/M_{\odot})=10.2-10.6$: the $R_{e}-M_{\ast}$ relation for galaxies below $M_{p}$ has shallower slope of $\alpha=0.10-0.14$, compared to those of more massive galaxies with relatively steeper slopes of $\beta=0.63-0.68$. The deviation from a single power-law at $M_{p}$ results in a higher likelihood for a smoothly broken power-law model than that for a single power-law model. The likelihood difference, when marginalized over all parameters, is reflected in the Bayes-factor evidence (Table~\ref{table:resultfitparams}). The redshift evolution of the best-fit parameters of the $R_{e}-M_{\ast}$ (Figure~\ref{fig:brokenpwl_evol}) shows that the pivot stellar mass ($M_{p}$) of the $R_{e}-M_{\ast}$ relation of quiescent population moderately decreases by $\sim0.4$~dex from $\log(M_{p}/M_{\odot})=10.6$ at $z\sim0.9$ to $\log(M_{p}/M_{\odot})=10.2$ at $z\sim0.3$. Moreover, we find that the intrinsic scatter of the $R_{e}-M_{\ast}$ slightly decreases from  $\sigma_{\log r}=0.19$ dex at $z\sim0.9$ to $\sigma_{\log r}=0.15$ dex at $z\sim0.3$, while the pivot radius is nearly constant with $r_{p}\sim2$~kpc, over the redshift range of $0.2<z<1.0$. 

\par For star-forming galaxies, the $R_{e}-M_{\ast}$ relation exhibits a change in slope at the pivot mass of  $\log(M_{p}/M_{\odot})=10.6-10.8$: the $R_{e}-M_{\ast}$ relation for galaxies below $M_{p}$ has shallower slope of $\alpha=0.16-0.19$, compared to those of more massive galaxies with slopes of $\beta=0.29-0.62$. Similar to quiescent galaxies,  the $M_{p}$ for star-forming galaxies weakly decreases by $\sim0.2$~dex from $\log(M_{p}/M_{\odot})=10.8$ at $z\sim0.9$ to $\log(M_{p}/M_{\odot})=10.6$ at $z\sim0.3$. Over the redshift range of $0.2<z<1.0$, we find no significant redshift evolution in the intrinsic scatter nor in the pivot radius, and they are consistent with  $\sigma_{\log r}\sim0.17$ and $r_{p}\sim5$~kpc, respectively. Interestingly, above the pivot stellar mass, the relation for star-forming galaxies clearly steepens with decreasing redshift (i.e., $R_{e}\propto M_{\ast}^{0.29}$ at $z\sim0.9$ and $R_{e}\propto M_{\ast}^{0.62}$ at $z\sim0.3$),  indicating that the size growth of these massive ($M_{\ast}>M_{p}$) star-forming galaxies depends much more on galaxy stellar mass at later cosmic time \citep[e.g.,][]{Paulino-Afonso2017,Mosleh2020}.

\par \editone{Finally, Figure~\ref{fig:brokenpwl_evol}a shows that, at a given redshift, the pivot stellar masses for both galaxy populations are nearly comparable to the stellar mass at which half of the galaxy population is quiescent. This finding corroborates the idea that the pivot stellar mass of the size-mass relation marks the mass above which both stellar mass growth and the size growth transition from being star formation dominated to being (dry) merger dominated, in good agreement with \cite{Mowla2019a} and \cite{Mosleh2020}.  Combining this observation with the evolution in power-law slope at the high-mass end of star-forming galaxies implies that the role of (dry) mergers in the mass and size growth is increasingly important for more massive galaxies and at lower redshifts, which has also been found in hydrodynamical simulations \citep[e.g.,][]{Rodriguez2016, Qu2017,Furlong2017, Clauwens2018,Pillepich2018,Davison2020}.}


\par To summarize, we determine size-mass relations 
using the large sample of HSC galaxies down to $\log(M_{\ast}/M_{\odot})=10.2~(9.2)$ at $z<1~(0.6)$ by taking into full account the uncertainties on size and cross-contamination between galaxy populations. This allows us to probe the shape of the $R_{e}-M_{\ast}$ distributions beyond the simple average relation and leads to one of the main conclusions in this work: in all redshift bins, the $R_{e}-M_{\ast}$ relations for both quiescent and star-forming galaxies show very strong Bayesian evidence promoting a broken power-law model with a clear change of power-law slope at a pivot mass $M_{p}$. Remarkably, the $M_{p}$ also nearly corresponds to the stellar mass at which 50\% of galaxy population is quiescent. Also, the relation at the low-mass end ($M_{\ast}<M_{p}$) for quiescent galaxies is shallow ($R_{e}\propto M_{\ast}^{0.1}$) similar to (or even shallower than) that of star-forming galaxies, suggesting that some of these \emph{low-mass} quiescent galaxies appear to have sizes comparable to those of star-forming galaxies of similar stellar mass at the same epoch. These results might encode important information on the formation paths of low- and high-mass galaxies. We will further discuss the implications of these findings in Section~\ref{sec:discussion}.

\begin{figure*}
	\centering
	\includegraphics[width=1\textwidth]{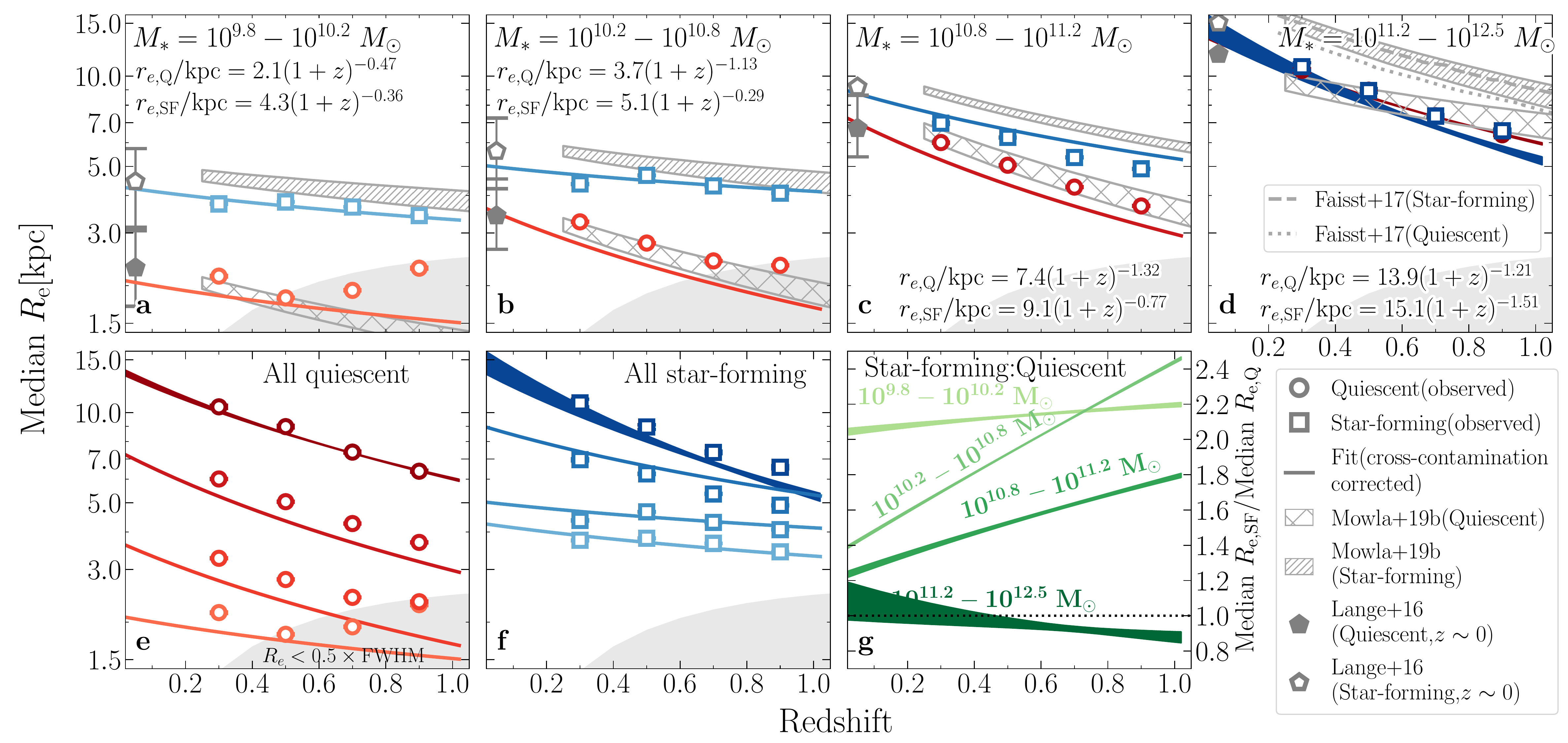}
	\caption{Evolution of galaxy sizes. (a) to (d) show the median size $R_{e}$ as a function of redshift for quiescent (open circle) and for star-forming (open square) galaxies at different stellar masses in bin of $d\log M_{\ast}=0.5 $ dex (light to dark colors). The uncertainties of median sizes are $\sim0.01-0.02$ kpc, and are smaller than the symbols. In each panel, the filled shaded region corresponds to the best-fit for the median sizes of the form $R_{e}=B_{z}\times(1+z)^{-\beta_{z}}$ and the inter-68th percentile after the cross-contamination between quiescent and star-forming populations is taken into account (as a result, some of the best-fit relations deviate from the observed median $R_{e}$; see text). The hatched regions are fits (and the inter-68th percentile range) to the median size evolution from \cite{Mowla2019b}. The median $R_{e}$ (with error bars) based on samples of early- and late-type galaxies at $z\sim0$ from GAMA \citep{Lange2015} are indicated in (a) to (d) as filled and open grey pentagon symbols, respectively.
	For a direct comparison, (e) and (f) show again the median size evolution (and best-fit results) for quiescent and star-forming galaxies, respectively, in all mass bins. (g) shows the ratio of sizes (inferred from the fits) of star-forming to those of the quiescent population as a function of redshift for four mass bins. The best-fit results for the most massive ($M_{\ast}>10^{11.4}~M_{\odot}$) quiescent (grey dotted curve) and star-forming galaxies (grey dashed curve) from \cite{Faisst2017} are shown in (d). The median sizes of all, quiescent, star-forming galaxies for different stellar mass bins are given in Table~\ref{table:mediansize_zevol}. }
	\label{fig:sizeevol4}
\end{figure*}
\subsection{Redshift Evolution of the Sizes}
\label{subsec:mediansize_evol}
In Figure~\ref{fig:sizeevol4} we show the median size evolution for galaxies in $0.5$ dex bins of stellar mass for quiescent and star-forming galaxies from $z=1.0$ to $z=0.2$.  We use the biweight estimator for the location and scale of a distribution to compute the median size and its scatter. The uncertainty on median $R_{e}$ is derived from bootstrap resampling.
\par We follow the same method as describe in Section~\ref{sec:fitting_sizemass} to fit the size evolution of galaxies by taking the cross-contamination between galaxy populations and catastrophic redshift failures into account. Because quiescent galaxies have a stellar mass distribution that is shifted to higher stellar masses compared to star-forming galaxies, we use the same method to account for this difference by assigning a weight to each galaxy that is inversely proportional to the number density. We use the number densities for quiescent and star-forming galaxies from the \cite{Tomczak2014} stellar mass functions. This ensures that each mass range (for both quiescent and star-forming galaxies) carries equal weight in the fit. We then assume  a lognormal distribution $N(\log (r_{e}), \sigma_{\log r_{e}})$, where $\log r_{e}$ is the mean and $\sigma_{\log r_{e}}$ is the dispersion, and we then parameterize $r_{e}$ as 
\begin{equation}
   r_{e}=B_{z}\times(1+z)^{-\beta_{z}}.
\label{eq:size_zevol}
\end{equation} 
Table~\ref{table:mediansize_zevol} provides the median $R_{e}$ and the best-fit parameters that describe its redshift evolution for all, quiescent, and star-forming galaxies at $0.2 < z < 1.0$.  


\par In Figure~\ref{fig:sizeevol4} we show the best-fit $R_{e}$ as a function of redshift ($R_{e}-z$) for each subsample after taking both cross-contamination between quiescent and star-forming populations and catastrophic redshift failures into account. Focusing on quiescent and star-forming galaxies in each mass bin (panels a to d), as expected, we find offsets between the observed median sizes and the best-fit $R_{e}$, particularly for quiescent galaxies  with $\log(M_{\ast}/M_{\ast})\lesssim11$ and at higher redshift bins, such that, at fixed redshift, the best-fit $R_{e}$ is smaller than the observed median size. 
\par To further quantify the relative sizes of star-forming and quiescent galaxies and their evolution with cosmic time, we use the best-fit $R_{e}-z$ relations to compute ratios of average $R_{e}$ of star-forming to those of quiescent galaxies ($R_{e,\mathrm{SF}}/R_{e,\mathrm{Q}}$) as a function of redshift (Figure~\ref{fig:sizeevol4}g). First of all, at fixed redshift and stellar mass over the range of $\log (M_{\ast}/M_{\odot})\lesssim11.2$, star-forming galaxies are on average larger than their quiescent counterparts. Second, over the  stellar mass range of $\log(M_{\ast}/M_{\odot})\lesssim11.2$, the relative size of the two populations decreases toward lower redshifts, except perhaps galaxies with $9.8<\log(M_{\ast}/M_{\odot})<10.2$. For example, for galaxies with $10.2<\log(M_{\ast}/M_{\odot})<10.8$, the star-forming galaxies are on average larger than those of quiescent counterparts roughly by a factor of 2.4 at $z\sim1$. Both of these populations grow their sizes but quiescent galaxies do so at a faster rate ($R_{e}\propto(1+z)^{-1.13}$ compared to $R_{e}\propto(1+z)^{-0.29}$ for star-forming galaxies) such that at $z=0.2$, star-forming galaxies are on average larger than their quiescent counterparts roughly by a factor of 1.6. 

\par \editone{Owing to sufficiently large volume and sample sizes for both quiescent and star-forming galaxies at $\log(M_{\ast}/M_{\odot})>11.2$ probed by the HSC, we provide statistically robust measurements of the size evolution for the most massive galaxies out to $z=1$. The median size evolution of quiescent 
and star-forming galaxies at these masses are statistically identical at all redshifts (Figure~\ref{fig:sizeevol4}d). One could expect that the effect of cross-contamination is increasingly important for the these massive galaxies due to our limitation to distinguish between truly quiescent galaxies and dusty star-forming galaxies at these very high masses using rest-frame colors (see Section~\ref{sec:urz_selection} and Appendix~\ref{appendix:misclass}).} After we take this effect into account,  we find that the average sizes of the most massive quiescent at $z>0.6$ are a factor $\sim1.1$ larger than those of star-forming galaxies of similar mass, while at lower redshift massive star-forming galaxies have average sizes comparable to  those of the counterpart quiescent galaxies ($R_{e,\mathrm{SF}}/R_{e,\mathrm{Q}}\sim1$). In contrast to the trend for lower masses, we find that the sizes of these massive star-forming galaxies evolve from $z=1$ to $z=0.2$ with faster rate as $R_{e}\propto(1+z)^{-1.51}$ relative to the counterpart quiescent galaxies ($R_{e}\propto(1+z)^{-1.21}$), and the normalization (i.e., $R_{e}$ measured at $z=0$) of the $R_{e}-z$ relations for both populations is consistent within uncertainties ($B_{z}=15.1\pm1.4$~kpc and $13.9\pm0.3$~kpc for star-forming and quiescent galaxies, respectively).

\par \editone{ As shown in Figure~\ref{fig:sizeevol4}, our median sizes at fixed stellar mass are smaller by $\sim0.1$ dex than those of \cite{Mowla2019b}, particularly for star-forming galaxies with $\log(M_{\ast}/M_{\odot})>10.8$ and at lower redshifts.  This size offset is consistent with the results from our tests using a set of simulated galaxies (Appendix~\ref{appendix:verification_sizes}), where we found that, even after we applied the correction for systematic biases in size measurements, sizes of the HSC galaxies, particularly for intrinsically large ones, could still be underestimated. For instance, galaxies with $21<i<23$ in the Wide layer and having sizes larger than $\sim5\arcsec$ (roughly corresponding to $\sim17$~kpc at $z=0.2$) could be systematically underestimated by $\lesssim0.1$~dex. Therefore, we attribute the discrepancy between our results and those of \cite{Mowla2019b} to the effect of surface brightness dimming near the outskirts of the HSC massive galaxies. Despite all of these effects, we will demonstrate in the following section that, over  the entire stellar mass range we probed, the rates of size evolution for both quiescent and star-forming galaxies are in good agreement (within uncertainties) with those from previous works utilizing space-based observations.}



\par Finally, in Figure~\ref{fig:sizeevol4}, we  additionally show the comparison between the size measurement for galaxies at $z=0.01-0.1$ from GAMA \citep{Lange2015}, which are in good agreement (within the uncertainties) with our best-fit size evolution extrapolated to $z\sim0$ for both quiescent and star-forming populations 

\startlongtable
\begin{deluxetable*}{ccccccc}
\tablecaption{Median Sizes of Galaxies as a Function of Galaxy Mass and Redshift, and Redshift Dependencies, Parameterized by $R_{e}=B_{z}\times(1+z)^{-\beta_{z}}$}
\label{table:mediansize_zevol}
\tablehead{ \colhead{Median Stellar mass} & \colhead{$z_{\mathrm{med}}$} & \colhead{All Median $R_{e}$(kpc)} && \colhead{Quiescent Median $R_{e}$(kpc)} & & \colhead{Star-forming Median $R_{e}$(kpc)} \\}
\startdata
\multirow{6}{*}{$\log(M_{\ast}/M_{\odot})=9.8-10.2$}& $0.3$ & $3.47\pm0.01$ & & $2.15\pm0.01$ & & $3.75\pm0.01$  \\
 & $0.5$ & $3.47\pm0.01$ & & $1.82\pm0.01$ & & $3.81\pm0.01$ \\
 & $0.7$ & $3.40\pm0.01$ & & $1.93\pm0.01$ & & $3.66\pm0.01$ \\
 & $0.9$ & $3.37\pm0.00$ & & $2.29\pm0.02$ & & $3.43\pm0.00$ \\
& $B_{z}$ &$3.39\pm0.01$ & &  $2.10\pm0.02$ & & $4.28\pm0.02$ \\
& $\beta_{z}$ & $0.08\pm0.01$ & & $0.47\pm0.02$ & & $0.36\pm0.01$ \\
\hline
\multirow{6}{*}{$\log(M_{\ast}/M_{\odot})=10.2-10.8$}& $0.3$ & $3.79\pm0.01$ & & $3.27\pm0.01$ & & $4.37\pm0.02$  \\
 & $0.5$ & $3.88\pm0.01$ & & $2.78\pm0.01$ & & $4.67\pm0.01$ \\
 & $0.7$ & $3.57\pm0.01$ & & $2.42\pm0.01$ & & $4.31\pm0.01$ \\
 & $0.9$ & $3.62\pm0.01$ & & $2.34\pm0.01$ & & $4.07\pm0.01$ \\
& $B_{z}$ &$3.86\pm0.02$ & &  $3.70\pm0.02$ & & $5.05\pm0.02$ \\
& $\beta_{z}$ & $0.19\pm0.01$ & & $1.13\pm0.01$ & & $0.29\pm0.01$ \\
\hline
\multirow{6}{*}{$\log(M_{\ast}/M_{\odot})=10.8-11.2$}& $0.3$ & $6.18\pm0.02$ & & $6.01\pm0.03$ & & $6.96\pm0.06$  \\
 & $0.5$ & $5.48\pm0.01$ & & $5.04\pm0.02$ & & $6.24\pm0.02$ \\
 & $0.7$ & $4.66\pm0.01$ & & $4.27\pm0.01$ & & $5.36\pm0.02$ \\
 & $0.9$ & $4.19\pm0.01$ & & $3.69\pm0.01$ & & $4.91\pm0.02$ \\
& $B_{z}$ &$7.64\pm0.05$ & &  $7.42\pm0.06$ & & $9.07\pm0.12$ \\
& $\beta_{z}$ & $1.02\pm0.01$ & & $1.32\pm0.02$ & & $0.77\pm0.02$ \\
\hline
\multirow{6}{*}{$\log(M_{\ast}/M_{\odot})=11.2-12.5$}& $0.3$ & $10.48\pm0.13$ & & $10.45\pm0.14$ & & $10.76\pm0.32$  \\
 & $0.5$ & $8.97\pm0.07$ & & $8.97\pm0.08$ & & $8.94\pm0.15$ \\
 & $0.7$ & $7.36\pm0.04$ & & $7.38\pm0.05$ & & $7.36\pm0.07$ \\
 & $0.9$ & $6.41\pm0.04$ & & $6.37\pm0.04$ & & $6.57\pm0.09$ \\
& $B_{z}$ &$13.85\pm0.26$ & &  $13.88\pm0.34$ & & $15.09\pm1.38$ \\
& $\beta_{z}$ & $1.28\pm0.04$ & & $1.21\pm0.05$ & & $1.51\pm0.17$ \\
\hline
\enddata
\end{deluxetable*}

\begin{figure}
	\centering
	\includegraphics[width=0.45\textwidth]{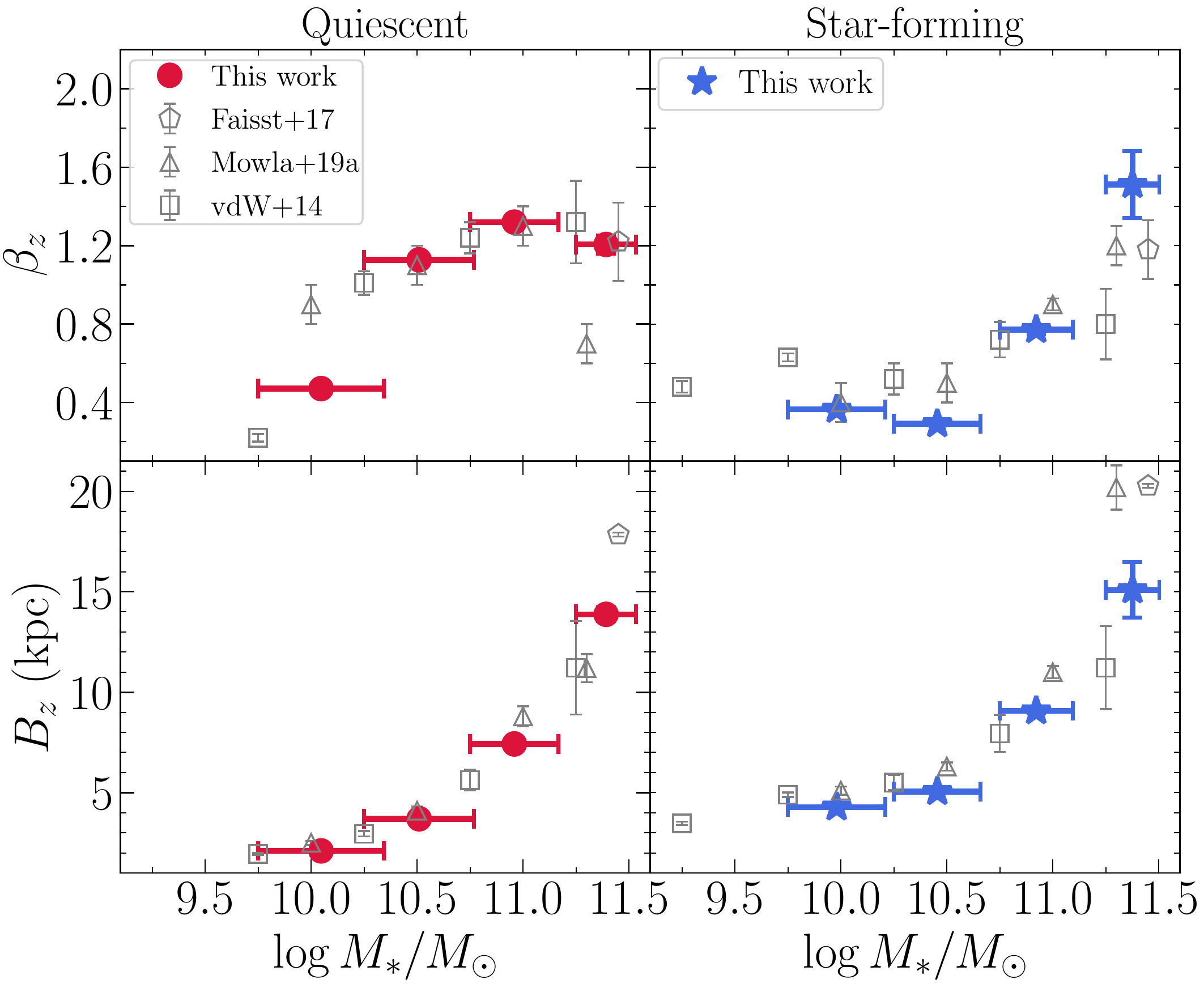}
	\caption{The stellar mass-dependence of the best-fit parameters for the median size evolution as given in Equation~\ref{eq:size_zevol} for quiescent (red) and star-forming (blue) galaxies. Top: the power-law slope ($\beta_{z}$)  of the size evolution as a function of stellar mass. Bottom: the normalization ($B_{z}$) of the size evolution as a function of stellar mass. }
	\label{fig:redshiftevol_massdep}
\end{figure}
\subsubsection{The Stellar-Mass Dependence of Size Evolution}
\par To better visualize the dependence of the redshift evolution of the sizes on stellar mass, in Figure~\ref{fig:redshiftevol_massdep}, we show the power-law slope ($\beta_{z}$) and normalization ($B_{z}$) of the $R_{e}-z$ relation as a function of stellar mass for quiescent and star-forming galaxies. First, regardless of galaxy star formation activity, a strong trend is immediately apparent: more massive galaxies undergo significantly faster size evolution with redshift, as indicated by the increasing power-law slope ($\beta_{z}$) with increasing stellar mass. Second, on average, the sizes of quiescent galaxies evolve with redshift faster than those of star-forming galaxies of similar stellar mass, with the exception for galaxies in the most massive bin ($\log(M_{\ast}/M_{\odot})>11.2$).
\par Focusing on quiescent galaxies with $10.2<\log(M_{\ast}/M_{\odot})<11.2$, the average sizes of more massive galaxies evolve with redshift faster than those of the lower mass counterparts. However, the rate of size evolution for more massive galaxies with $\log(M_{\ast}/M_{\odot})>10.8$ becomes nearly independent with stellar mass. On the other hand, the  normalization $B_{z}$ of the $R_{e}-z$ relation moderately increases with increasing stellar mass from $B_{z}=2.1$ kpc at $9.8<\log(M_{\ast}/M_{\odot})<10.2$ to $B_{z}=13.9$ kpc at $\log(M_{\ast}/M_{\odot})>11.2$.
\par For star-forming galaxies, the average size of galaxies with $9.8<\log(M_{\ast}/M_{\odot})<10.8$ evolves with redshift as $R_{e}\propto(1+z)^{-0.36}$, nearly independence with stellar mass. In contrast, above this mass range, the average sizes of galaxies evolve with redshift faster for more massive galaxies, and this trend continues even for most massive galaxies with $\log(M_{\ast}/M_{\odot})>11.2$, consistent with the results from previous works \citep[e.g.,][]{Mowla2019b,Paulino-Afonso2017}. Finally, the normalization $B_{z}$ of the $R_{e}-z$ relation moderately increases with increasing stellar mass from $B_{z}=4.3$ kpc for galaxies with $9.8<\log(M_{\ast}/M_{\odot})<10.2$ to $B_{z}=15.1$ kpc for galaxies with $\log(M_{\ast}/M_{\odot})>11.2$.


\par In summary, although both quiescent and star-forming galaxies grow in their stellar mass and size over cosmic time, we observed strong evidence for differential size evolution with mass: more massive galaxies exhibits a more rapid size evolution, regardless of being quiescent or star-forming galaxies. For the subsample classified by their $u-r$ and $r-z$ rest-frame colors, at fixed stellar mass over the range of  $\log(M_{\ast}/M_{\odot})\sim 10-11$, the average size of quiescent galaxies evolves more rapidly than that of star-forming galaxies.  
\section{Discussion}
\label{sec:discussion}
\par The main finding of our study of $\sim1,500,000$ galaxies with $i<24.5$ from the three layers of the HSC  PDR2 (Wide and Deep+UltraDeep) at $0.2<z<1.0$ is that the $R_{e}-M_{\ast}$ distributions of both quiescent and star-forming galaxies show strong Bayesian evidence in preference of a broken power-law model over a single power-law model --
the $R_{e}-M_{\ast}$ steepens above a pivot mass $M_{p}$, while it flattens below $M_{p}$. At a given redshift, the $M_{p}$ for both quiescent and star-forming galaxies is also similar to the mass where the fraction of quiescent galaxies reaches 50\% (Figure~\ref{fig:brokenpwl_evol}a).

\par \editone{Previous observational studies of the galaxy size-mass relation have been based mostly on deep pencil-beam surveys
\citep[see e.g.,][]{vanderWel2014,Whitaker2017,Huang2017,Mowla2019a, Mosleh2020} and provided a global view on the evolution of galaxy sizes across a wide range in redshift, stellar mass, and star formation activity. However, given the small angular coverage of these surveys, previous studies have typically been hampered by small sample sizes of massive galaxies  (i.e., $\log(M_{\ast}/M_{\odot})\gtrsim11$) at low redshift ($0.1\lesssim z\lesssim1$); thus the size-mass relation at the most massive end over this redshift range has not been well constrained.  For instance, the slope of the size-mass relation could be biased to shallower (steeper) values if these massive galaxies, which are not sufficiently sampled, on average have larger (smaller) size than the extrapolation from the relation of lower mass galaxies (see Section~\ref{sec:lmasslt11}). Additionally, the size-mass relations of quiescent and star-forming galaxies are traditionally parameterized using a single power-law function, which is not sufficient to characterize the relations over a broad range of galaxy stellar mass, as we have clearly demonstrated here by comparing Bayes factor evidences \citep[see also e.g.,][]{Cappellari2013,Norris2014,Lange2015,Whitaker2017,Hill2017,Zhang2019,Mowla2019a,Mosleh2020,Nedkova2021}. For instance, \cite{vanderWel2014} analyzed structural parameter measures from CANDELS imaging and found evidence for the steepening of the $R_{e}-M_{\ast}$ relation for massive star-forming galaxies out to $z=1$, but their sample contains too few of such objects to perform a robust broken power-law fit. }

\par On the other hand, our measurements of the galaxy size-mass relation depend on the reliability of the size measurements using the ground-based HSC $i-$band imaging and the quiescent/star-forming galaxy classification based on the $urz$ selection. To gauge the reliability, in the following section we will compare our results to previous works. In Sections~\ref{sec:discuss_starforming} and~\ref{sec:discuss_quiescent}, we will further discuss the implications of all these findings for the evolutionary paths of star-forming and quiescent galaxies



\subsection{Our Results in Context}
\label{sec:compare_previousworks}

\begin{figure}

	\centering
	\includegraphics[width=0.48\textwidth]{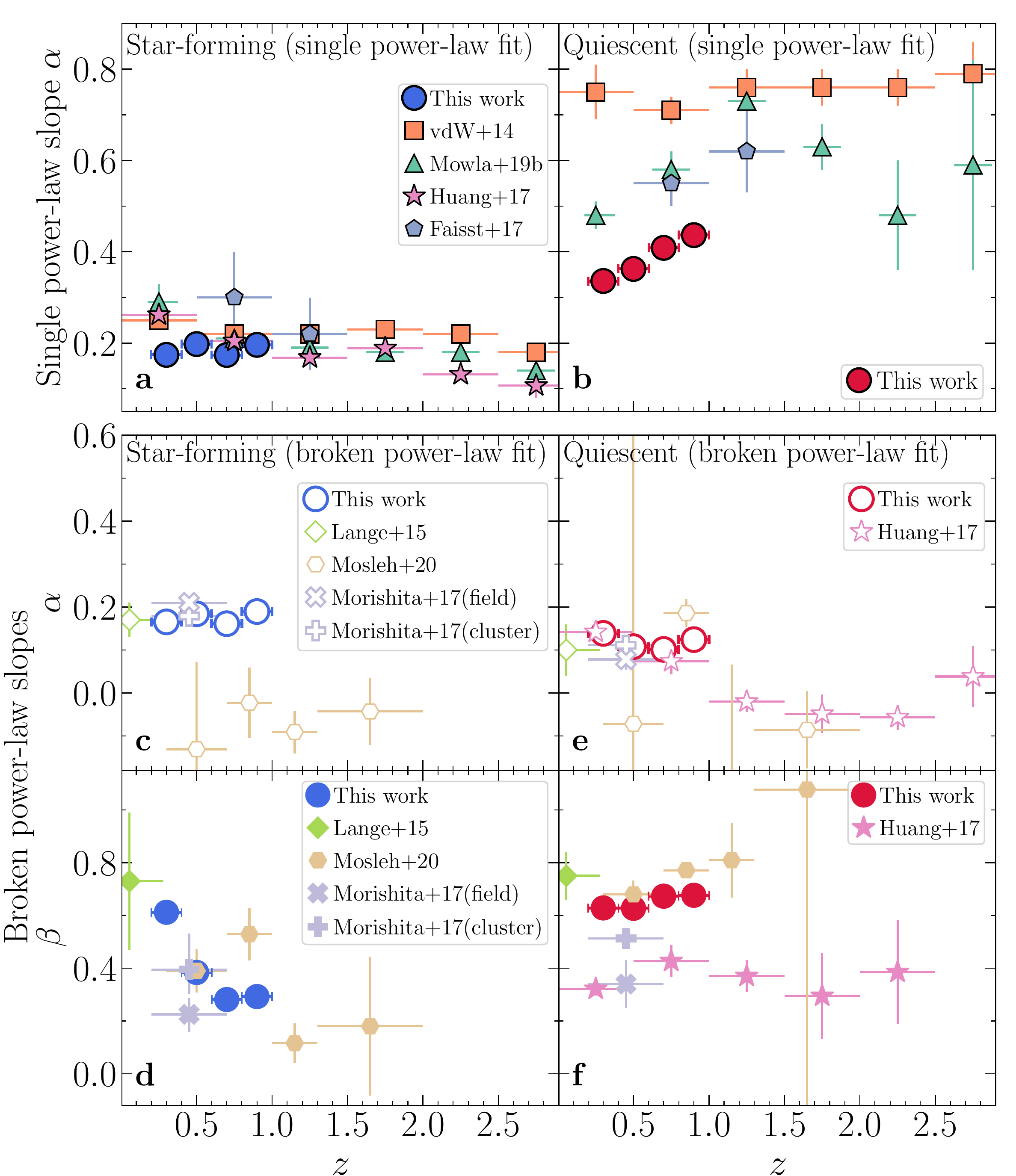}
	\caption{Redshift evolution of the best-fit power-law slopes of the $R_{e}-M_{\ast}$ relations for star-forming (left panels) and quiescent (right panels) galaxies. (a) and (b) show the redshift evolution of the slope $\alpha$ derived by fitting a single power-law to the relations using the data from HSC (circle), and other efforts as indicated. (c) to (f) show the redshift evolution of broken power-law slopes: low-mass end slope $\alpha$ ((c) and (e)) and high-mass end slope $\beta$ ((d) and (f)). }
	\label{fig:powerlawslope_others}
\end{figure}

\label{sec:compare_z1works}

\par To facilitate the comparison between our results with those from other studies, in Figure~\ref{fig:powerlawslope_others}, we show the power-law slopes of the $R_{e}-M_{\ast}$ relations as a function of redshift taken from this work, the 3D-\emph{HST}+CANDELS  \citep{vanderWel2014,Huang2017,Mosleh2020}, COSMOS-DASH \citep{Mowla2019b}, COSMOS/UltraVISTA \citep{Faisst2017}, the deep Hubble Frontier Fields imaging and slitless spectroscopy from the Grism Lens-Amplified Survey from Space \citep[GLASS;][]{Morishita2017}, and the GAMA survey \citep{Lange2015}.  We first focus on the comparisons of power-law slopes derived by fitting a single power-law to the relation. For star-forming galaxies, our single power-law slope of $\alpha_{\mathrm{single}}\sim0.2$ (Figure~\ref{fig:powerlawslope_others}a) is in good agreement (within the uncertainties) with those of \cite{vanderWel2014}, \cite{Mowla2019b}, \cite{Faisst2017}, and \cite{Huang2017}, demonstrating that the slope of the relation for star-forming galaxies is nearly constant at least out to $z\sim3$. For quiescent galaxies, our single power-law slopes of $\alpha_{\mathrm{single}}\sim0.34-0.42$ are shallower than other studies (Figure~\ref{fig:powerlawslope_others}b). The difference is likely driven by the fact that we fit the $R_{e}-M_{\ast}$ relation down to low-mass galaxies ($\log(M_{\ast}/M_{\odot})=9.0$ at $z<0.4$), which are better characterized by a shallow power-law slope, compared to more massive galaxies. In contrast, both \cite{vanderWel2014} and \cite{Mowla2019b} avoid the flatter part of the $R_{e}-M_{\ast}$ relation for quiescent galaxies below $M_{\ast}=10^{10.3}~M_{\odot}$ and fit the relation above this mass.  We therefore expect that the single power-law slopes of quiescent galaxies from those studies will be better in agreement with our broken power-law slopes at the high-mass end ($M_{\ast}>M_{p}$; see further discussion below).


\par Our results derived by fitting a smoothly broken power-law model to $R_{e}-M_{\ast}$ relations can be directly compared to those of \cite{Huang2017}, who utilized the size measurements from \cite{vanderWel2012} and adopted fainter selection limits for the galaxy sample in Deep region of CANDELS ($H_{160}=25.2$~mag) and the Hubble Ultra-Deep Field (HUDF; $H_{160}=26.7$~mag) region. As a result, \cite{Huang2017} is able to extend the $R_{e}-M_{\ast}$ relations to much lower stellar masses than those of \cite{vanderWel2014} and show that the median $R_{e}-M_{\ast}$ relation of the quiescent population flattens below $\log(M_{\ast}/M_{\odot})\sim10$, confirming the \cite{vanderWel2014} result at the low-mass end. In order to perform a direct comparison with the results of \cite{Huang2017}, we fit a smoothly broken power-law function to median $R_{e}-M_{\ast}$ relations for their late- and early-type galaxy samples (Figure 10 in their study) using \textsc{curvefit} of \textsc{scipy}. \editone{As shown in Figure~\ref{fig:powerlawslope_others}e, our fitting method of a size-mass relation, which takes into full account both random and systematic uncertainties in galaxy size and cross-contamination between galaxy populations, provides us a robust constraint on power-law slope $\alpha$ for quiescent galaxies at $0.2<z<1.0$ of $\alpha=0.10-0.14$. This result is well-matched to those of \cite{Huang2017} ($\alpha\sim0.07-0.14$). Results from the \cite{Huang2017} and those presented in this work show that, while the slopes of the size-mass relation at the low-mass end for quiescent galaxies is flat or even negative at $z\gtrsim1.5$, it becomes steeper ($\alpha\sim0.1$) at lower redshift and consistent with the result from the local universe \citep[e.g.,][green diamond in Figure~\ref{fig:powerlawslope_others}e]{Lange2015}. Interestingly, at $z\lesssim0.5$, the $R_{e}-M_{\ast}$ relation for these low-mass quiescent galaxies nearly coincides with that of the star-forming galaxies,} \edittwo{suggesting a connection between the quenching of star formation and the size evolution for low-mass galaxies with $M_{\ast}<M_{p}$ \citep[e.g.,][]{Barro2013,Woo2015,Belli2015,Wu2018,Morishita2017,Woo2019,Suess2021,Chen2020}.}

\par \editone{On the other hand, for quiescent galaxies above the pivot mass, the broken power-law slope of \cite{Huang2017} ($\beta\sim0.32-0.43$) is shallower than our values of $\beta=0.63-0.68$ (Figure~\ref{fig:powerlawslope_others}f), but this difference is  mostly due to difference in mass ranges probed between the two studies: \cite{Huang2017} sample is less massive ($\log(M_{\ast}/M_{\odot})\lesssim11$) than our sample (and also \cite{vanderWel2014} and \cite{Mowla2019b}). This will be more evident when we compare the single power-law slopes from \cite{Mowla2019b}, who also included a galaxy sample with $\log(M_{\ast}/M_{\odot})>11.3$, to our broken power-law slopes at the high-mass end (rather than our single power-law slopes). Those values are in better agreement (within the uncertainties). Moreover, the difference between the \cite{Huang2017} and the results presented in this paper could possibly arise from the differences in the depth and resolution of \emph{HST} and HSC images. We have found that HSC-based galaxy sizes are on average $0.01-0.04$ dex smaller than the $HST-$based, and the size offset is larger for galaxies at fainter magnitudes and with higher S\'{e}rsic indices (Appendix~\ref{appendix:verification_sizes} and Figure~\ref{fig:comparesize_candels}). Another possibility of the difference could be due the difference in galaxy classification: \cite{Huang2017} divided their sample into late- and early-type galaxies on the basis of S\'{e}rsic index or specific star formation rate rather than rest-frame colors. Even though there is a general correspondence between these different proxies for quiescent and star-forming galaxies, it is not perfect. }


\par \editone{In Figure~\ref{fig:powerlawslope_others}, we also compare the HSC power-law slopes of size-mass relations for both quiescent and star-forming galaxies to those from \cite{Morishita2017}, who measured $R_{e}-M_{\ast}$ relations of cluster and field galaxies at $0.2<z<0.7$ utilizing deep Hubble Frontier Fields imaging and slitless spectroscopy from GLASS. These authors split the sample at $\log(M_{\ast}/M_{\odot})=9.8$ and separately fitted the relations for the low- ($7.8<\log(M_{\ast}/M_{\odot})<9.8$) and high-mass ($\log(M_{\ast}/M_{\odot})\ge 9.8$) galaxies. For low-mass quiescent galaxies, \cite{Morishita2017} reported shallow power-law slopes for both field ($\alpha=0.08$) and cluster ($\alpha=0.11$) galaxies, while the slopes for more massive galaxies are steeper with more pronounced change in slope for those residing in clusters ($\beta=0.51$) compared to the field environments ($\beta=0.34$). For the star-forming population, the slopes for low-mass field and cluster galaxies are also shallow and consistent with each other ($\alpha=0.21$  and $\alpha=0.18$ for field and cluster galaxies, respectively), while the steepening of the slope at the high-mass end is more evident for galaxies residing in clusters ($\beta=0.40$) compared to the field environments ($\beta=0.23$). Our broken power-law slopes at the low-mass end ($\alpha$) for both quiescent and star-forming galaxies are in excellent agreement with those of \cite{Morishita2017}, regardless of environments (field/cluster galaxies). Our broken power-law slopes at the high-mass end ($\beta$) are also in good agreement with those of \citeauthor{Morishita2017} for high-mass galaxies, particularly for those residing in clusters.  }

\par \editone{Recently, \cite{Mosleh2020} measured the size-mass relations based on sizes containing 50\% of the total stellar mass ($r_{50,\mathrm{mass}}$) for galaxies at $0.3<z<2$ from CANDELS/3D-\emph{HST}. Figure~\ref{fig:powerlawslope_others}d shows that, for star-forming galaxies over the similar redshift range, our values for the high-mass end slopes are in good agreement with the result of \cite{Mosleh2020} ($\beta=0.39-0.53$). These authors also found that the slope $\beta$ steepens as decreasing redshift since $z\sim2$. Based on the light-based galaxy sizes, we confirm this finding of \cite{Mosleh2020} that the evolution in the high-mass end slope $\beta$ is statistically significant and also persists over the redshift range of $0.2<z<1.0$, thanks to the sufficiently large survey volume probed the HSC to include massive star-forming galaxies at this redshift.  On the other hand, \cite{Mosleh2020} reported nearly flat or even a negative low-mass end slope for star-forming galaxies at $0.3<z<1.0$ (Figure~\ref{fig:powerlawslope_others}c). This difference is most probably attributed to the difference between mass-based sizes and light-based sizes. For instance, \cite{Suess2019ApJ} found that, at $1.0\le z\le2.5$, the slopes of the relations based on half-mass radii are $\sim0.1-0.3$~dex shallower than those of the relations based on half-light radii, broadly consistent with the difference between the low-mass end slopes from the HSC and from \cite{Mosleh2020}.}

\par \edittwo{Moreover,  \cite{Mosleh2020} reported a $\sim0.3$~dex decrease in the pivot mass of the $R_{e}-M_{\ast}$ relation for star-forming galaxies from $\log(M_{p}/M_{\odot})=10.69\pm0.12$ at $0.7<z<1.0$ to $\log(M_{p}/M_{\odot})=10.37\pm0.17$ at $0.3<z<0.7$, while the pivot mass for quiescent galaxies decreases by $\sim0.4$~dex from $\log(M_{p}/M_{\odot})=10.61\pm0.03$ at $0.7<z<1.0$ to $\log(M_{p}/M_{\odot})=10.22\pm0.20$ at $0.3<z<0.7$. The \cite{Mosleh2020} results are in good agreement with our finding that $\log(M_{p})$ of the relations for quiescent and star-forming galaxies decrease by $\sim0.4$ dex and $\sim0.2$ dex, respectively, from $z\sim1$ to $z\sim0.2$ (Figure~\ref{fig:brokenpwl_evol}).}



\par \edittwo{For quiescent galaxies in our highest redshift bin ($0.8<z<1.0$), the  pivot stellar mass of $\log(M_{p}/M_{\odot})=10.6$ is only 0.4 dex above the mass completeness limit of $\log(M_{\mathrm{lim}}/M_{\odot})=10.2$. One could raise a question of whether we have many galaxies below the pivot mass to constrain the power-law slope of the $R_{e}-M_{\ast}$ relation at the low-mass end.  To test the reliability of our result of shallow power-law slope, we adopt a fainter selection limit of $i=25.5$ and re-estimate a stellar mass completeness limit using a galaxy sample from the Deep+UltraDeep layer. At this faint magnitude, our analysis using a set of simulated galaxies indicates that we are able to recover sizes of galaxies with $24.5<i<25.5$ with 5\%-10\% level accuracy, particularly for those smaller than $\sim1\arcsec$ (corresponding to $\sim$ 8 kpc at $z\sim1$). Because low-mass quiescent galaxies below the pivot mass ($\log(M_{p}/M_{\odot})=10.6$) at this high redshift tend to have sizes smaller than this limit, the size measurements of these galaxies should be less prone to the systematic biases due to surface brightness dimming. For our testing purpose, we restrict the redshift range of quiescent galaxies to $0.8<z<0.9$, where our sample of quiescent galaxies are mass-complete down to $\log(M_{\ast}/M_{\odot})=9.6$. We then re-fit the $R_{e}-M_{\ast}$ relation, and find the the relation at the low-mass end for quiescent galaxies is flatter  ($\alpha\sim0$) than that ($\alpha\sim0.1$) derived using the full sample of quiescent galaxies at $0.8<z<1.0$, but it is in excellent agreement with \cite{Huang2017}. We therefore conclude that our finding of the shallow $R_{e}-M_{\ast}$ relation of quiescent galaxies at the highest redshift bin is not driven by small sample size of low-mass quiescent galaxies.}

\par \editone{Finally, our results also concur with previous studies for galaxies in the local Universe \cite[e.g.,][]{Shen2003,Mosleh2013,Lange2015,Zhang2019}, in the sense that the size-mass relation for quiescent galaxies at $z\sim0$ is better characterized by a smoothly broken power-law relation with a clear change of slope at a pivot mass ($\log(M_{p}/M_{\odot})\sim10.1-10.6$), and the slope of the relation below the pivot mass shallower  (or nearly flat, $\alpha\sim0$) than those at the higher mass ($\beta\sim0.6-0.8$). The trend is similar for star-forming galaxies, but with the shallower high-mass end slopes ($\beta\sim0.4-0.7$), compared to quiescent galaxies.  Our results from HSC can be directly compared with those of \cite{Lange2015}, who measured half-light radii (elliptical semimajor axis) by fitting 2D S\'{e}rsic light profiles, as we do here. As shown in Figures~\ref{fig:powerlawslope_others}c and d, the \citeauthor{Lange2015} results of power-law slopes (both low- and high-mass galaxies) for both populations are in good agreement with ours.}

\begin{figure}
	\centering
	\includegraphics[width=0.45\textwidth]{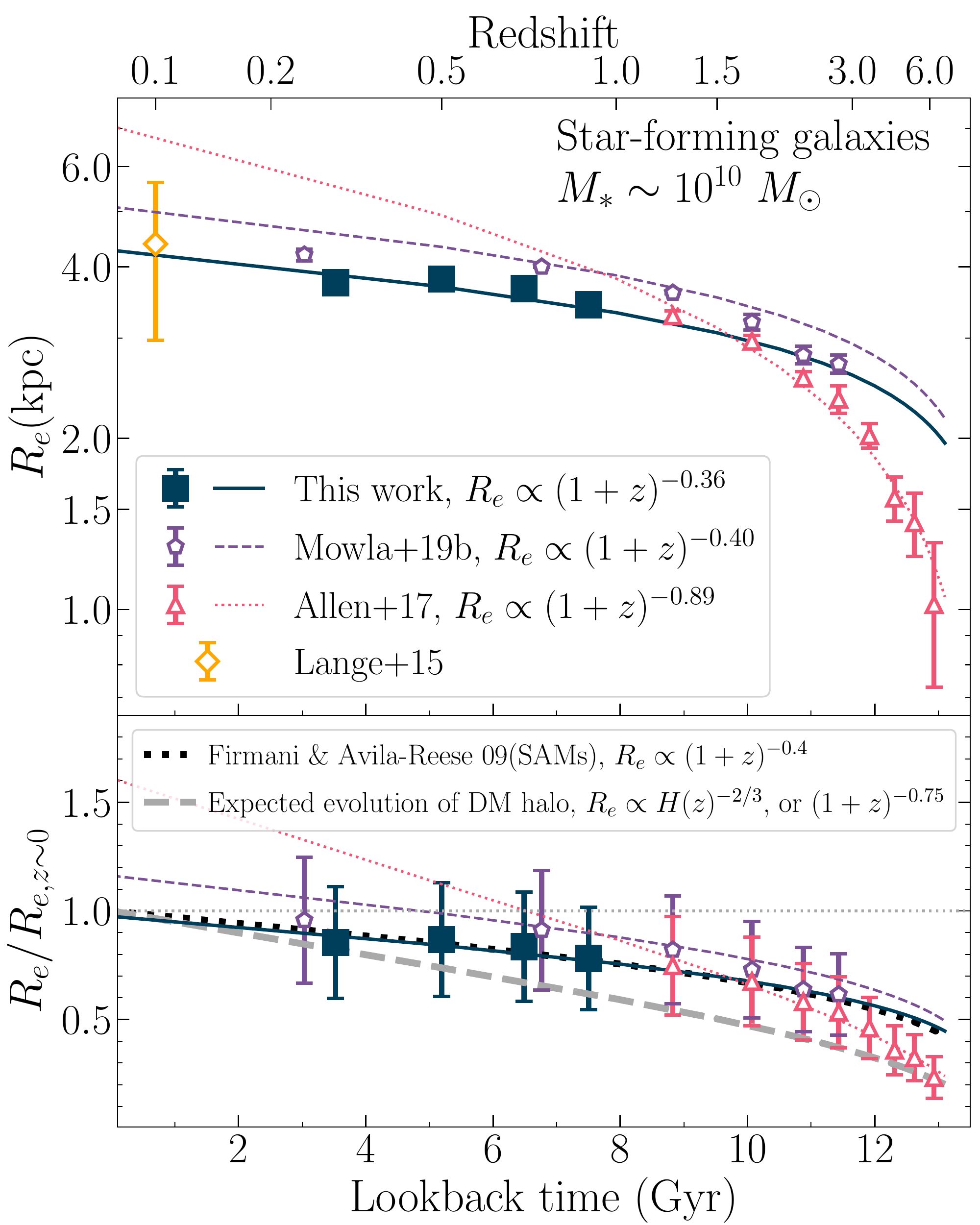}
	\caption{Top: size evolution of star-forming galaxies with $M_{\ast}\sim10^{10}~M_{\odot}$ (blue filled square and solid curve). The result at $0.1<z<3$ from 3D-\emph{HST}+CANDELS \citep{Mowla2019b}, at $1<z<7$ from ZFOURGE \citep{Allen2017}, and at $z\sim0$ from GAMA \citep{Lange2015} are also shown. Bottom: the evolution of the ratio of the median sizes at different redshifts ($R_{e}/R_{e,z\sim0}$). In the bottom panel, the grey thick dashed curve indicates the predicted size of galactic disks based on the evolution of  dark matter halos. The black dotted curve indicates the predicted size evolution of disks taken from the semi-analytic models (SAMs) of \cite{Firmani2009}, which is shallower than the one corresponding to the halos but in good agreement with the observations from HSC and \cite{Mowla2019b}.}
	\label{fig:redshiftevol_m10}
\end{figure}
\subsubsection{The Redshift Evolution of $R_{e}$}
In Figure~\ref{fig:redshiftevol_m10}, we show the size evolution of star-forming galaxies with stellar mass $M_{\ast}\sim10^{10}~M_{\odot}$ from the present day and up to $z\sim7$ taken from this work and previous studies, including GAMA \cite[$z\sim0$;][]{Lange2015}, CANDELS \cite[$0.1<z<3$;][]{Mowla2019b}, and the FourStar Galaxy Evolution Survey (ZFOURGE) \cite[$1<z<7$;][]{Allen2017}.  \citeauthor{Allen2017} showed that, at $1<z<7$, the growth of a mass-complete ($M_{\ast}>10^{10}~M_{\odot}$) sample of star-forming galaxies appears to be rapid following $R_{e}\propto(1+z)^{-0.89}$.
This rate differs from the size evolution found  for star-forming galaxies at $z<1$ -- the average size of star-forming galaxies evolves more slowly with redshift, following $R_{e}\propto(1+z)^{\beta_{z}\sim -0.4}$, broadly consistent with the previous studies of \cite{Lilly1998, Ravindranath2004,Barden2005,Trujillo2006} \cite[see also][for a similar conclusion based on the half-mass radii]{Suess2019ApJL}. 

\par Figure~\ref{fig:redshiftevol_m10} further shows that the average size evolution of $(1+z)^{\beta_{z}\sim-0.04}$ observed at $z<1$, is slower than the prediction from the evolution of the parent dark matter halo of a galaxy (although consistent within the error bars): assuming that the galaxy size, in particular disk scale length, is proportion to the virial radius of its parent dark matter halo, the galaxy size is given by $R_{e}\propto (1+z)^{-0.75}$ \cite[e.g.,][]{Mo1998,Bouwens2002}. According to this theoretical expectation, a galaxy size increases by a factor of $\sim2$ from $z=1$ to the present day, which is a larger increase than what we observed in our study \cite[see also][]{Dutton2011}.  The difference between the observed evolution and the theoretical expectation from \cite{Mo1998} could be explained by the fact that semi-analytic models (SAMs) based on the \cite{Mo1998} formalism refer to ``static'' population of disk galaxies, in particular baryonic disks, at a given redshift; i.e., the models do not follow the evolution of individual disk galaxies with time. On the contrary, \cite{Firmani2009} followed the evolution of individual disks  inside growing $\Lambda$CDM halos and incorporated the evolutionary processes of gas infall and transformation of local gas into stars in their SAMs. The prediction from the \cite{Firmani2009} model is that for $z<2.5$ galaxies at fixed stellar mass, the effective radius of a galactic disk evolves as $R_{e}\propto (1+z)^{-0.4}$ for all masses they studied ($M_{\ast}=10^{9}-10^{11}~M_{\odot}$), in good agreement with observations \cite[e.g.,][]{Barden2005,Mowla2019b}, including ours.



\subsubsection{The Stellar-to-halo Mass Ratios}
\label{sec:derive_smhm}
\par \editone{Regardless of being quiescent or star-forming galaxies, we observed that a size-mass relation has the form of a broken power-law with a clear change of slope at a pivot stellar mass, which also nearly coincides with the stellar mass at which half of the galaxy population is quiescent (Figure~\ref{fig:brokenpwl_evol}). This form of the size-mass relation is reminiscent of the form of the stellar-to-halo mass relation \citep{Mowla2019a}. Motivated by all of these, we use our best-fit $R_{e}-M_{\ast}$ relation from Figure~\ref{fig:smfitbkpwlquisf} to estimate the stellar-to-halo mass ratio (SMHM), $\log(M_{\ast}/M_{\mathrm{halo}})$, as a function of stellar mass. Theoretically, this ratio measures the formation efficiency of galaxies, namely, how efficiently baryons accreted into the halo are cooled and subsequently converted into stars. Following an approach similar to \cite{Mowla2019a}, we estimate halo mass using virial radii, ($R_{\mathrm{vir}}$; i.e., $M_{\mathrm{halo}}\propto R_{\mathrm{vir}}^{3}$). Here we are assuming that a ratio between galaxy size and the virial radius of the halo are related through $R_{\mathrm{vir}}=R_{e}/\gamma$, where we adopted $\gamma=0.02$ from \cite{Shibuya2015} for star-forming galaxies. For quiescent galaxies, we adopted $\gamma=0.01$, which is $\sim0.2$~dex lower than that of star-forming galaxies (\citealt{Huang2017} and see also \citealt{Somerville2018,Lapi2018,Zanisi2020})}.  
\begin{figure}
	\centering
	\includegraphics[width=0.46\textwidth]{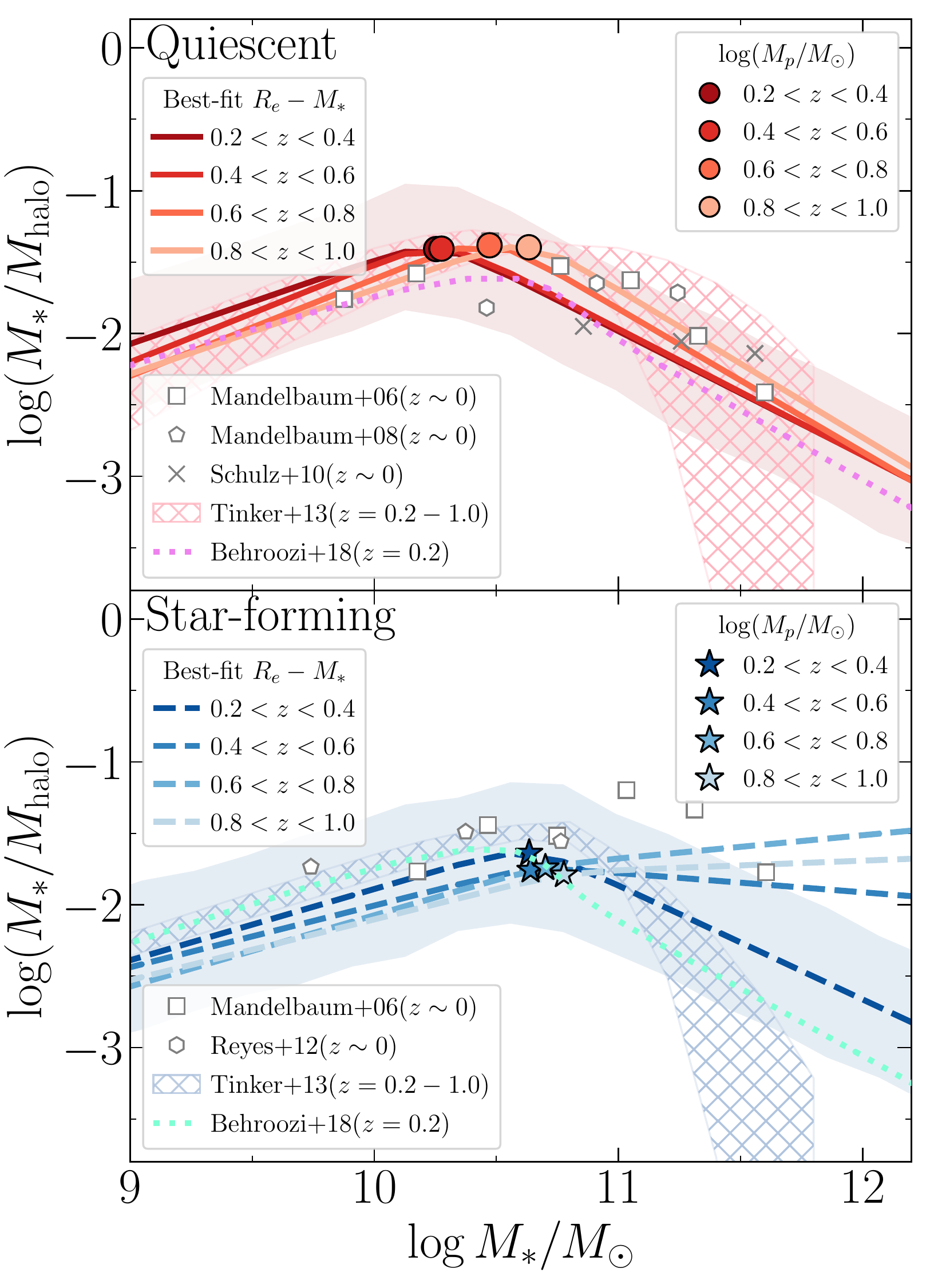}
	\caption{The stellar-to-halo mass ratios ($\log(M_{\ast}/M_{\mathrm{halo}})$) derived from our best-fitting $R_{e}-M_{\ast}$ relations for quiescent (solid curves) and star-forming galaxies (dashed curves) in four redshift  bins. The $M_{\mathrm{halo}}$ is determined from virial radii, simply assumed to be scaled as $R_{\mathrm{vir}}=\gamma^{-1} R_{e}$. Shaded region represents the inter-68th range of $\log(M_{\ast}/M_{\mathrm{halo}})$, accounting for $\sigma_{\log~r}$ of the $R_{e}-M_{\ast}$ relation (only shown for galaxies at $z=0.2-0.4$ for clarity).  The  $\log(M_{\ast}/M_{\mathrm{halo}})$ derived using abundance matching technique from \cite{Behroozi2018} for quiescent and star-forming galaxies at $z=0.6$ and the results derived from several galaxy-galaxy weak lensing studies are also shown. The $M_{p}$ of the $R_{e}-M_{\ast}$ relations (circles and stars) for quiescent galaxies (and star-forming galaxies at $z<0.4$) corresponds to where $\log(M_{\ast}/M_{\mathrm{halo}})$ is maximum.  }
	\label{fig:smhm_ratio}

\end{figure}

\par \editone{Figure~\ref{fig:smhm_ratio} shows that our inferred $\log(M_{\ast}/M_{\mathrm{halo}})$ for both quiescent and star-forming galaxies at $0.2<z<1.0$ reach a maximum of $\log(M_{\ast}/M_{\mathrm{halo}})\sim0.02-0.04$ around the $M_{p}$ of the $R_{e}-M_{\ast}$ relation and decline rapidly above the $M_{p}$, with the exception for star-forming galaxies at $z>0.4$. For star-forming galaxies, the slope of $\log(M_{\ast}/M_{\mathrm{halo}})-\log(M_{\ast})$ relation above $M_{p}$ steepens as decreasing redshift, such that the slope at $0.4<z<1.0$ are nearly flat and become similar to that of the quiescent counterparts at $z<0.4$. 
Overall, our inferred  $\log(M_{\ast}/M_{\mathrm{halo}})-\log(M_{\ast})$ relation is broadly consistent with the constraints from previous works \citep[e.g.,][and reference therein]{Wechsler2018}. For instance, Figure~\ref{fig:smhm_ratio} shows that, the slopes and pivot masses of our  $\log(M_{\ast}/M_{\mathrm{halo}})-\log(M_{\ast})$ relations are consistent with the (inverted) SMHM  relations derived from abundance matching technique of \cite{Behroozi2018}, and those derived from the joint analysis of galaxy-galaxy weak lensing and clustering from \cite{Tinker2013} at $z=0.2-1.0$, with the exception for the (evolving) slopes at the high-mass end for star-forming galaxies. Also, our inferred  $\log(M_{\ast}/M_{\mathrm{halo}})-\log(M_{\ast})$ relations are also broadly in agreement with those derived from galaxy-galaxy weak lensing for early- and late-galaxies in the local Universe \citep{Mandelbaum2006,Mandelbaum2008,Schulz2010,Reyes2012}.  Here we do not intend to perform a comprehensive comparison with previous studies of SMHM relations.  The broad agreement between our results derived by assuming a simple scaling relation between $R_{e}$ and $R_{\mathrm{vir}}$, and those studies, particularly for the functional form of the  $\log(M_{\ast}/M_{\mathrm{halo}})-\log(M_{\ast})$ relations, is encouraging. We will discuss the implication of our finding on the interplay between structure formation and galaxy formation in Sections~\ref{sec:discuss_starforming} and ~\ref{sec:discuss_quiescent}}.


\par \editone{To summarize, the good agreement between the HSC size-mass relations and those from the deeper and spaced-based observations by \cite{Huang2017} and \cite{Morishita2017}, particularly for low-mass galaxies, strongly suggests that we have robust size measurements for galaxies down to $\log(M_{\ast}/M_{\odot})=10.2~(9.2)$ at $z<1~(0.6)$, thanks to an excellent seeing condition of the HSC images. In addition,  the SMHM relations derived from the $R_{e}-M_{\ast}$ relations are broadly consistent with the constraints from different methods, particularly in terms of pivot mass, low-mass end slope, and high-mass end slope. All of these further provide us with confidence that our observation of the shallow  $R_{e}-M_{\ast}$ relation below the pivot mass is not driven by a systematic bias in our size measurement.  In addition, the larger sample of massive galaxies ($\log(M_{\ast}/M_{\odot})\gtrsim11$) at $0.2<z<1.0$ than the previous deep surveys enable us to derive the tighter constraint on the size-mass relation for most massive galaxies at this redshift range. Our results of the size-mass relations bridge the gap between the local Universe \citep[e.g.,][]{Shen2003, Mosleh2013, Lange2015, Zhang2019} and the high redshift \citep[$z\lesssim3$][]{vanderWel2014,Huang2017,Whitaker2017, Mowla2019b,Mosleh2020, Nedkova2021}.}

\subsection{Implications for the Formation of Star-forming Galaxies}
\label{sec:discuss_starforming}
\par Our finding of nearly constant pivot stellar mass of $\log(M_{p}/M_{\odot})\sim10.7$ for star-forming galaxies (Figure~\ref{fig:brokenpwl_evol}) over the redshift $0.2<z<1.0$ is in excellent agreement with \cite{Shen2003}, who found that the observed $R_{e}-M_{\ast}$ relation for late-type galaxies from SDSS changes both slope and dispersion $\sigma_{\log R_{e}}$ at a characteristic stellar mass of $M_{\ast}\sim10^{10.6}~M_{\odot}$. To explain this observation, \cite{Shen2003} used a simple theoretical model for the formation of a late-type galaxy, which generally consists of a rotationally supported thin disk \citep[e.g.,][]{Mo1998,Renzini2020} and central dispersion-dominated bulge \cite[e.g.,][]{Wyse1997,Kormendy2004, DeLucia2011,Lang2014}. They demonstrated that a successful model requires that the material in the galaxy has specific angular momentum similar to that of its halo. In their model, feedback from star formation is the main driver for the decrease of the star formation efficiency in low-mass galaxies  \cite[e.g.,][]{Dekel1986,Bower2012,Hopkins2012}, consistent with the decreasing in  $\log(M_{\ast}/M_{\mathrm{halo}})$ for star-forming galaxies below the pivot mass (Figure~\ref{fig:smhm_ratio}).
\par Moreover, we observed that the intrinsic scatter in the $R_{e}-M_{\ast}$ relation for star-forming galaxies remains nearly constant over $0.2<z<1.0$ ($\sigma_{\log R_{e}}\sim0.17$ dex, see Figure~\ref{fig:brokenpwl_evol}). This result is also in good agreement with \cite{vanderWel2014}, who found no evidence of redshift evolution of the intrinsic scatter of $R_{e}-M_{\ast}$ ($\sigma_{\log R_{e}}\sim0.16-0.19$ dex) for star-forming galaxies since $z=3$. This intrinsic scatter of galaxy size is comparable to, but perhaps somewhat smaller than, the scatter of $0.22-0.25$ dex in the halo spin parameter  \cite[$\lambda$;][]{Bullock2001,Maccio2008}. This trend is compatible with the expectations of the theoretical model of \cite{Mo1998} in which galaxy sizes are controlled by specific angular momentum acquired by tidal torques during cosmological collapse by both baryons and dark matter such that $R_{e}=\gamma R_{\mathrm{vir}}$ (see Section~\ref{sec:derive_smhm} and reference therein). 
\par \editone{In addition,  we observed that the rate of size evolution ($\beta_{z}$) of star-forming galaxies continues to increase with increasing stellar mass over the range of $\log(M_{\ast}/M_{\odot})\gtrsim10.8$, in contrast to that of quiescent galaxies over the same mass range, which is nearly stellar mass-independent (Figure~\ref{fig:redshiftevol_massdep}). The faster size evolution of more massive star-forming galaxies might be explained by the accretion at large radii of stars via mergers, which results in faster size growth relative to their lower-mass counterparts \cite[e.g.,][]{Oser2010,Lackner2012,Cooper2013,Qu2017,Furlong2017,Clauwens2018,Pillepich2018}, although star-forming galaxies have a lower fraction of stellar mass formed ex-situ compared to quiescent galaxies at fixed stellar mass \citep{Rodriguez2016,Davison2020}. For instance, \cite{Davison2020} used the EAGLE hydrodynamical simulation to show that, at fixed stellar mass, more extended galaxies have higher ex-situ mass fraction, supporting the idea that more extended galaxies are more extended because they have accreted more material preferentially onto their outskirts. Second, the increasing in the rate ($\beta_{z}$) of size evolution for star-forming galaxies with increasing stellar mass over the range of $\log(M_{\ast}/M_{\odot})\gtrsim10.8$ is consistent with the increasing in the power-law slope at the high-mass end of the $R_{e}-M_{\ast}$ relation (and also the SMHM relation) with decreasing redshift. The evolution is such that at $z=0.2$ the relation becomes more consistent with that of counterpart quiescent galaxies (Figures~\ref{fig:smfitbkpwlquisf} and~\ref{fig:smhm_ratio}), in agreement with the prediction from hydrodynamical simulations that the contribution of (dry) mergers to the size and mass growth for these massive galaxies increases at later cosmic time \citep[e.g.,][]{Rodriguez2016,Qu2017}.}


\par On the other hand, for star-forming galaxies with $9.8<\log(M_{\ast}/M_{\odot})<10.8$, the rate of size evolution appears to be nearly independent with stellar mass, possibly implying that bulges in star-forming galaxies play an important role in the size evolution for these galaxies. As our galaxy half-light radius is measured based on fitting a single-component S\'{e}rsic profile to the light distribution of the whole galaxy, our measurement will result from the combination of disk and bulge components. Also, in galaxies with larger bulge fractions, the half-light radii of the combined bulge and disk is less than that of the disk. In addition, observations of nearby galaxies from SDSS showed that higher mass galaxies tend to have higher bulge-to-total stellar mass ratios (B/T), in contrast to the lower mass ones \cite[e.g.,][]{Weinzirl2009,Bernardi2010,Bluck2014,Bruce2014,Lang2014}. This observation is also consistent with the prediction from SAM by \cite{Tonini2016} who showed that the bulges of disk galaxies with stellar mass above $\log(M_{\ast}/M_{\odot})\sim 10$ are mostly formed through the accumulation of stars in the galaxy center \cite[referred to as instability-driven bulges, akin to secular (or pseudo-)bulges;][]{Kormendy2004,Sellwood2014,Izquierdo-Villalba2019}, and these bulges tends to increase with increasing galaxy stellar mass. Combining this trend with our finding of the slower size evolution for star-forming galaxies with $9.8<\log(M_{\ast}/M_{\odot})<10.8$, it suggests that the formation of bulges in star-forming galaxies slows down the observed size evolution at this stellar mass and redshift range. In future work, we will perform single and double (bulge-disk decompositions) S\'{e}rsic fits to the surface brightness profiles for a sample of relatively well resolved galaxies from HSC, such as low-redshift massive galaxies. The derived multi-component effective radii will provide us with a better description of the size of these galaxies than those inferred from single S\'{e}rsic models and will give us a better interpretation of the size-mas relation \citep[e.g.,][]{Reis2020}.

\subsection{Implications for the Formation of Quiescent Galaxies}
\label{sec:discuss_quiescent}
\subsubsection{Low-mass Quiescent Galaxies}
\par \editone{In this work, we showed the size-mass relation at the low-mass end ($M_{\ast}<M_{p}$) for quiescent galaxies at \hbox{$0.2<z<1$} has a shallow power-law slope ($\alpha\sim0.1$), but it is significantly steeper those at high redshift ($1.5\lesssim z \lesssim  3$), which has nearly flat or even slightly negative slope (\citealt{Huang2017} and Figure~\ref{fig:powerlawslope_others}). The evolution is such that the power-law slopes at the low-mass end for both quiescent and star-forming galaxies at $z<0.5$ are nearly identical, suggesting that some of these low-mass quiescent galaxies on average have sizes comparable to those of star-forming galaxies at a similar stellar mass and redshift. This observation is consistent with \cite{Morishita2017} who showed that low-mass ($\log(M_{\ast}/M_{\odot})<9.8$) quiescent cluster galaxies with larger sizes are also bluer in rest-frame $U-V$ colors, similar to the field star-forming galaxies at the time close to the epoch of observation, whereas the smaller quiescent cluster galaxies are redder and are consistent with having been in place for many Gyr.} 

\par Returning to the observed evolution in a power-law slope $\alpha$ for low-mass quiescent galaxies, if star formation in the majority of galaxies below $M_{p}$ is suppressed by their environments \citep[e.g.,][]{Hogg2003,Quadri2012,Moutard2018,Socolovsky2018}, the implications of our observation are the following: 1) physical mechanisms responsible for the observed size growth and quenching of star formation for these low-mass quiescent galaxies are related; and 2) this mechanism evolves with cosmic time such that the effect is stronger toward lower redshift ($z\lesssim1.5$). The latter interpretation is supported by previous observations that the effect of quenching related to galaxy environment (also quantified by ``environmental quenching efficiency'') evolves with redshift, environmental density, and stellar mass \cite[e.g.,][]{Peng2010,Peng2012,Kovac2014,McGee2014,Balogh2016,Nantais2016,Kawinwanichakij2017,Guo2017,Ji2018,Papovich2018,Moutard2018,vanderBurg2020,Chartab2020}. In particular, \cite{Kawinwanichakij2017} showed that for low-mass galaxies ($M_{\ast}\lesssim10^{10}~M_{\odot}$), the environmental quenching efficiency is very low at $z\gtrsim1.5$, but rapidly increases with decreasing redshift such that environmental quenching can account for nearly all low-mass quiescent galaxies, which appear primarily at $z\lesssim1.0$. Taking all of these together, our result is compatible with the picture in which \editone{low-mass quiescent galaxies, particularly those with large sizes, are recently drawn from the star-forming population through an environmental quenching mechanism without significantly transforming their structure \citep[e.g.,][]{Morishita2017,Wu2018, Matharu2020,Suess2021}}. This observation points to environmental quenching processes including ram pressure striping \cite[e.g.,][]{Gunn1972, Abadi1999} and starvation \citep[also called ``strangulation''; e.g.,][]{Larson1980, Balogh1997} as both of these gas-stripping processes will primarily modify the color and SFR of a galaxy without transforming the galaxy morphology \citep[e.g.,][]{Weinmann2006,vandenBosch2008}\footnote{Even though this is true for the morphology of galaxies as traced by stellar mass, for the morphology as traced by light in any passband, even in near-IR, the young blue stars will outshine the red bulge and make the star-forming disks more prominent. This will lead to significant changes in the visual appearance of the morphology when the star formation in the galaxy ceases \citep[e.g.,][]{Fang2013}. }. Nevertheless, irrespective of the nature of the quenching mechanisms, the addition of recently quenched low-mass galaxies with large sizes to the pre-existing quiescent population is also a manifestation of progenitor bias whose contribution to the average size growth of low-mass quiescent galaxies rises toward lower redshift \citep[e.g.,][]{Damjanov2019, Belli2019,Suess2021}.





\subsubsection{High-mass Quiescent Galaxies}



\par \editone{We have demonstrated that the galaxy stellar mass-to-halo mass ratios ($\log(M_{\ast}/M_{\mathrm{halo}}$) inferred from our best-fit $R_{e}-M_{\ast}$ relation, assuming a simple scaling relation between $R_{e}$ and $R_{\mathrm{vir}}$, is qualitatively in good agreement with the constraints from a number of different methods, particularly in terms of the pivot mass of the $\log(M_{\ast}/M_{\mathrm{halo}})-\log(M_{\ast})$ relation and the slopes below and above this mass (Figure~\ref{fig:smhm_ratio}). From the stellar-to-halo mass relation point of view, the decreasing of $\log(M_{\ast}/M_{\mathrm{halo}})$ as increasing stellar mass for galaxies above the pivot mass can be explained by the decrease in the efficiency of star formation due to AGN feedback \citep[e.g.,][]{Silk1998,Croton2006}, while the host dark matter halos of massive quiescent galaxies continue growing hierarchically.  In contrast, $\log(M_{\ast}/M_{\mathrm{halo}})$ ratios of star-forming galaxies above $M_{p}$ do not strongly decline mainly because they continue to grow their stellar masses through star formation \citep[e.g.,][]{Tinker2013,Rodriguez-Puebla2015}. Taking these aspects together, a smoothly broken power-law form of quiescent galaxies size-mass relation at $0.2<z<1.0$ is compatible with a picture in which massive quiescent galaxies undergo a two-phase formation scenario}: mass and size growth during a first phase ($2<z<6$) dominated by in-situ star formation from infalling cold gas. In addition to infalling gas, there could also be growth through major mergers at early epochs ($z>2$). There is growing observational evidence for major mergers in high$-z$ massive quiescent galaxies \citep[e.g.,][]{Sawicki2020}. This first phase is followed by an extended phase ($z\lesssim2$) during which mass and size growth is primarily due to accretion in the outskirts of galaxies via mergers
\cite[e.g.,][]{Bundy2009,Bezanson2009,Trujillo2011,vanDokkum2010,Patel2013,McLure2013,Szomoru2012,Greene2012, Greene2013,Belli2019}. Previous works, which utilized hydrodynamical simulations to reproduce the observed size and mass growth of massive elliptical galaxies by stellar accretion, have also consistently support this two-phase formation scenario \cite[e.g.,][]{Naab2009, Hopkins2010,Oser2010,Oser2012,Hilz2013,Wellons2015}. Additionally, \cite{Rodriguez-Puebla2017} used the abundance matching technique and the observational constraints to infer evolutionary tracks of galaxies on $R_{e}-M_{\ast}$ relations for progenitors halos with different virial mass at $z=0$. The authors found that, after the star formation in these progenitor galaxies is quenched, they grow in size much faster than in their stellar mass, resulting in a steeper slope of the relation at low redshift, implying that the upturn in the size-mass relation at the high-mass end can be explained by minor mergers \citep[e.g,][]{vanDokkum2010,vanDokkum2015,Tacchella2015b,Barro2017,Hill2017,Woo2019, Nelson2019, Wilman2020,Chen2020,Suess2021}.
\par \editone{Additionally, we found that the pivot mass of the size-mass relation for quiescent galaxies decreases by $0.4$ dex from $z=1$ to $z=0.2$, broadly in agreement with \cite{Leauthaud2012}, who found a similar trend in both pivot stellar mass and pivot halo mass of the SMHM relations since $z=1$ to $z=0.2$. These observations are compatible with ``downsizing" trend of galaxies -- a pattern in which the sites of active star formation shift from high-mass galaxies at earlier times to lower mass galaxies at later epoch \citep[e.g.,][]{Cowie1996,Brinchmann2000,Juneau2005,Bundy2006,Noeske2007,Fontanot2009}.}

\subsection{The Impact of the Cross-Contamination of Massive Galaxies to the Size-Mass Relations}
\label{sec:lmasslt11}
\par \editone{Lastly, we consider how our result would be impacted if we exclude galaxies with $\log(M_{\ast}/M_{\odot})>11$. Comparing to lower mass galaxies, these massive counterparts could significantly be affected by higher cross-contamination between $urz$-selected quiescent and star-forming galaxies  (Figure~\ref{fig:contamfrac_vs_mass}), even though we have attempted to account for this bias. To quantify the effect, we fit $R_{e}-M_{\ast}$ distributions by including only galaxies with stellar mass of  $\log(M_{\ast}/M_{\odot})<11$ and find that the relations for star-forming galaxies above $M_{p}$ 
are relatively steeper -- $\beta=0.86\pm0.06,0.43\pm0.03,0.31\pm0.05,$ and $0.72\pm0.19$ at $z\sim0.3,0.5,0.7$, and 0.9 -- compared to those based on the whole sample of star-forming galaxies. Nevertheless, the pivot mass remains unchanged, with the exception of that for star-forming galaxies at $z\sim0.9$ with $\log(M_{p}/M_{\odot})=11.0$, which is $\sim0.2$ dex larger than those based on the whole sample. We also do not find significant change in the other parameters ($\alpha, r_{p}$, and $\sigma_{\log~r}$). Therefore, even without including massive galaxies  with $\log(M_{\ast}/M_{\odot})>11$ to the analysis, our main conclusion of a broken power-law form for the size-mass relation of star-forming galaxies at $0.2<z<1.0$ remains unchanged, albeit with larger uncertainties in slope $\beta$.}
\par \editone{However, the systematic uncertainties arising from missing these massive star-forming galaxies contribute $0.03-0.4$ dex to the error budget in their average sizes at fixed stellar mass. Our finding here underscores the importance of imaging surveys covering large cosmic volumes to sufficiently sample massive star-forming galaxies, thereby significantly reducing the uncertainties in the best-fit parameters of the size-mass relation at the high-mass end. Additionally, as we already noted in Section~\ref{sec:urz_selection}, even though we have used rest-frame $U-V$ and $V-J$ colors from the deeper and multiwavelength catalogs to calibrate our $urz$ selection and incorporate contaminating fractions when fitting size-mass relation, any remaining cross-contamination might impact our best-fit parameters of the size-mass relation and the interpretation of their evolution.} Similar analysis of the $UVJ-$selected samples of quiescent and star-forming galaxies using the $u$-band photometry from CLAUDS \citep{Sawicki2019} and \emph{Spitzer}/IRAC photometry, in addition to the HSC photometry, will be presented by George et al. (in prep.).




\section{Summary}
\label{sec:conclusion}
\par In this paper we presented the galaxy size-mass distribution over the redshift range $0.2<z<1.0$ using a sample of $\sim1,500,000$ galaxies with $i < 24.5$ (corresponding to $\log(M_{\ast}/M_{\odot})>10.2$ at $z=1$ and $>9.2$ at $z=0.6$). We utilized photometric redshift, stellar masses, and rest-frame properties determined using a subset of the data from the HSC PDR2, covering $100~\mathrm{deg}^{2}$ of the HSC Wide and Deep+UltraDeep layers in five broad-band filters ($grizy$).  We measured galaxy sizes from HSC $i-$band imaging (median seeing of $0\farcs6$) by fitting single-component S\'{e}rsic profiles to two dimensional light distributions, with corrections for observational biases (due to PSF bluring and surface brightness dimming) and (redshift- and stellar mass-dependent) color gradients. 
We classified our galaxy sample into quiescent and star-forming galaxies on the the basis of their rest-frame $u-r$ and $r-z$ colors. Our results can be summarized as follows:
%
\begin{enumerate}
  \item Quiescent galaxies are on average smaller than star-forming galaxies at a fixed stellar mass (in the range of $9<\log (M_{\ast}/M_{\odot}) < 11$) and redshift, corroborating the results from previous studies. The large sample size, in combination with accurate size measurements, not only enables us to confirm the previous results but also demonstrate that, the $R_{e}-M_{\ast}$ relations for both quiescent and star-forming galaxies at $0.2<z<1.0$ have very strong Bayesian evidence promoting the form of a smoothly broken power-law, which is characterized by a change in the slope at a
  pivot stellar mass $M_{p}$, over a single power-law model. The $M_{p}$ of the relation also nearly coincides with the stellar mass at which half of the galaxy population is quiescent (Figure~\ref{fig:brokenpwl_evol}).
 
  \item The $R_{\ast}-M_{\ast}$ relation for quiescent galaxies is characterized by a clear change in power-law slope at an evolving pivot stellar mass of $\log(M_{p}/M_{\odot})=10.2$ at $z=0.2$ and $\log(M_{p}/M_{\odot})=10.6$ at $z=1.0$. Below the pivot mass, the relation is shallower, with $R_{e}\propto M_{\ast}^{0.1}$, and the relation is steeper above $M_{p}$ with $R_{e}\propto M_{\ast}^{0.6-0.7}$. 
  
  \item The $R_{e}-M_{\ast}$ for star-forming galaxies  is characterized by a change in the power-law slope at a pivot stellar mass of $\log(M_{p}/M_{\odot})\sim10.7$. Below $M_{p}$, the relation is shallower with $R_{e}\propto M_{\ast}^{0.2}$, similar to those of quiescent galaxies. Above $M_{p}$, the relation is steeper and exhibits evolution with redshift: the relation steepens as decreasing redshift from $R_{e}\propto M_{\ast}^{0.3}$ at $z=1$ to $R_{e}\propto M_{\ast}^{0.6}$ at $z=0.2$.

  \end{enumerate}
 
 \par Taken together, our analysis, combined with results in the literature, paint a picture in which large low-mass ($M_{\ast}<M_{p}$) quiescent galaxies have been recently drawn from star-forming population at the time close to the epoch of observation, presumably through  environment-specific quenching mechanisms, such as ram-pressure stripping or starvation, without significantly transforming their morphology. Also, the contribution of the environmental effect to the observed size growth for low-mass quiescent galaxies becomes more important at lower redshifts ($z\lesssim1.0$). 
 \par On the other hand, the similarity between the pivot stellar mass of the size-mass relation and the stellar mass at which half of the galaxy population is quiescent, corroborating the idea that the pivot mass marks the mass above which both stellar mass growth and the size growth transition from being star formation dominated to being (dry) merger dominated. The steepening of power-law slopes at the high-mass end of star-forming galaxies with decreasing redshift implies the increasing role of minor mergers in the growth of these massive galaxies at later cosmic times.
 \par Additionally, we quantified the redshift evolution of the median sizes of quiescent and star-forming galaxies over the range of \hbox{$0.2<z<1.0$} and arrived at the following results:
 \begin{enumerate}
     \item Although both quiescent and star-forming galaxies grow in their stellar mass and size over cosmic time, more massive galaxies undergo significantly faster size evolution, but the quiescent galaxies do so at faster rate compared to their star-forming counterparts, particularly over the stellar mass range of \hbox{$\log(M_{\ast}/M_{\odot})=10-11$} (Figure~\ref{fig:sizeevol4} and~\ref{fig:redshiftevol_massdep}). 
    \item  In contrast to lower mass galaxies, the average size of the most massive star-forming galaxies with \hbox{$\log(M_{\ast}/M_{\odot})\gtrsim11$} evolve with faster rate \hbox{($R_{e}\propto(1+z)^{-1.51}$)} than that of quiescent counterparts \hbox{($R_{e}\propto(1+z)^{-1.21}$)}, such that by $z=0$, these most massive star-forming galaxies are able to ``catch up'' with their quiescent counterparts, resulting in the comparable sizes for these massive galaxies, with $15.1\pm1.4$ kpc and $13.9\pm0.3$ kpc for star-forming and quiescent galaxies, respectively.
 \end{enumerate}
 \par The size-mass distributions and their evolution from the HSC dataset presented here provide a solid framework for galaxy formation models and by giving observational constraints on the physical mechanisms governing stellar mass assembly and structure formation. The non-evolving power-law slope at the low-mass end and intrinsic scatter of the size-mass relation for star-forming galaxies present strong constraints on the formation of galaxies through baryon condensation at the centers of dark matter halos and feedback processes, while the evolution in the power-law slopes at the high-mass end present constraints on the contribution of mergers to the growth of massive galaxies.

\acknowledgements
The authors would like to thank the anonymous referee for a
comprehensive and constructive report that allowed us to improve the overall quality of the manuscript. We are grateful for the support from the World Premier International Research Center Initiative (WPI Initiative), MEXT, Japan. LK is supported by JSPS KAKENHI Grant Number JP20K14514. JS is supported by JSPS KAKENHI Grant Number JP18H01251 and the World Premier International Research Center Initiative (WPI), MEXT, Japan. The research of AG, ID, and MS is supported by the National Sciences and Engineering Council (NSERC) of Canada.
We would like to thank to Alvio Renzini for his helpful comments and suggestions.
\par The Hyper Suprime-Cam (HSC) collaboration includes the astronomical communities of Japan and Taiwan, and Princeton University. The HSC instrumentation and software were developed by the National Astronomical Observatory of Japan (NAOJ), the Kavli Institute for the Physics and Mathematics of the Universe (Kavli IPMU), the University of Tokyo, the High Energy Accelerator Research Organization (KEK), the Academia Sinica Institute for Astronomy and Astrophysics in Taiwan (ASIAA), and Princeton University. Funding was contributed by the FIRST program from Japanese Cabinet Office, the Ministry of Education, Culture, Sports, Science and Technology (MEXT), the Japan Society for the Promotion of Science (JSPS), Japan Science and Technology Agency (JST), the Toray Science Foundation, NAOJ, Kavli IPMU, KEK, ASIAA, and Princeton University. 

This paper makes use of software developed for the Large Synoptic Survey Telescope. We thank the LSST Project for making their code available as free software at  http://dm.lsst.org

\par This paper is based [in part] on data collected at the Subaru Telescope and retrieved from the HSC data archive system, which is operated by Subaru Telescope and Astronomy Data Center at National Astronomical Observatory of Japan. Data analysis was in part carried out with the cooperation of Center for Computational Astrophysics, National Astronomical Observatory of Japan.
\par The Pan-STARRS1 Surveys (PS1) and the PS1 public science archive have been made possible through contributions by the Institute for Astronomy, the University of Hawaii, the Pan-STARRS Project Office, the Max-Planck Society and its participating institutes, the Max Planck Institute for Astronomy, Heidelberg and the Max Planck Institute for Extraterrestrial Physics, Garching, The Johns Hopkins University, Durham University, the University of Edinburgh, the Queen’s University Belfast, the Harvard-Smithsonian Center for Astrophysics, the Las Cumbres Observatory Global Telescope Network Incorporated, the National Central University of Taiwan, the Space Telescope Science Institute, the National Aeronautics and Space Administration under Grant No. NNX08AR22G issued through the Planetary Science Division of the NASA Science Mission Directorate, the National Science Foundation Grant No. AST-1238877, the University of Maryland, Eotvos Lorand University (ELTE), the Los Alamos National Laboratory, and the Gordon and Betty Moore Foundation.

\software{Astropy \citep{astropy:2013, astropy:2018}, \textsc{Dynesty} \citep{Speagle2020}, \textsc{Photutils} \citep{Bradley2019}, EAZY \citep{Brammer2008}, \textsc{Lenstronomy} \citep{Birrer2015,Birrer2018}, \textsc{Mizuki} \citep{Tanaka2015}, hscPipe \citep{Bosch2018}}

\appendix

\begin{figure}
	\centering
	\includegraphics[width=0.8\textwidth]{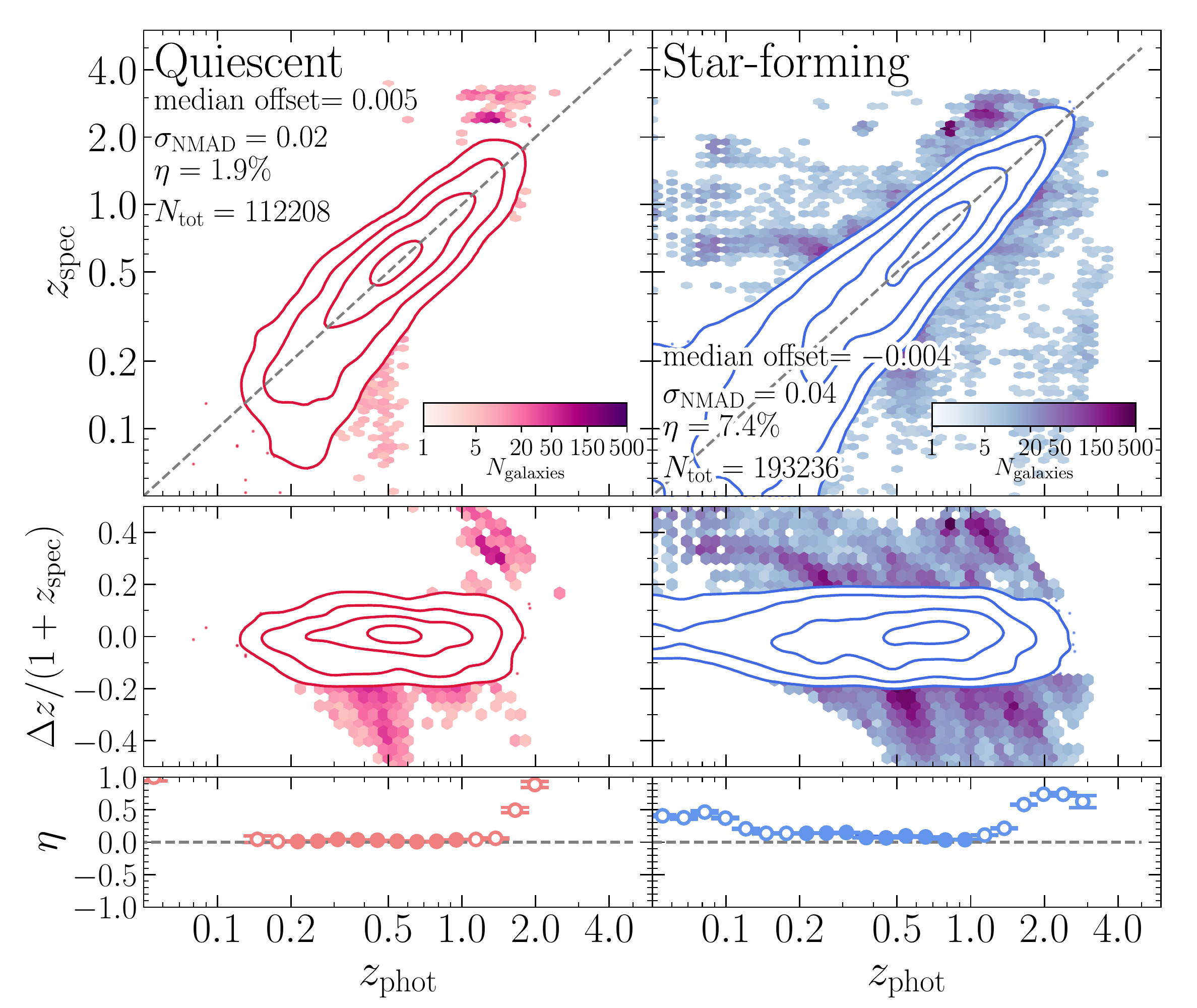}
	\caption{Top: comparison of photometric redshifts from \textsc{Mizuki} ($z_{\mathrm{phot}}$) to available spectroscopic redshifts ($z_{\mathrm{spec}}$) in HSC for quiescent (left), and star-forming (right) galaxies at $0<z_{\mathrm{spec}}<4$ with $i<24.5$. Middle:  $\Delta z=z_{\mathrm{spec}}-z_{\mathrm{phot}}$, normalized to $1+z_{\mathrm{spec}}$ as a function of $z_{\mathrm{phot}}$.  The quiescent and star-forming galaxies are classified using the $urz$ selection (Section ~\ref{sec:urz_selection}). The contours indicate the  distribution of galaxies from $1\sigma$ to $4\sigma$. The one-to-one relation is plotted in each panel as a dashed grey line. The background colorscale in top and middle rows represent the catastrophic redshift failures and are defined as those objects with $\left |\Delta z \right |/(1+z_{\mathrm{spec}}) >0.15$ \citep{Ilbert2013}. The $\sigma_{\mathrm{NMAD}}$ denotes 1.48 times the median absolute deviation of $\Delta z /(1+z_{\mathrm{spec}})$. Bottom: the fraction of catastrophic redshift failures ($\eta$) in bins of $z_{\mathrm{phot}}$, which is also magnitude-dependent (here we only show it for $i<24.5$ for illustration purpose). The indicated median offsets, $\sigma_{\mathrm{NMAD}}$, $\eta$, and the number of galaxies ($N_{\mathrm{tot}}$) are the estimates for the galaxies we use in this study ($0.2<z_{\mathrm{phot}}<1$). }
	\label{fig:zspec_zphot}
\end{figure}

\section{HSC Photometric Redshift Uncertainties}
\label{appendix:photoz_test}
\subsection{Comparison with Spectroscopic Redshifts}
\par We compare the photo-$z$ with spectroscopic redshifts (hereafter spec-$z$ or $z_{\mathrm{spec}}$) using the compilation of publicly available spectroscopic redshifts in the HSC fields provided by the HSC PDR2 database \citep{Aihara2019}.  The public spectroscopic redshifts catalog includes redshifts from zCOSMOS DR3 \citep{Lilly2009}, UDSz \citep{Bradshaw2013,McLure2013}, 3D-\emph{HST} \citep{Skelton2014,Momcheva2016}, FMOS-COSMOS \citep{Silverman2015, Kashino2019}, VVDS \citep{LeFevre2013}, VIPERS PDR1 \citep{Garilli2014}, SDSS DR12 \citep{Alam2015}, the SDSS IV QSO catalog \cite{Paris2018}, GAMA DR2 \citep{Liske2015}, WiggleZ DR1 \citep{Drinkwater2010}, DEEP2 DR2 \citep{Davis2003, Newman2013}, DEEP3 \citep{Cooper2011,Cooper2012}, and PRIMUS DR1 \citep{Coil2011, Cool2013}. For our purpose, we select sources with secure spectroscopic redshifts by using a homogenized spectroscopic confidence flag from the HSC PDR2 database. Focusing on our galaxy sample at $0.2<z<1.0$, $i<24.5$, and also complete in stellar mass (see Section~\ref{sec:derive_masscomplete}), the median offset (bias) between the $z_{\mathrm{spec}}$ and $z_{\mathrm{phot}}$ ($\Delta z=z_{\mathrm{spec}}-z_{\mathrm{phot}}$) is 0.001. The normalized median absolute deviation (the scatter; $\sigma_{z/(1+z)}$), defined as:
\begin{equation}
     \sigma_{\mathrm{NMAD}}=1.48\times\mathrm{median}\left ( \left | \frac{\Delta z-\mathrm{median}(\Delta z) }{1+z_{\mathrm{spec} }} \right | \right ),
\end{equation}
is 0.06. We find $\sim7\%$ of the sample being catastrophic redshift failures, which are defined as those objects with $\left| \Delta z \right|  / (1+z_{\mathrm{spec}})>0.15$ \citep{Ilbert2013}. 
\par In Figure~\ref{fig:zspec_zphot} we further explore the dependence of median offset and scatter in $\Delta z/(1+z_{\mathrm{spec}})$ on galaxy type, classified using the rest-frame $u-r$ and $r-z$ colors (see Section ~\ref{sec:urz_selection}). Our comparison between $z_{\mathrm{phot}}$ and $z_{\mathrm{spec}}$ for stellar mass-complete samples of quiescent and star-forming galaxies at $0.2< z_{\mathrm{phot}}<1$ reveals that the median offset and scatter in $\Delta z/(1+z_{\mathrm{spec}})$ for quiescent galaxies are 0.004 and $\sigma_{\mathrm{NMAD}}=0.02$, with a catastrophic redshift failure of $\sim3\%$. For star-forming galaxies, the median offset and scatter in $\Delta z/(1+z_{\mathrm{spec}})$ are $-0.005$ and $\sigma_{\mathrm{NMAD}}=0.04$, with a catastrophic redshift failure of $\sim7\%$. This finding is in agreement with \cite{Quadri2010} who demonstrated that, on average, star-forming galaxies will have less accurate photometric redshifts than quiescent galaxies in broadband optical/NIR survey out to at least $z\sim2$. This is because quiescent galaxies have a strong break in their SEDs near 4000~\AA, and if one can pinpoint the location of this break using the observed photometry, the redshift will be tightly constrained. In contrast, star-forming galaxies, have weaker features in their SEDs over the range of observed wavelengths.
\par We additionally investigate the major source of catastrophic redshift failures coming from the confusion of the Lyman break (at 912~\AA) with the Balmer/4000~\AA~break due to the limited wavelength coverage. This degeneracy results in galaxies that are preferentially scattered down from $2<z<3$ to lower redshifts ($z\lesssim1$) or vice versa \citep[e.g.,][]{Sawicki1997,Massarotti2001,Sun2009}. This effect is particularly significant when $u-$band photometry or near-IR photometry is not available to rule out the possibility of a galaxy being a low-redshift case \citep[e.g.,][]{Ellis1997,Fernandez-soto1999,Benitez2000,Dahlen2008,Abdalla2008,Rafelski2009,Tanaka2015}. In Figure~\ref{fig:zspec_zphot}, we find an overdensity of galaxies located at  $z_{\mathrm{spec}}\sim3$ and $0.5\lesssim z_{\mathrm{phot}}\lesssim1$. This effect is presented in both quiescent and star-forming samples.  Nevertheless, because  of our imposed magnitude cut of $i<24.5$ AB mag, the fraction of galaxies with such catastrophic redshift failures is less than $15\%$ of the galaxy sample in each redshift bin over the range of $0.2<z_{\mathrm{phot}}<1$ (bottom panel of the Figure), and these catastrophic redshift failures should not significantly affect our results. However, as we will explain below, we also take catastrophic redshift failures into account when fitting size-mass relations.
\subsection{The Derivation of the Fraction of Catastrophic Redshift Failures}
\par Figure~\ref{fig:zspec_zphot} also shows an excess of catastrophic redshift failures such that $z_{\mathrm{phot}}>z_{\mathrm{spec}}$. If this was not properly account for, it would lead to an excess of intrinsically large galaxies at this redshift bin and hence bias the inferred the redshift evolution of galaxy sizes. To account for this, we use the following method. First, in narrow ($\Delta i=1.5$ mag) bins of $i-$ band magnitude from $i=18$ mag to $i=24.5$ mag, we measure the fraction of catastrophic redshift failures ($\eta$) in narrow ($\Delta z_{\mathrm{phot}}=0.05$\footnote{Even though our adopted $\Delta z_{\mathrm{phot}}=0.05$ is comparable to typical uncertainties of our photo-$z$'s ($\sigma_{\mathrm{NMAD}}=0.02-0.04$), we have experimented using $\Delta z_{\mathrm{phot}}=0.1$, and find that $\eta(i,z_{\mathrm{phot}})$ and the best-fitted $R_{e}-M_{\ast}$ relation after applying the correction using $\eta(i,z_{\mathrm{phot}})$ are not appreciably affected by the choice of binning of $\Delta z_{\mathrm{phot}}$.}) bins of $z_{\mathrm{phot}}$ for quiescent and star-forming galaxies, separately (see the bottom panel of Figure~\ref{fig:zspec_zphot}). Second, for each magnitude bin and galaxy type,  we fit the third-degree polynomial to $\eta$ as a function of $z_{\mathrm{phot}}$ to obtain  $\eta(i,z_{\mathrm{phot}})$ for quiescent ($\eta_{\mathrm{Q}}$) and star-forming ($\eta_{\mathrm{SF}}$) galaxies.  Throughout this paper, we account for the catastrophic redshift failures by incorporating the $\eta_{\mathrm{Q}}$ and $\eta_{\mathrm{SF}}$ into the estimation of likelihood for each population when we fit the size-mass distribution and its redshift evolution (Section~\ref{sec:fitting_sizemass}).

\section{Systematic and Random Uncertainties of HSC size measurements}
\label{appendix:size_uncertainties}
\subsection{Random Uncertainty Estimates using MCMC}
\label{appendix:mcmc}
\par \textsc{Lenstronomy} additionally provides us an option to infer the parametric confidence interval using a MCMC framework for Bayesian parameter inference \citep[\textsc{emcee};][]{Foreman-Mackey2013}. However, it is very computationally expensive to infer confidence interval for every source due to our large sample size.
Thus, we only take the best-fit inference using the Particle Swarm Optimizer \citep[PSO;][]{Kennedy1995} for our analysis and estimate  uncertainties on the fit parameters using MCMC for a subset of our galaxy sample and then assign uncertainties to all galaxies in the sample. This procedure is similar to that adopted by \cite{vanderWel2012}, and we will describe the procedure in more details below.

\par \edittwo{We randomly select $\sim60000$ galaxies from both HSC Wide and Deep+UltraDeep layers using galaxies from our final sample (Section~\ref{sec:lenstronomy}), which passed all of the selection criteria including, fits with median and scatter of 1D residual (the difference between model and observed light profile) less than 0.05 mag pixel$^{-2}$ and 0.03 mag pixel$^{-2}$, respectively}. \editone{This randomly selected sample accounts for $4\%$ of our galaxy sample, and we then refer to them as a sample $\textbf{\textit{p}}$}. \edittwo{We have also verified that the properties of a sample $\textbf{\textit{p}}$ with MCMC results match to those of the entire sample we use in the analysis.} Second, we utilize \textsc{Lenstronomy} with MCMC to fit the surface brightness profile of this galaxy sample and infer full posterior distribution of each fitted parameter. We then compute the $16-84$ percentile range of the posterior distribution of each fitted parameter. This provides an uncertainty estimate for each parameter of a galaxy $i$:
\begin{equation}
   \delta_{i}=\delta{p_{i}} = (\delta m_{i}, \delta \log n_{i}, \delta \log R_{e,i}, \delta q_{i})
\end{equation}
\noindent where $m_{i}$ is the magnitude of a galaxy $i$. This sample $\textbf{\textit{p}}$ serves as a database which we use to estimate the uncertainties in the \textsc{Lenstronomy} parameters for all galaxies in our sample. 
For each galaxy $i$ in the sample $\textbf{\textit{p}}$, we define $p_{i}$ as 
\begin{equation}
p_{i}= (m_{i}/\sigma(\textbf{\textit{m}} ), \log n_{i}/ \sigma(\log \textbf{\textit{n}}), \log R_{e,i}/ \sigma(\log \textbf{\textit{R}}_{e})  )   
\label{eq:p_i}
\end{equation}
\noindent where $\sigma$ denotes the standard deviation in the respective parameters. We introduce these factor in order to make differences in parameters values comparable and produce dimensionless, normalized distances. Here we find no correlation between $q$ (and its uncertainty) and the uncertainties in any of the parameters. Therefore, we  follow \cite{vanderWel2014} by assuming that the measurement uncertainties depend on $m$, $n$, and $R_{e}$, but not on $q$, and we do not include axis ratio $q$ when defining $\textbf{\textit{p}}$.

\par Second, for each galaxy $g_{i}$ (defined similarly as $p_{i}$) in the target sample $\textbf{\textit{g}}$ to which we want to assign measurement uncertainties, we identify the 25 most similar galaxies in the sample galaxy $p_{i}$.  To do that, we compute the normalized distances $p_{i}-g_{j}$ in the three dimensional parameter space spanned by $m$, $n$, and $R_{e}$, where $p_{i}$ represent the galaxies in the sample $\textbf{\textit{p}}$  (or database), and $g_{i}$ represent the galaxies to which we want to assign measurement uncertainties. We take the average $\delta p_{i}$ of the 25 ``nearest" galaxies $p_{i}$ in the database as the measurements uncertainties of all parameters ($m$, $n$, $R_{e}$, $q$) for the target galaxy $g_{i}$. We repeat this computation for all galaxies $\textbf{\textit{g}}$ in HSC Wide and Deep+UltraDeep, separately, to map out the measurement uncertainties throughout parameters space sample by $\textbf{\textit{p}}$, and we will refer to the random uncertainty estimated using MCMC as $\sigma_{\mathrm{rand},\log R_{e}}$. 
\begin{figure*}
	\centering
	\includegraphics[width=1\textwidth]{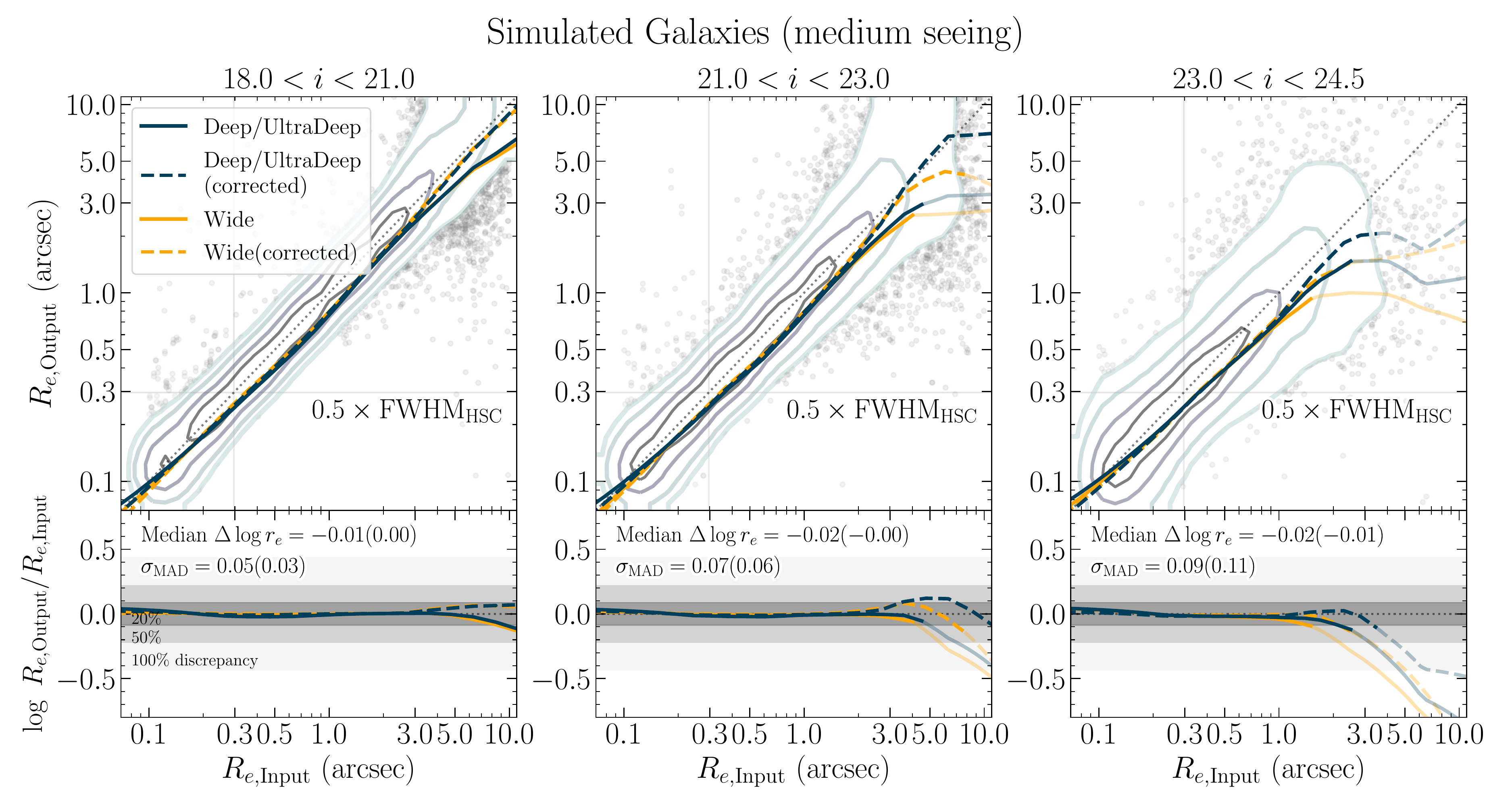}
	\caption{The comparison between the intrinsic effective radii ($R_{e,\mathrm{Input}}$) for the set of simulated galaxies convolved with the medium PSF ($0{\farcs}6$), and those recovered by \textsc{Lenstronomy} ($R_{e,\mathrm{Output}}$) in three magnitude bins: $18<i<21$, $21<i<23$, and $23<i<24.5$ (from left to right). The contours indicate the distribution from $1\sigma$ to $2.5\sigma$ (with a spacing of $0.5\sigma$) after applying correction for systematic biases (see Appendix~\ref{appendix:sizecorr}). The datapoints and contours are the results from simulated galaxies in all three layers of the HSC (Wide, Deep+UltraDeep).  The running median for simulated galaxies in the Deep+UltraDeep and Wide layers are indicated as dark blue and yellow curves, respectively. The measurements before and after and after applying the correction for the systematic biases are indicated as solid and dashed curves, respectively. The median and scatter of the offset between $R_{e,\mathrm{Input}}$ and $R_{e,\mathrm{Output}}$ (combined three survey layers) before applying the correction for the systematic biases are indicated in the bottom panel, whereas those after the correction are in parentheses.  The grey regions in the bottom row show 20\%, 50\%, and 100\% discrepancies. The curve of running median with lighter shaded color indicates the size discrepancy exceeding 20\%. Furthermore, the vertical and horizontal solid lines in the top rows indicate median HWHM of PSF size. On average, the median size difference for all galaxies with $18 < i < 24.5$ is $\lesssim0.02$ dex (better than $5\%$)  and the scatter is $\sigma_{\mathrm{MAD}}=0.03-0.1$ dex. }
	\label{fig:comparesize_sim}
\end{figure*}
\begin{figure}
	\centering
	\includegraphics[width=0.8\textwidth]{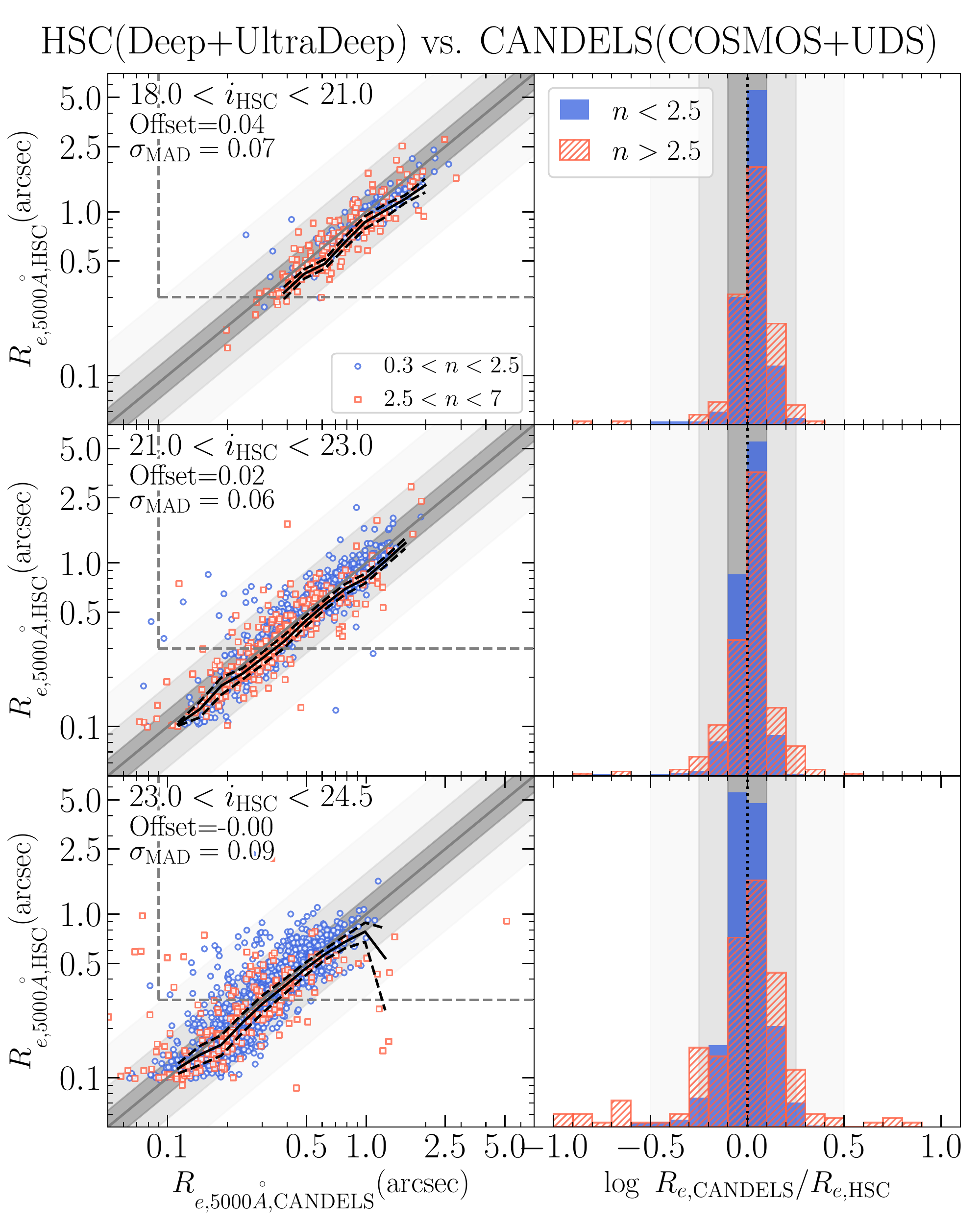}
	\caption{Left: the comparison of effective radii measured in the WFC3/F160W ($H_{160}$) bandpass taken from \cite{vanderWel2012} and those measured in the HSC $i-$ band for galaxies in three magnitude bins: $18<i<21$, $21<i<23$, and $23<i<24.5$ (from top to bottom).  Both of the size measurements are corrected for color gradients to a common rest-frame wavelength of 5000~\AA~(see Section~\ref{sec:colorgrad}). Orange and blue points show galaxies with $0.3<n<2.5$ and $2.5<z<7.0$, the grey regions show 20\%, 50\%, and 100\% discrepancies, and the black lines show running median with scatter (dashed black lines). Furthermore, the dashed horizontal (vertical) line shows median HWHM of PSF size for the HSC (CANDELS). Right: the normalized histograms of $\log~R_{e,\mathrm{CANDELS}} /R_{e,\mathrm{HSC}}$ for galaxies with $n<2.5$ (filled blue) and $n>2.5$ (hatched orange). The median offset and scatter ($\sigma_{\mathrm{MAD}}$) of $\log~R_{e,\mathrm{CANDELS}} /R_{e,\mathrm{HSC}}$ for all galaxies ($0.3<n< 7)$ in each magnitude bin are indicated in the left panel. On average, the median size difference for all galaxies with $18 < i < 24.5$ is $0.01-0.04$ dex or $<10\%$, with galaxies in HSC smaller than in CANDELS, and the scatter is $<0.1$ dex.}
	\label{fig:comparesize_candels}
\end{figure}

\subsection{ Systematic Uncertainty Estimated using Simulated Galaxies}

\label{appendix:verification_sizes}
\par \edittwo{The measurement of galaxy morphology is subjected to biases. For instance, small and compact galaxies could be affected by the PSF (leading to an overestimation of $R_{e}$), while large and extended galaxies suffer surface brightness dimming in the outskirts \cite[leading to an underestimation of $R_{e}$; e.g.,][]{Bernstein2002a,Bernstein2002b,Benitez2004,Haussler2007,Cameron2007,Carollo2013,Cibinel2013}. Therefore, it is crucial to investigate possible biases and correct for them using simulated galaxies with known light distributions.} 
\par To do that, we follow the method as described in \cite{Carollo2013}, and we refer the reader to this paper for additional details \citep[and see also][]{Cibinel2013,Tarsitano2018}. In brief,  we use \textsc{Lenstronomy} to construct a \edittwo{2D} single-component S\'{e}rsic model on a grid of points in magnitude, size, S\'{e}rsic index, and (projected) axis ratio parameter space: $\log (R_{e}/\mathrm{arcsec})=[-1,-0.5,0, 0.5, 1]$, $m=[18.5, 19.2, 19.8, 20.5, 21.1, 21.8, 22.4, 23.1, 23.7, 24.4,25]$, $\log n=[-0.52,-0.18,0.16,0.50,0.85]$, and $q=[0,0.33,0.66,1]$. For each grid node, we randomly generate simulated galaxies to have: radii and S\'{e}rsic indices within $\pm30\%$ of the nominal radius and index $n$ at the grid point, magnitude within $\pm0.25$ mag and difference in axis ratio equal to $\pm0.05$. We convolve all model galaxies described above with three PSF sizes, corresponding to the best ($0{\farcs}5$), medium ($0{\farcs}6$), and worse ($0{\farcs}9$) HSC seeing in the $i$ band. In total, this results in $\sim100,000$ galaxy models per each seeing condition. For each set of the seeing condition, we convolve a model galaxy with randomly selected HSC PSF images with that seeing condition. This allows us to mimic the noise in the reconstruction of the PSF in the real data, when performing the convolution. We then insert a PSF-convolved model galaxy into randomly chosen empty regions extracted from the real HSC imaging. We repeat this procedure for a set of simulated galaxies in three layers of the HSC (Wide and Deep+UltraDeep), separately. Finally, we run \textsc{Lenstronomy} on our set of simulated galaxies using the same method as for the HSC galaxy sample. We find that simulated galaxies with input magnitude of $18<i<24.5$ in HSC Wide and Deep+UltraDeep layers with reasonable fitting results (the difference between model and observed light profile is less than 0.05 mag pixel$^{-2}$, and the scatter of the difference is less than 0.03 mag pixel$^{-2}$) are accounted for $60\%-70\%$ of the original simulated galaxy samples, consistent with the results from real HSC galaxies (Section~\ref{sec:lenstronomy}).

\par Figure~\ref{fig:comparesize_sim} shows the comparison between the intrinsic effective radii for the set of simulated galaxies convolved with the medium PSF and those recovered by \textsc{Lenstronomy} in three magnitude bins: $18<i<21$, $21<i<23$, and $23<i<24.5$. For all sets of simulated galaxies in three layers of the HSC, our comparison shows that we have a small systematic offset between the recovered and the intrinsic sizes, $\log(R_{e,\mathrm{Output}}/R_{e,\mathrm{Input}})$. \editone{This offset correlates with the magnitude and is negligible ($\lesssim0.01$ dex) for galaxies brighter than $i=23$, and does not strongly depend on the depth of the HSC, with the exception for fainter galaxies or those having intrinsic sizes larger than $\sim5\arcsec$ (corresponding to $\sim17$~kpc at $z=0.2$ and $\sim41$~kpc at $z=1.0$).} On average, the median $\log(R_{e,\mathrm{Output}}/R_{e,\mathrm{Input}})$ for all galaxies with $18 < i < 24.5$ is $\lesssim0.02$ dex ($5\%$ level accuracy or better). The scatters in $\log(R_{e,\mathrm{Output}}/R_{e,\mathrm{Input}})$ increases with increasing magnitude: $\sigma_{\mathrm{MAD}}=0.03-0.05$ for bright galaxies ($18<i<23$) to $\sigma_{\mathrm{MAD}}=0.1$ for fainter galaxies. \editone{Moreover, for simulated galaxies in the faintest magnitude bin ($23<i<24.5$) and with intrinsic sizes larger than $\sim2\arcsec$ (corresponding to $\sim7$~kpc at $z=0.2$ and $\sim16$~kpc at $z=1.0$), the recovered sizes tend to be underestimated, which is likely due to surface brightness dimming in the outskirts of these galaxies. This trend is more pronounced for simulated galaxies in the Wide layer (yellow solid curve in Figure~\ref{fig:comparesize_sim}). For example,  the median size offset for simulated galaxies with $23.0<i<24.5$ and $R_{e,\mathrm{Input}}\sim2\arcsec-3\arcsec$ is $|\log(R_{e,\mathrm{Output}}/R_{e,\mathrm{Input}})|\sim0.3-0.4$ dex, whereas the offset is smaller for the simulated galaxies in the Deep+UltraDeep layer ($\sim0.1-0.2$~dex; blue solid curve in the Figure). In the following subsection, we will account for these systematic biases using the simulated galaxies.}


\par We make two cautionary remarks: first, we generate simulated galaxies to populate \emph{uniformly} a broad grid in the five-dimensional $m - R_{e} - n - q-\mathrm{PSF}$ space. Real galaxies are not uniformly distributed in this parameter space. This introduces the risks that regions of parameter space with large systematic effects are unjustifiably ignored by design. Second, we deliberately ignore dust or stellar population segregation effects, as our models are constructed smooth and neglect dust attenuation. 

\par As a further check,  we measure structural parameters of objects in the HSC-Deep+UltraDeep footprint and compare them to the measurements of the same objects (cross-matched within $3\arcsec$) in the CANDELS field (COSMOS and UDS fields) by \cite{vanderWel2012}.  As the $HST-$based galaxy sizes are measured in the WFC3/F160W ($H_{160}$) bandpass, whereas our sizes are measured in the HSC $i-$band, we begin with correcting both size measurements to a common rest-frame wavelength of 5000\AA~(see Section~\ref{sec:colorgrad}). Figure~\ref{fig:comparesize_candels} shows the comparison of sizes and the histograms of log-ratio ($\log~R_{e,\mathrm{CANDELS}} /R_{e,\mathrm{HSC}}$; right panel) for galaxies with S\'{e}rsic indices $n<2.5$ and $n>2.5$ in three bins of magnitude: $18<i<21$, $21<i<23$, and $23<i<24.5$.  On average, the median size difference for all galaxies with $18 < i < 24.5$ is $\log~R_{e,\mathrm{CANDELS}} /R_{e,\mathrm{HSC}}=0.01-0.04$ dex, and the scatter is $0.06-0.1$ dex. These offset are such that galaxies in HSC systematically smaller than in CANDELS. \editone{The size offset ($\log~R_{e,\mathrm{CANDELS}} /R_{e,\mathrm{HSC}}$) and its scatter increase for galaxies at fainter magnitudes and higher S\'{e}rsic indices.}



\subsubsection{Correcting for Biases in Size Measurements}
\label{appendix:sizecorr}
Although the systematic offset between the recovered and input sizes is negligible ($\lesssim0.01$ dex) for bright ($i<23$) galaxies, it can be more significant for fainter magnitudes and larger galaxies (up to 0.4 dex). We therefore derive the corrections required for the structural parameters within the multidimensional parameter space ($m_{\mathrm{tot}},R_{e},n,q,\mathrm{PSF}$) using the same set of simulated galaxies as described in Appendix~\ref{appendix:verification_sizes}, in an identical fasion as in \cite{Carollo2013} and \cite{Tarsitano2018}. For our galaxy sample with $i<24.5$ mag, the corrections in effective radius and S\'{e}rsic index are on average less than $5\%$ and $10\%$, respectively. 
\par \editone{After we apply the correction for systematic biases to the size measurements, we again compare the corrected $\log(R_{e,\mathrm{Output}})$ and $\log(R_{e,\mathrm{Input}})$. Figure~\ref{fig:comparesize_sim} shows that the median offsets between $\log(R_{e,\mathrm{Output}})$ and corrected $\log(R_{e,\mathrm{Input}})$ for simulated galaxies in both Wide and Deep+UltraDeep layers are significantly reduced, particularly for those with large intrinsic sizes. For instance,
in bin of $21<i<23$, the median $|\log(R_{e,\mathrm{Output}}/R_{e,\mathrm{Input}})|$ for simulated galaxies in the Wide layer are reduced from $0.2-0.5$~dex to $\lesssim0.1$~dex ($25\%$ level accuracy or better) for those with intrinsic sizes of $R_{e,\mathrm{input}}=3\arcsec-8\arcsec$. For the same level of accuracy, we are able to recover sizes of fainter ($23<i<24.5$) simulated galaxies with intrinsic sizes smaller than $3\arcsec$ for the Wide layer and smaller than $4\arcsec$ for the Deep+UltraDeep layer. }


\par \editone{On the other hand, even after the correction, sizes of intrinsically large galaxies with $21<i<23$ and $R_{e,\mathrm{input}}\gtrsim8\arcsec$ are still systematically underestimated by
$\sim0.2-0.3$~dex, particularly for the Wide layer. At fainter magnitudes ($23<i<24.5$), sizes of simulated galaxies with $5\arcsec<R_{e,\mathrm{Input}}<10\arcsec$ are systematically underestimated by
$\sim0.2-0.5$~dex for the Deep+UltraDeep layer, and this offset can be up to $\sim0.5$~dex in the Wide layer. Even though our HSC galaxies, which are fainter than $i=21$ and having sizes larger than $\sim5\arcsec$, account for only about $\sim0.3$\% of the total galaxy sample, the systematic biases due to surface brightness dimming in the outskirts of galaxies could impact our results, particularly for massive star-forming galaxies (Section~\ref{subsec:mediansize_evol}). Regardless, we argue that our method to correct for the systematic bias enables us to recover sizes with 5\% level accuracy or better for most of the HSC galaxies in both Wide and Deep+UltraDeep layers over the magnitude range of $18<i<24.5$.  }




\par Finally, even we have already utilized a set of simulated galaxies to correct our size measurements for various systematic biases, including PSF-blurring and the different depths of the HSC, some level of systematic uncertainties could still remain as we discussed above. To further account for the uncertainties, we define the scatter in $\log(R_{e,\mathrm{Output}}/R_{e,\mathrm{Input}})$ (using corrected $R_{e,\mathrm{Output}}$) as the systematic uncertainty of a galaxy size, $\sigma_{\mathrm{sys},\log R_{e}}$. We measure $\sigma_{\mathrm{sys},\log R_{e}}$ in bins of $i-$band magnitude, fit the second-order polynomial, and obtain,  
\begin{equation}
    \sigma_{\mathrm{sys},\log R_{e}}=0.0051 i^{2} -0.1996 i + 1.986  
\end{equation}
\noindent The total uncertainty, $\sigma_{\log R_{e},j}$, on each source $j$ is then given by
\begin{equation}
 \sigma_{\log R_{e},j}^{2}=  \sigma_{\mathrm{sys},\log R_{e},j}^{2}+ \sigma_{\mathrm{rand},\log R_{e},j}^{2}
 \label{eq:total_re_err}
\end{equation}
where $\sigma_{\mathrm{rand},\log R_{e},j}$ is the random uncertainty estimated using MCMC for a subset of our galaxy sample (Appendix~\ref{appendix:mcmc}). We then incorporate the total uncertainty when we fit the size-mass relation and the redshift evolution of size. We have also verified that there is no significant dependence of $\log(R_{e,\mathrm{Output}}/R_{e,\mathrm{Input}})$ on other input $n$ and $q$: for simulated galaxies with $18<i<24.5$, $|\log(R_{e,\mathrm{Output}}/R_{e,\mathrm{Input}})|$ is nearly constant and consistent with  $\lesssim0.1$ dex over the range of $n$ and $q$ we probed in this work. 
\label{fig:size_beforeafter_corr}

\section{Quiescent and Star-forming Galaxies Cross-Contaminating Fraction}
\label{appendix:misclass}
\par In this paper, we classify our galaxy sample into quiescent and star-forming populations based on their rest-frame SDSS $u-r$ versus $r-z$ ($urz$ selection). Here we aim to make a comparison between $urz$ selection and $UVJ$ selection.  As we have discussed in Section~\ref{sec:urz_selection} that our HSC photometry does not allow us to robustly derive the rest-frame $J-$band magnitude, we therefore cross-matched our sample of HSC galaxies in COSMOS field (the Deep+UtraDeep layer) to UltraVISTA catalog \citep{Muzzin2013} within 3\arcsec. \edittwo{Similarly, we perform cross-matching between our sample in SXDS field (the Deep+UtraDeep layer) and AEGIS field (the Wide layer) to UKIDSS UDS catalog (Almaini et al. in prep.) and the NMBS catalog \citep{Whitaker2011}, respectively. This results in 82,745 galaxies at $0.2<z<1.0$}, allows us to obtain more robust rest-frame $J-$ band magnitudes from these external catalogs, and hence classify a subsample of HSC galaxies into quiescent and star-forming using $UVJ$ selection, independent of the method used in this paper. To do that, we follow \cite{Williams2009} to separate the star-forming and quiescent galaxies. 

\par To make a comparison between $urz$ selection and $UVJ$ selection, for a sample of quiescent galaxies at a given redshift and stellar mass bin, we calculate \edittwo{the fraction of $urz$ quiescent galaxies that are classified as $UVJ$ star-forming galaxies and refer to this as the contaminating fraction of $urz$ quiescent galaxies ($f_{\mathrm{cont,Q}}$).  On the other hand, we calculate the fraction of $UVJ$ quiescent galaxies that are classified as $urz$ quiescent galaxies and refer to this as the recovering fraction of $urz$ quiescent galaxies ($f_{\mathrm{recov,Q}}$). Similarly, we calculate the fraction of $urz$ star-forming galaxies that are classified as $UVJ$ quiescent and refer to this as $f_{\mathrm{cont,SF}}$, while the fraction of $UVJ$ star-forming galaxies that are classified as $urz$ star-forming galaxies are referred to as $f_{\mathrm{recov,SF}}$.} By doing these, we are assuming that the $UVJ$ selection and the rest-frame colors from UltraVISTA, UKIDSS UDS, and \edittwo{NMBS} datasets provide a cleaner separation of both populations \citep{Muzzin2013,Whitaker2011}.
\begin{figure*}
	\centering
	\includegraphics[width=0.9\textwidth]{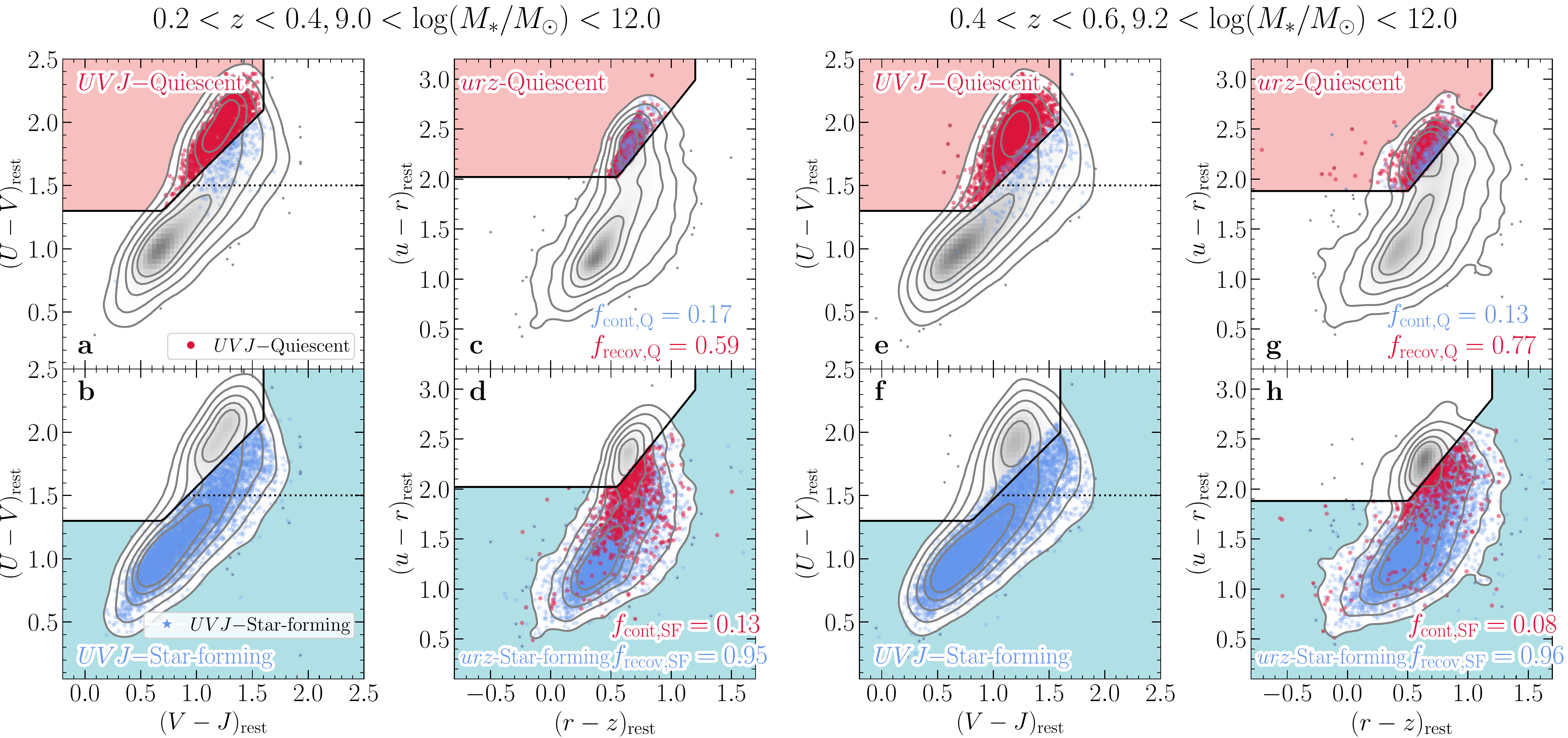}
	\caption{The comparison between the classification of quiescent and star-forming galaxies  at $0.2<z<0.4$ ((a) to (d)), and at $0.5<z<1.0$ ((e) to (h)) based on the $(U-V)_{\mathrm{rest}}$ and $(V-J)_{\mathrm{rest}}$ colors ($UVJ$ diagram; (a), (b), (e), and (f)) and those based on the $(u-r)_{\mathrm{rest}}$ versus $(r-z)_{\mathrm{rest}}$ colors ($urz$ diagram; (c), (d), (g), and (h)). The galaxy sample are from COSMOS and SXDS fields of the HSC Deep+UltraDeep and AEGIS field of the HSC Wide with stellar masses above the completeness limit for quiescent galaxies in each redshift bin. The  $(U-V)_{\mathrm{rest}}$ and $(V-J)_{\mathrm{rest}}$ are from UltraVISTA \citep{Muzzin2013}, UKIDSS UDS (Almaini et al., in prep.), and the NMBS \citep{Whitaker2011} catalogs. The grey contours indicated the distribution of rest-frame colors for all galaxies from $1\sigma$ to $3\sigma$, with spacing of $0.5\sigma$.   Galaxies with $(U-V)_{\mathrm{rest}}$ and $(V-J)_{\mathrm{rest}}$ colors within the selection boundary (black solid line; \cite{Williams2009}) of quiescent and star-forming are shown in dark red circles and blue stars, respectively. The dotted horizontal line in $UVJ$ diagram indicates the separation of star-forming galaxies into unobscured star-forming ($(U-V)_{\mathrm{rest}}<1.5$) and dusty star-forming galaxies ($(U-V)_{\mathrm{rest}}>1.5$)
	\citep{Fumagalli2014}. 	Galaxies located in the upper left region of the $urz$ selection boundary as indicated by shaded region in (c), (d), (g), and (h) are classified as $urz$ quiescent, whereas those outside the region are $urz$ star-forming galaxies (see Figure~\ref{fig:urz_calib}). \edittwo{In each $urz$ diagram, we indicate the contaminating fraction ($f_{\mathrm{cont}}$) and recovering fraction ($f_{\mathrm{recov}}$) of  quiescent and star-forming galaxies (see text).}}
	\label{fig:urz_selection_vs_uvj}
\end{figure*}

\begin{figure*}
	\centering
	\includegraphics[width=0.9\textwidth]{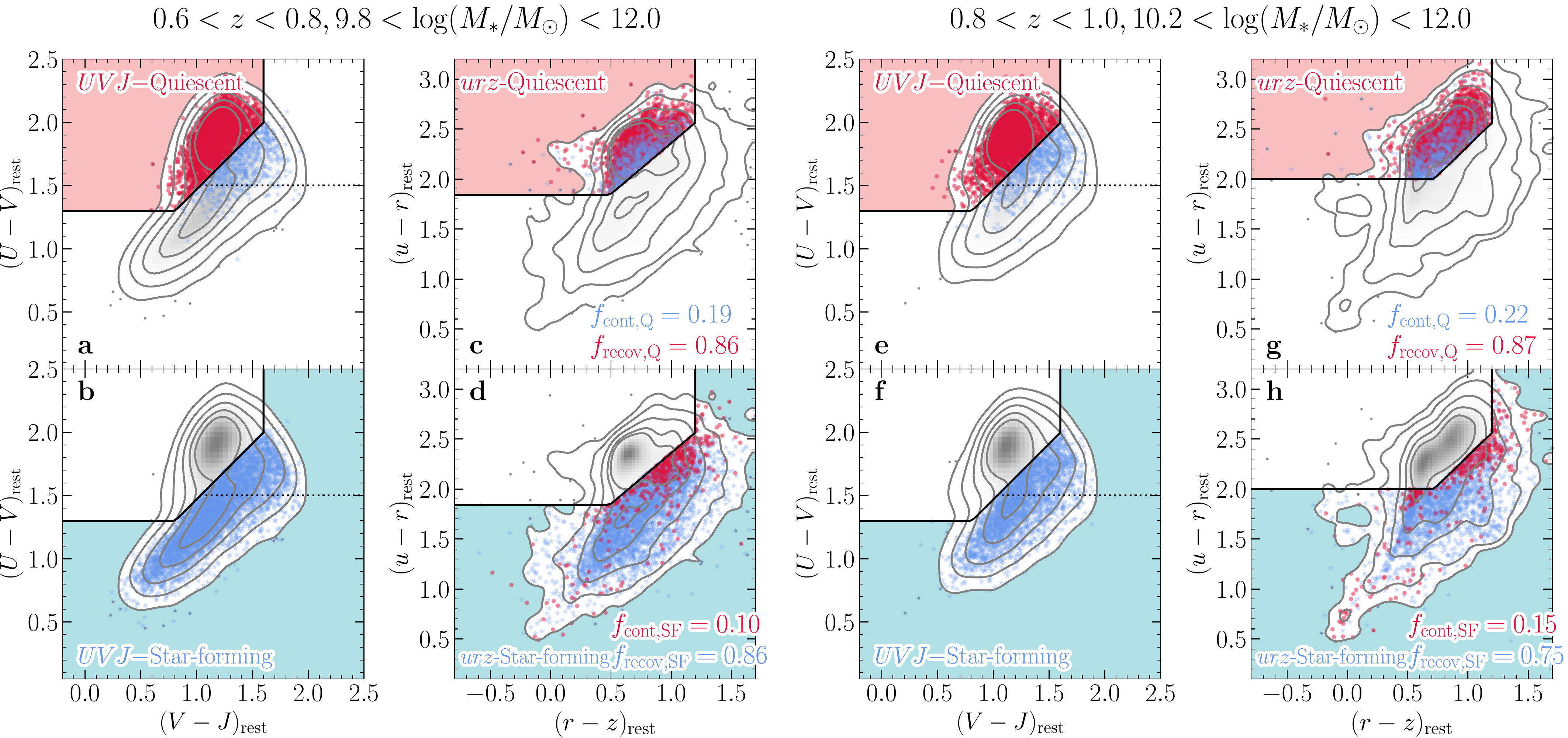}
	\caption{Same as Figure~\ref{fig:urz_selection_vs_uvj} but for those galaxies  at $0.6<z<0.8$ ((a) to (d)), and at $0.8<z<1.0$ ((e) to (h)).}
	\label{fig:urz_selection_vs_uvj2}
\end{figure*}

\begin{figure}
	\centering
	\includegraphics[width=0.6\textwidth]{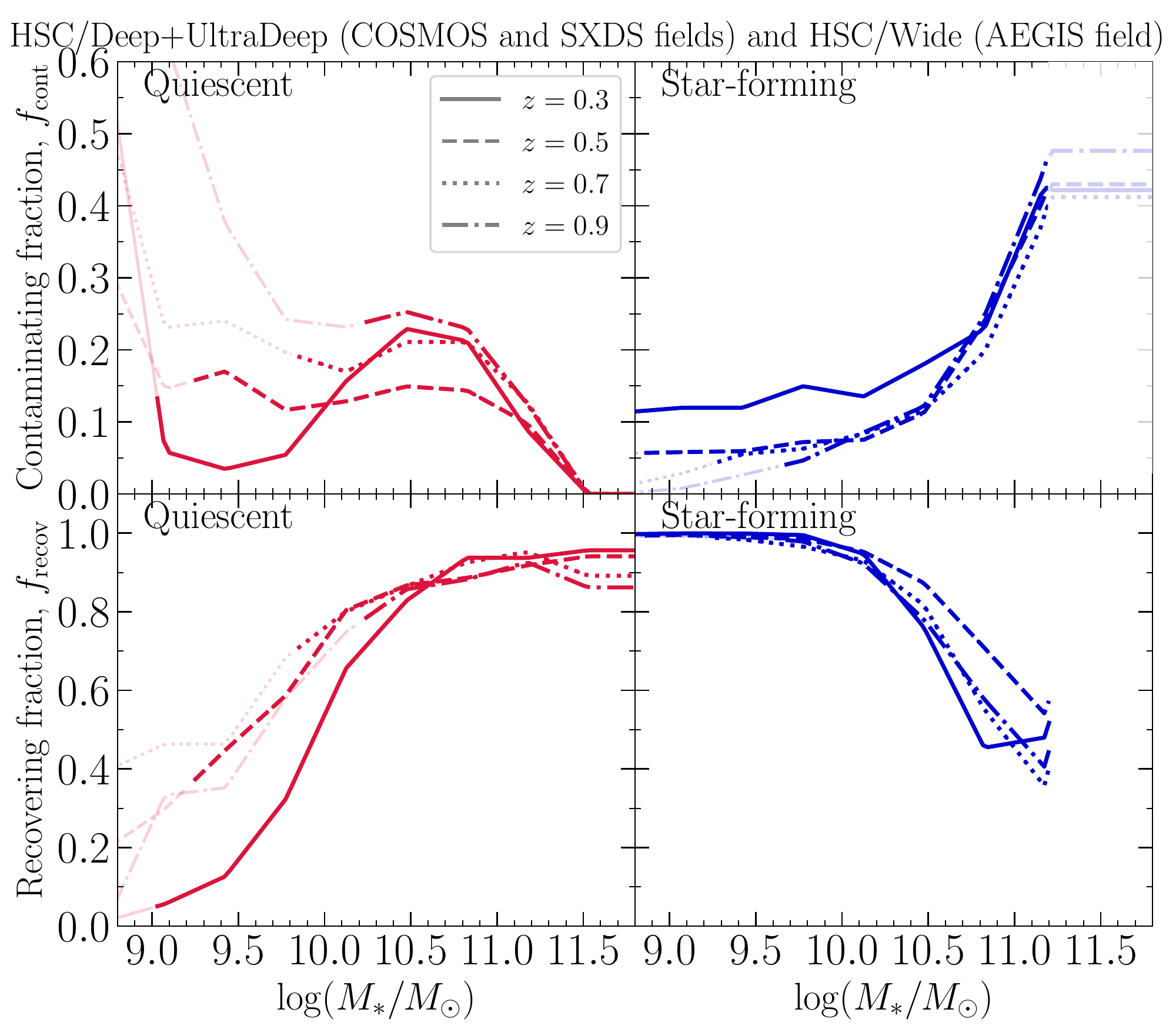}
	\caption{Top: the contaminating fraction, $f_{\mathrm{cont}}$, for quiescent (left panel; $f_{\mathrm{cont,Q}}$) and star-forming (right panel; $f_{\mathrm{cont,SF}}$) galaxy sample as a function of stellar mass in four redshift bins, as labeled in the Figure. The selection of quiescent and star-forming populations is based $urz$ selection. \edittwo{$f_{\mathrm{cont,Q}}$ is defined as the fraction of $urz$ quiescent galaxies that are classified as $UVJ$ star-forming galaxies, while $f_{\mathrm{cont,SF}}$ is defined as the fraction of $urz$ star-forming galaxies that are classified as $UVJ$ quiescent.} The $f_{\mathrm{cont}}$ for galaxies with stellar masses above the mass completeness limit at each redshift bin is indicated as darker color, whereas $f_{\mathrm{cont}}$ for those below the limit is indicated as lighter color. We take the cross-contamination between galaxy populations into account by incorporating $f_{\mathrm{cont,Q}}$ and $f_{\mathrm{cont,SF}}$  into the fitting of the size-mass distributions (Section~\ref{sec:fitting_sizemass}). Due to very small number ($N\lesssim10$) of  $UVJ$ star-forming galaxies with $\log(M_{\ast}/M_{\odot})\gtrsim11.2$, we are unable to robustly estimate $f_{\mathrm{cont}}$ for star-forming galaxies at these $M_{\ast}$, and we therefore assume the $f_{\mathrm{cont}}$ for these massive star-forming galaxies to the value at lower $M_{\ast}$. Bottom: similar to the top panel but for the recovering fraction for both populations. \edittwo{$f_{\mathrm{recov,Q}}$ is defined as the fraction of $UVJ$ quiescent galaxies that are classified as $urz$ quiescent galaxies, while $f_{\mathrm{recov,SF}}$ is defined as the fraction of $UVJ$ star-forming galaxies that are classified as $urz$ star-forming galaxies.}}
	\label{fig:contamfrac_vs_mass}
\end{figure}

\par Figure~\ref{fig:urz_selection_vs_uvj} shows the location of our quiescent and star-forming galaxies at $0.2<z<0.4$ and at $0.4 < z < 0.6$ (with stellar masses above the completeness of quiescent galaxies at a give redshift) on the $UVJ$ diagram and the $urz$ diagram. We also show the result for galaxies at $0.6<z<0.8$ and $0.8<z<1.0$ in Figure~\ref{fig:urz_selection_vs_uvj2}. Over these redshift ranges and for galaxies with stellar masses above the completeness limit at a given redshift, we demonstrate that our calibrated $urz$ selection (see Section~\ref{sec:urz_selection}) provides a clean separation of quiescent and star-forming galaxies similar to the $UVJ$ selection -- we recover $60\%-90\%$ of $UVJ$ quiescent galaxies, while around $10\%-20$\% of $urz$ quiescent galaxies are $UVJ$ star-forming galaxies. Most of these contaminants also have $(U-V)_{\mathrm{rest}}>1.5$, consistent with being dusty star-forming galaxies according to the selection criteria of \cite{Fumagalli2014}. On the other hand, over the same stellar mass and redshift ranges, we recover around 75\%-95\% of star-forming galaxies with $8\%-15$\% of contamination from $UVJ$ quiescent galaxies 
\par Finally, we calculate the contaminating fraction ($f_{\mathrm{cont}}$) and recovering fraction ($f_{\mathrm{recov}}$) for both quiescent and star-forming galaxies in bins of stellar mass ($\Delta \log M_{\ast}=0.35$~dex) and in four redshift bins: $0.2<z<0.4$, $0.4<z<0.6$, $0.6<z<0.8$, $0.8<z<1.0$. At all redshifts and stellar mass bins, Figure~\ref{fig:contamfrac_vs_mass} shows that the  $f_{\mathrm{cont}}$ for quiescent galaxies with stellar mass above the completeness limit of each redshift bin is roughly below $25\%$ (top panel of this Figure). For star-forming galaxies with $\log(M_{\ast}/M_{\odot})<11$, the $f_{\mathrm{cont}}$ is below $20\%$ but strongly increases with stellar mass to $\sim40\%-50\%$ for more massive galaxies. However, for the most massive star-forming galaxies with $\log(M_{\ast}/M_{\odot})\gtrsim11$, the number of $UVJ$ star-forming galaxies from COSMOS, SXDS, and AEGIS fields are very small ($N\lesssim10)$, and we are unable to robustly estimate $f_{\mathrm{cont}}$ for star-forming galaxies at these $M_{\ast}$. We therefore assume the $f_{\mathrm{cont}}$ for these massive star-forming galaxies to the value at lower $M_{\ast}$. The bottom panel of Figure~\ref{fig:contamfrac_vs_mass} showed that the $f_{\mathrm{recov}}$ quiescent  galaxies increases with increasing stellar mass, in contrast to the star-forming population. This is expected as the massive end of the galaxy stellar mass function is almost entirely comprised of quiescent galaxies, whereas star-forming galaxies vastly dominates quiescent galaxies at the low-mass end \citep[e.g.,][]{Blanton2009,Peng2010,Moustakas2013, Kawinwanichakij2020}.

\section{The Impact of the Underestimation of Stellar Masses of Star-forming Galaxies on their Size-Mass Relations}
\label{appendix:discuss_masscorr}
\par Growing observational evidence has shown that stellar masses of star-forming galaxies derived from integrated (spatially-unresolved) photometry are systematically lower than the total stellar masses derived from summing individual pixels from the stellar mass map (spatially-resolved) \citep[e.g.,][]{Zibetti2009, Sorba2015,MarinezGarcia2017,Sorba2018}. \cite{Sorba2015} presented spatially-resolved, pixel-by-pixel SED fitting on nearby galaxies  and compared the stellar mass estimates from spatially-resolved model fits to those from spatially-unresolved, integrated light. The authors showed that the unresolved mass estimates for star-forming galaxies tend to be underestimated by $13-25$ percent ($0.06-0.12$ dex), and the difference between the unresolved and resolved mass estimates is strongly correlate with the galaxy's specific star formation rate (sSFR). These authors concluded that this effect results from the outshining effect in the SED of galaxies, i.e., light from massive, young stars dominates a galaxy's SED with significant higher luminosities than the older stellar populations, particularly at optical wavelengths. As a result, the SED fitting procedures preferentially favour matching the large amount of flux coming from the younger stellar population and missing the relatively low amount of flux coming from the older stellar populations (which dominate the stellar mass content of a galaxy). This could potentially underestimate the mass-to-light ratio ($M_{\ast}/L$) \citep[see][ for further discussions on outshining and its effects]{Sawicki1998,Papovich2001,Maraston2010,Pforr2012}.  

\par Unfortunately, the SED fitting using HSC five-band optical photometry does not provide us a reliable estimate of sSFR. Therefore, to obtain an estimate of the bias in the stellar mass estimates due to outshining for our sample of star-forming galaxies, we use the best-fit of the evolution of star-forming galaxy main sequence (MS) from \cite{Speagle2014} (their Equation 28) to estimate typical sSFRs. We then use these sSFRs to derive the correction for unresolved stellar mass estimates by using the empirical relation from \cite{Sorba2018} (their Equation 6). From $z=0.2$ to $z=1.0$, we find the median corrections for star-forming galaxies of $d\log(M_{\ast}/M_{\odot})=0.08-0.11$ dex. To further quantify the effect of the underestimation on stellar masses of star-forming galaxies on the size-mass relations, we apply this correction to stellar mass estimates from \textsc{Mizuki} for the star-forming galaxy sample and measure the size-mass relation. 
\par After we applied the corrections, the slopes of size-mass relations for star-forming galaxies above the pivot mass $M_{p}$ are $\beta=0.56,0.34,0.23,0.26$ for galaxies at $z=0.3,0.5,0.7,0.9$, which are shallower than the slopes without the correction ($\beta=0.61, 0.38,0.28,0.29$). On the other hand, we find no significant change in the power-law slope $\alpha$ for star-forming galaxies below $M_{p}$ and the other parameters of a smoothly broken power-law ($r_{p}$, $M_{p}$, and $\sigma_{\log~r}$). \edittwo{In summary, the underestimation in stellar masses of star-forming galaxies due to the outshining effects does not alter our main conclusion. Throughout this paper, we use stellar mass estimates from \textsc{Mizuki} without any correction. }

\bibliography{references}
\end{document}